\documentclass[12pt]{oujdathese}
\usepackage{amsmath}
\usepackage{amssymb}
\usepackage{amsfonts}
\usepackage{graphicx}
\usepackage[latin1]{inputenc}
\usepackage[francais]{babel}
\usepackage[T1]{fontenc}
\usepackage{slashbox}
\usepackage{epsfig}
\begin{document}
\title{RECHERCHE DE CORR\'{E}LATIONS TEMPORELLES DES MUONS COSMIQUES AVEC MACRO \\
ET PERTE D'\'{E}NERGIE DES NUCL\'{E}ARITES}
\author{MOUSSA Abdelilah}
\ordre{145/09} 
\discipline{Physique}%
\specialite{{Physique Théorique et Physique des Particules}}%
\date{27/06/2009}%
\president{M J. E. DERKAOUI }{Pr. Faculté des Sciences d'Oujda}%
\jurun{M M. BEN EL MOSTAFA}{Pr. Faculté des Sciences d'Oujda}%
\jurdeux{M D. BENCHEKROUN}{Pr. Faculté des Sciences Ain Chok, Casablanca}%
\jurtrois{M H. DEKHISSI}{Pr. Faculté des Sciences d'Oujda}%
\jurquatre{M A. FAHLI}{Pr. Faculté des Sciences Ben M'sik, Casablanca}%
\jurcinq{M G. GIACOMELLI}{Pr. Université de Bologna, Italie}%
\jursix{M F. MAAROUFI}{Pr. Faculté des Sciences d'Oujda}%
\maketitle
%
\begin{center}
{\LARGE {Remerciement}}
\end{center}
\par
J'aimerai tout d'abord remercier Mr. Hassan Dekhissi, mon directeur de thèse. Son enthousiasme, sa patience et sa disponibilité ainsi que sa gentillesse ont fortement contribué à mener à bien mon travail et à faire des mes années de thèse une période agréable et enrichissante.\par 
Mes sincères remerciements vont aux membres de jury. Mr. J. E. Derkaoui m'a fait l'honneur de présider le jury, je le remercie également pour son aide et ses critiques constructives. Ma gratitude va à Mr. M. BEN EL MOSTAFA, Mr. A. FAHLI et à Mr. D. BENCHEKROUN, qui ont accepté d'être rapporteurs de cette thèse. Merci à Mme F. Maaroufi, qui a également accepté de siéger parmi ce jury, pour ses encouragements et ses conseils. \par 
Je souhaite remercier le professeur G. Giacomelli de m'avoir accueilli dans son groupe pour plusieurs périodes de stage. Ma reconnaissance à tous les membres des collaborations MACRO et SLIM de Bologna : L. Patrizii, S. Cecchini, M. Giorgini, V. Popa, V. Togo et plus particulièrement M. Sioli pour son aide et son encouragement.\par 
C'est avec beaucoup de plaisir que je souhaite remercier à présent tous les membres du LPTPM d'Oujda pour l'aide et la sympathie qu'ils m'ont témoigné durant ces années. \par 
L'INFN et le département de physique de Bologna m'ont apporté des aides financières durant mes séjours en Italie, que leurs membres trouvent ici l'expression de ma gratitude. \par 
Ma reconnaissance et ma sympathie à tous mes amis qui ont contribué de prés ou de loin pour la réalisation de ce modeste travail.\par 
Enfin, ma profonde reconnaissance à tous les membres de ma famille, qui m'ont apporté un soutien continu durant mes années de thèse. 

\pagenumbering{roman} \setcounter{page}{1}
\bigskip
 \pagebreak
\tableofcontents%
\pagebreak
\pagenumbering{arabic}
\pagestyle{myheadings}
\linespread{1.6}

\addcontentsline{toc}{chapter}{Introduction}
\chapter*{Introduction}	
La physique des particules est apparue pour traiter l'infiniment petit et chercher à comprendre l'organisation de la matière à son niveau le plus fondamentale. L'astrophysique à son tour s'ouvre à l'infiniment grand et tente de percer les mystères de l'organisation de l'univers dans son ensemble, de même, la cosmologie s'intéresse à l'infiniment ancien en essayant de reconstituer l'histoire de l'univers depuis ses premiers instants.\\ 
Autour de ces trois filières s'est développée une nouvelle discipline, l'astroparticule. Un domaine vaste et fortement interdisciplinaire, ayant pour objectif principale l'étude de toutes particules ou rayonnement venant de l'espace, la mise en évidence et la compréhension des phénomènes cosmiques de haute énergie. Il s'agit d'étudier (ou découvrir) des objets au sein desquels ont lieu des processus non thermiques, capables d'émettre des particules (notamment des rayons cosmiques (RC)) de haute énergie au dessus du GeV.\par 
En effet la radiation cosmique, mis en évidence par Victor Hess en 1912, est une composante principale de la galaxie, vu que sa densité d'énergie est comparable, si elle n'est pas supérieure, à celle des autres radiations présentes dans l'univers.\\ 
Les expériences souterraines jouent un rôle important dans l'étude des différentes caractéristiques des RC telles que la composition chimique, les mécanismes d'accélérations, l'astronomie des rayons gamma et des neutrinos en plus de la recherche des phénomènes rares (désintégration du proton, monopôles magnétiques, nucléarites, etc...).\par
L'expérience MACRO, située en Italie, constituait l'une des plus grandes expériences souterraines. Elle a permis de couvrir un vaste domaine de recherche dont on cite: la recherche des particules rares tels que les monopôles magnétiques \cite{MAC23} et les nucléarites \cite{nuc16}, l'étude des neutrinos atmosphériques \cite{INTRO1}\cite{INTRO1r} et de leur oscillations \cite{INTRO2}, l'étude des muons de haute énergie du bas \cite{INTRO3} \cite{dtr11} et l'étude de la composante pénétrante des RC.\par
Parmi les questions de recherche qui présentent un intérêt fondamental, on distingue l'étude de l'anisotropie temporelle du flux de muons cosmiques. Des modulations dans les distributions temporelles du temps d'arrivée de muons ont été observées par quelques expériences. 
Deux d'entre elles ont mis en évidence la présence d'un signal modulé avec une période de 4.8 h, exactement égale à la période orbitale de Cyg X-3 dans la région des X du spectre. D'autres ont observé des modulations dans les distributions du temps d'arrivée des muons dues à la présence d'une composante qui n'est pas aléatoire. Les résultats de ces expériences n'ont pas été confirmés par d'autres groupes et le problème reste confus.\par
Dans ce travail on se propose d'étudier les variations du flux des muons cosmiques, et ceci en utilisant les données collectées par le détecteur souterrain MACRO. Nous disposons d'un grand nombre de muons (environ 38 millions d'événements) détectés durant une période de plus de 6 ans.\par
Cette thèse est organisée en 5 chapitres. Dans le premier, nous passons sur un rappel sur les  rayons cosmiques, les sources et les mécanismes d'accélérations, leurs interactions dans l'atmosphère ainsi que la perte d'énergie des muons dans la roche. Dans le second, nous allons donner une vue sur quelques expériences souterraines et notamment l'expérience MACRO. \\
Le troisième chapitre sera consacré à l'étude des distributions des temps d'arrivée des muons cosmiques avec une énergie plus grande que 1.3 TeV au sommet de la montagne de Gran Sasso et collecté durant la période allant de 1995 jusqu'à 2000. Un grand échantillon de muons a été utilisé comparé à celui utilisé dans les travaux ultérieurs de MACRO. Cette étude nous permet de chercher l'existence de modulations dans le flux de muons cosmiques et de conclure sur la nature de la distribution des temps d'arrivée des RC afin de se situer par rapport aux résultats obtenus par les expériences. 

Dans le quatrième chapitre, nous présentons les résultats de l'étude des variations du flux de muons en utilisant deux approches complémentaires : 
\begin{itemize}
\item Recherche de clusters d'événements avec la méthode Scan statistics.
\item Recherche de variations périodiques avec la méthode de Lomb-Scargle.
\end{itemize}
La dernière partie de ce travail concerne l'étude de la perte d'énergie des nucléarites dans différents milieux, notamment dans l'atmosphère, afin d'évaluer la contribution de l'atmosphère dans le processus de perte d'énergie.

\chapter{Généralités sur les rayons cosmiques (RC)}
\section {Introduction}
\label{chap1}
Le rayonnement cosmique demeure un des problèmes centraux de l'astrophysique, suscitant de nombreuses questions bien souvent interconnectées. Son étude permettra d'établir des liens très riches entre les différentes parties des sciences de l'univers, touchant notamment à l'étude du milieu interstellaire, des champs magnétiques ou de l'interaction des particules avec les plasmas, à la modélisation des sources énergétiques telles que les noyaux actifs de galaxie ou les sursauts gamma, à l'astronomie gamma, la formation des étoiles, à la nucléosynthése spallative, etc. En effectuant plusieurs vols en ballon avec des électroscopes, le scientifique Victor Hess apporta la preuve, en 1912, que ce rayonnement vient de l'espace. Ce rayonnement fut surnommé "rayonnement cosmique" par Robert Millikan en 1925. C'est grâce à cette découverte que la physique des particules expérimentale prit son essor et son étude a conduit entre 1930 et 1950 à la découverte du positron, du muon, des mésons $\pi$ et $k$ ainsi que les baryons $\Sigma$, $\Lambda$ et $\Xi$.\\
En 1938, Pierre Auger et ses collaborateurs observèrent des coïncidences en temps entre plusieurs détecteurs de particules séparés de plusieurs mètres. Auger émis alors l'hypothèse selon laquelle les particules détectées au sol étaient les constituants de gerbes atmosphériques initiées par des rayons cosmiques dont l'énergie peut aller au delà de $10^{15}$ eV.\\
Aujourd'hui, les observations d'évènements autour de $10^{20}$ eV restent encore une énigme. Une partie de la compréhension des rayons cosmiques d'ultra haute énergie passe par l'étude de leurs interactions dans l'espace interstellaire et l'atmosphère.\\
Majoritairement constitué de particules chargées, le rayonnement cosmique frappe l'atmosphère terrestre de manière continue. Le flux de particules chargées qui heurtent l'atmosphère peut 
atteindre 1000 particules/$m^2$s.\\
Les rayons cosmiques peuvent être classifiés en trois types.\par 
\begin{itemize}
	\item Les RC primaires chargés, sont constitués d'environ 90$\%$ de protons, 9$\%$ de particules $\alpha$ et le reste par des 	noyaux atomiques tels que (C, O, N, Mg, Fe,...). Toutefois, cette composition varie avec l'énergie, par exemple, à des énergies de l'ordre de 400 TeV \footnote{1 GeV$=10^9$eV, 1 TeV$=10^{12}$eV, 1 PeV$=10^{15}$eV et 1 EeV$=10^{18}$eV.} on a environ
	12$\%$ de protons, 25$\%$ de particules $\alpha$, 26$\%$ de CNO, 15$\%$ de Si-S et 21$\%$ de noyaux avec Z$>$17.
	\item Les RC secondaires, sont constitués de mésons $\pi$, K et des p,n, etc..., résultant de l'interaction des primaires chargés avec des noyaux atomiques de l'atmosphère terrestre à une altitude moyenne de l'ordre de 20 km.
	\item Les RC tertiaires, sont constitués de muons, neutrinos, photons, gerbes électromagnétiques et des électrons résultant de la désintégration des mésons chargés et neutres des RC secondaires.
\end{itemize}
Du point de vue énergétique les rayons cosmiques représentent une composante fondamentale de l'univers. Leur densité énergétique est approximativement 1 $eV/cm^{3}$ comparée à la densité d'énergie de la lumière stellaire qui est de l'ordre de $0.6$ eV$/cm^{3}$. On dispose alors, d'un véritable échantillon de la matière extraterrestre. Si aux plus basses énergies, la propagation des rayons cosmiques chargés dans les champs magnétiques va uniformiser leurs directions d'arrivée, rendant ce rayonnement isotrope autour de la terre, aux plus hautes énergies, ceux-ci vont avoir une rigidité suffisante pour permettre de "pointer" leur source. Cette isotropie de la direction d'arrivée rend difficile la localisation des sources et par conséquent il est difficile d'établir l'origine galactique ou extragalactique des rayons cosmiques primaires. L'absence de toute anisotropie dans la direction de provenance peut être due à une distribution isotrope des sources, aux effets de propagation dans l'espace interstellaire ou à un manque de statistique. En effet, en traversant l'espace entre la source et la terre, les RC peuvent interagir avec les photons du fond cosmologique et le gaz interstellaire et peuvent subir également des déviations provoquées par les champs magnétiques galactiques et extragalactiques.\\
La problématique des rayons cosmiques se présente à la fois sous des aspects théoriques, phénoménologiques et observationnels. Des travaux mettent l'accent sur l'accélération des rayons cosmiques, tandis que d'autres se focalisent sur leur propagation, ou sur leurs interactions avec le milieu environnant, depuis leurs sources jusqu'à l'atmosphère terrestre.\\
La compréhension des phénomènes à l'origine de la production des rayons cosmiques, les mécanismes d'accélération et de propagation dans le milieu interstellaire et intergalactique, nécessite la connaissance de leur composition chimique, du spectre énergétique, de la distribution du temps et de la direction d'arrivée.\\
Peu de questions trouvent leurs réponses sur l'origine des RC, cependant on sait clairement que ceux de haute énergie proviennent de l'extérieur du système solaire. On ne connaît pas encore ni les sources responsables de l'émission des RC ni les mécanismes qui favorisent leur accélération à de très hautes énergies ($>10^{17}$ eV).
\section {Spectre énergétique des RC}
\begin{figure}[t!]
\begin{center}
\leavevmode
\epsfig{file=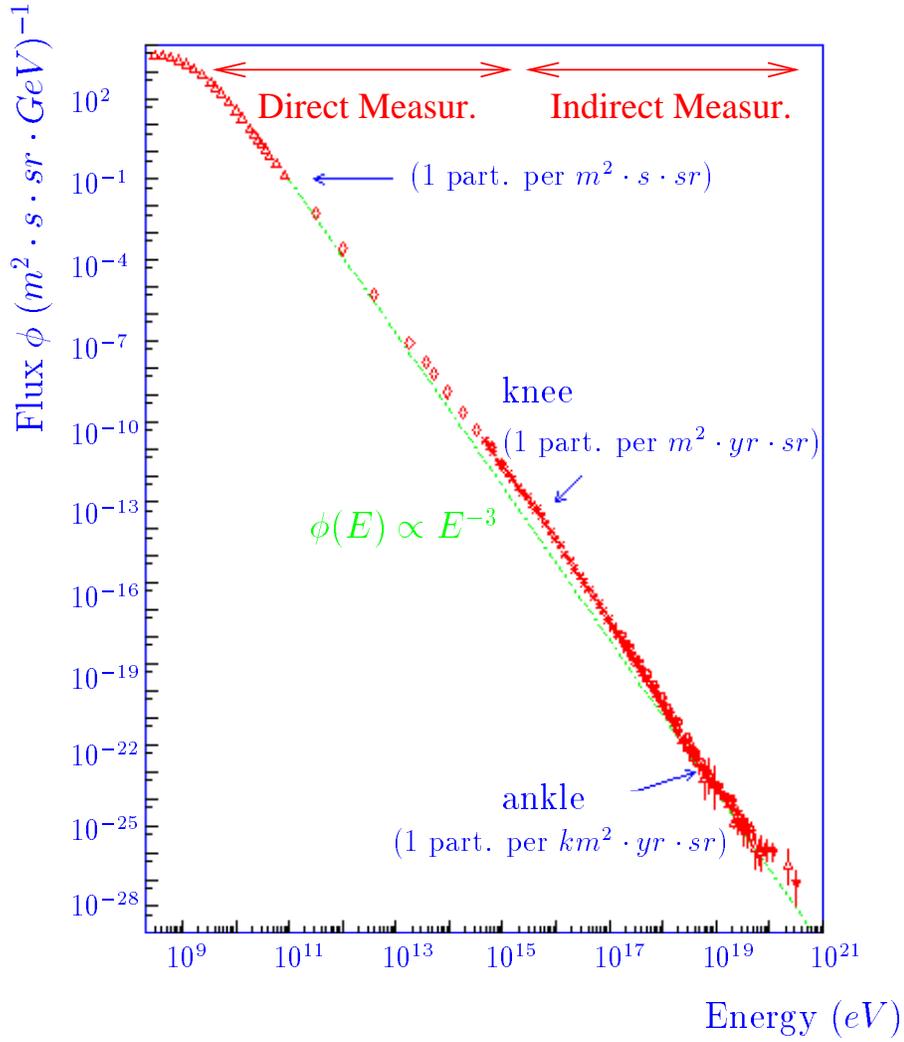,width=24cm}
\vskip -8cm
\caption{\em \textbf{Spectre des rayons cosmiques (flux en fonction de l'énergie). En \mbox{dessus} sont indiquées approximativement les régions couvertes par les mesures directes et indirectes.}}
\label{fig:RC1}
\end{center}
\end{figure}

Plus de 90 années d'observation des rayons cosmiques ont permis de construire leur spectre en énergie (figure (\ref{fig:RC1})) qui s'étend de manière remarquablement continue sur environ 13 décades en énergie (entre $\sim 10^8eV$ et $\sim 10^{21}$) et 32 décades en flux (entre quelques milliers de particules par m$^2$ et par seconde aux basses énergies et de l'ordre d'une particule par km$^2$ et par siècle pour les plus énergétiques).\\
Cet intervalle étant très large, il parait clairement que pour le couvrir expérimentalement, l'utilisation de diverses techniques de détection est exigée, du petit détecteur tel que la chambre à émulsion dont l'acceptance est de l'ordre de $0.1$ $m^{2}$ aux grands détecteurs dont la surface peut atteindre des $km^{2}$. Tout cela est également accompagné de nombreux problèmes que ce soit au niveau de l'analyse des données (normalisation des résultats produits par les différents détecteurs) ou au niveau de la technique expérimentale (détection des particules de très hautes énergies).\\
En effet le flux différentiel peut s'exprimer comme une simple loi de puissance \cite{RC1}:
\begin{equation}
\frac{dN}{dE}\propto E^{-\gamma}
\label{spectre}
\end{equation}
\noindent o\`u E est l'énergie par nucléon de la radiation, $\gamma $ est l'indice spectral qui dépend de l'énergie et N est l'intensité des RC.\\
En effet, le spectre n'est pas un pur spectre de puissance, c'est à dire l'exposant $\gamma$ dépend de l'énergie. Pourtant, dans une très large bande d'énergie il s'avère que l'équation (\ref{spectre}) donne une bonne approximation du spectre.
Comme on peut le voir sur la figure (\ref{fig:RC1}) il existe plusieurs régimes de ce spectre à savoir: 
\begin{itemize}
	\item En dessous de quelques MeV le flux est dominé par les particules solaires. Celles-ci constituent le vent solaire qui influence les rayons cosmiques d'énergie allant jusqu'à quelques GeV par le biais de la modulation solaire. 
	\item Entre $\sim$ 0.1 GeV et 100 TeV l'indice spectral est $\gamma\sim 2.7$. Ceci est assez bien expliqué par les mécanismes d'accélération classiques. Il faut cependant tenir compte du fait que jusqu'à une énergie de quelques GeV, la loi de puissance est modifiée par la modulation solaire.
	\item Entre 10$^{14}$ eV et 10$^{18}$ eV l'indice spectral devient $\gamma\sim 3.2$. La zone située autour de 3-5$\times$10$^{15}$ où à lieu le changement de la pente est appelée genou (Knee). Dans cette gamme d'énergies les flux sont tellement faibles (de l'ordre d'une particule par m$^2$ et par an) que l'on ne peut plus faire les mesures de manière directe. On doit alors observer les cascades atmosphériques. 
	\item Une autre rupture de pente apparaît à 10$^{18}$ eV communément appelée la cheville (ankle). Celle ci marque l'entrée dans le domaine des rayons cosmiques d'ultra haute énergie pour lesquels l'indice spectral se radoucit et devient $\gamma\sim 2.8$. Ici les flux sont tellement faibles (de l'ordre de 1 particule par km$^2$ et par siècle) que les résultats expérimentaux sont rares. Ils ne permettent pas à l'heure actuelle d'éclaircir le mystère de l'origine de telles particules. En effet il parait irréaliste de produire de telles énergies dans les mécanismes d'accélération classiques. Une solution pourrait venir d'objets extragalactiques comme les noyaux actifs de galaxies. Cependant, l'interaction de particules avec une telle énergie avec les photons du fond cosmologique (CMB) devrait provoquer une perte d'énergie par photoproduction de pions. C'est la coupure appelée GZK \cite{RC2}. 
L'expérience AGASA \cite{RC4} semble voir une anisotropie et un excès d'évènements au dessus de la coupure GZK par rapport à une source extragalactique, ce qui n'est pas le cas du détecteur HiRes \cite{RC5}. En tenant compte des erreurs systématiques importantes, leurs mesures restent cependant compatibles et ne fournissent donc pas d'indice significatif. Pour commencer à lever la voile il faudra attendre d'avoir une bonne statistique avec l'expérience AUGER \cite{RC6}.
\end{itemize}
Ces changements peuvent être dus à des modifications substantielles dans les mécanismes de production et d'accélération des RC de haute énergie, ou être liés à la propagation du rayonnement cosmique dans l'espace. Cette dernière idée a été développée par le 
modèle de diffusion des RC dans la galaxie dit "{\it leaky box}" \cite{RC6b}. Selon ce modèle, la galaxie est considérée comme un réservoir confinant des RC qui permet à ces derniers d'avoir des probabilités non nulles de s'enfuir. En effet, les particules produites dans notre galaxie traversent le milieu interstellaire avant d'atteindre la terre. Ce milieu est constitué de 
nuages de gaz neutre et ionisé et il est caractérisé par un champ magnétique galactique désordonné B ($\approx 3\times 10^{-6}$ gauss). Une particule de charge Z qui traverse ces nuages magnétiques subit des déflections et aura un rayon de courbure,
\begin{equation} 
 R[cm]=E[eV]/300\cdot B[gauss]Z
\end{equation}
Lorsque l'énergie augmente, le rayon augmente, ce qui signifie que beaucoup de protons (Z=1) atteignent des rayons supérieurs au rayon de la galaxie elle même. Cet effet tend à les faire
disparaître dans l'espace intergalactique et ils ne peuvent plus par conséquent contribuer au flux de RC. Naturellement il y a d'autres particules qui arrivent d'autres galaxies.\par
\begin{figure}
\centering
\includegraphics[height=18cm,width=14cm]{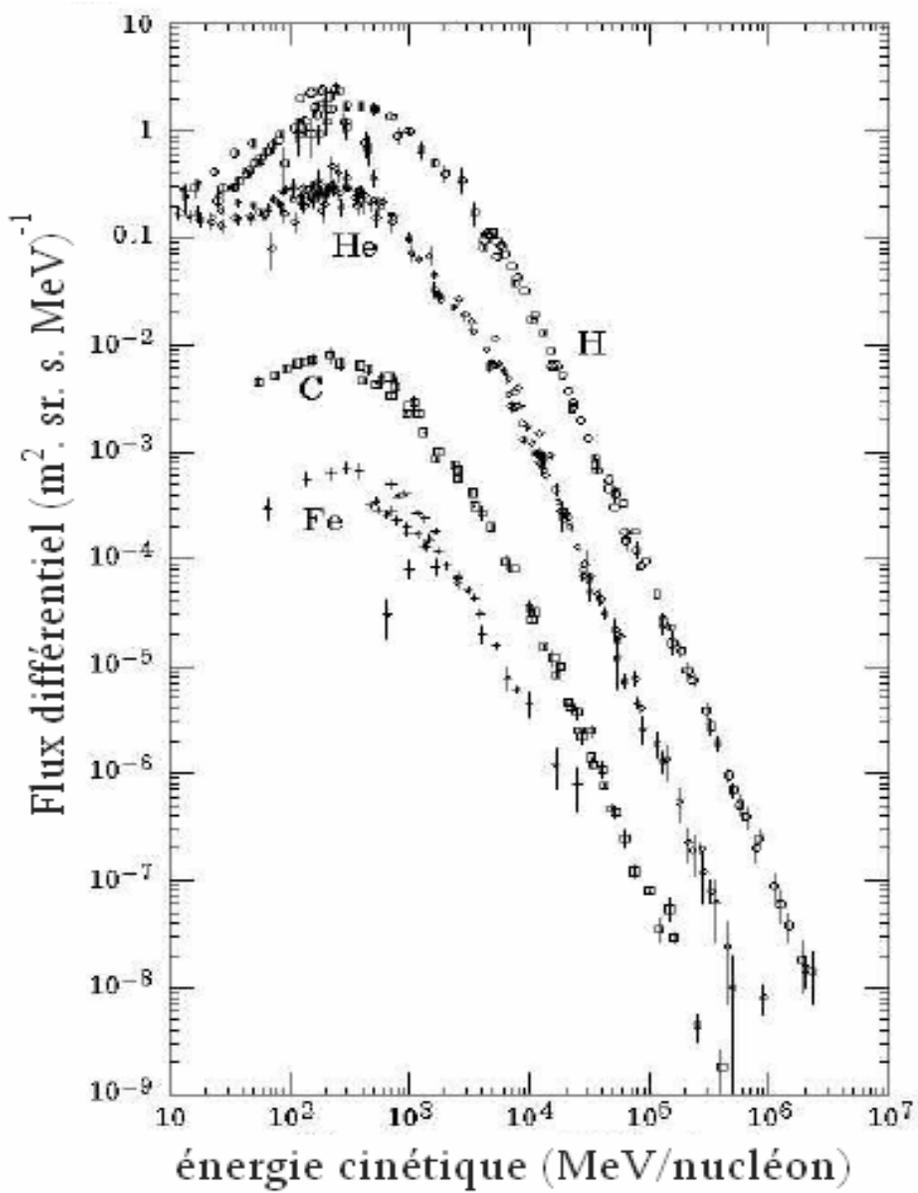}
\vskip -1cm
\caption{\em \textbf{Spectre énergétique individuel de l'hydrogène, l'hélium, du carbone et du fer.}}
\label{fig:RC2}
\end{figure}

A leur tour, les éléments chimiques simples qui constituent les RC suivent l'allure de la fonction exponentielle. Sur la figure (\ref{fig:RC2})  \cite{RC7}, on présente le spectre énergétique individuel des noyaux présents dans les RC et qui peut être décrit également par une loi exponentielle ${dN_i/ {dE}} = KE^{-\gamma_i}$ avec $\gamma _i$ et K qui changent selon
que l'énergie soit supérieure ou non à l'énergie du genou.
\section{Composition chimique des RC primaires}
\begin{figure}
\centering
\includegraphics[height=14cm,width=12cm]{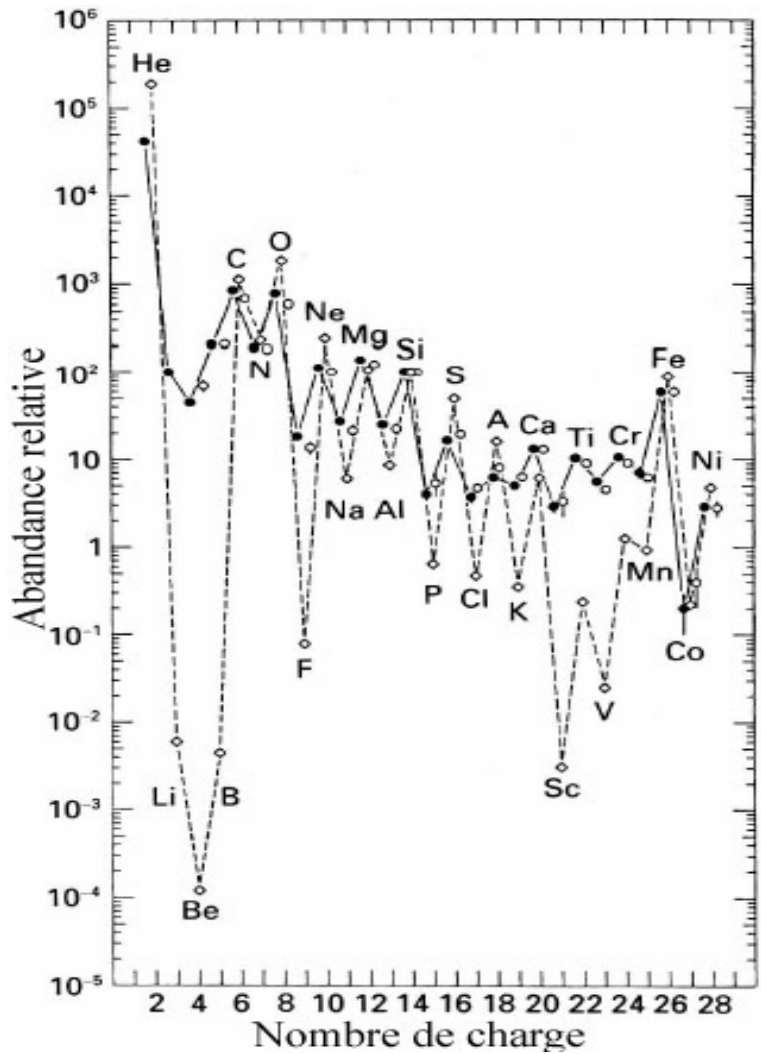}
\caption{\em \textbf{Abondance relative des éléments dans le rayonnement cosmique (ligne solide) et dans le système solaire (ligne en pointillée) normalisé à 100 pour le Si.}}
\label{fig:RC3}
\end{figure}

L'étude de l'abondance relative des éléments du système solaire (basée sur l'analyse des matériaux terrestres et des météorites et les mesures spectroscopiques des étoiles tel que le soleil) et celle des isotopes présents dans les rayons cosmiques permet d'obtenir des informations importantes sur la nature des sources émettrices et sur le mode de propagation de ces particules dans le milieu interstellaire. L'abondance chimique des éléments dans les RC est connue au moins jusqu'à quelques dizaines de TeV/nucléon à partir des observations perforées au niveau de la haute atmosphère à l'aide des ballons, des satellites,...). On peut voir sur la figure (\ref{fig:RC3}) les abondances relatives des différentes espèces dans le rayonnement cosmique comparées à celles du système solaire \cite{RC7}; les cercles pleins sont relatifs aux données à basse énergie (70 - 280) MeV/A, les cercles vides sont relatifs aux données des rayons cosmiques à haute énergie (1000 - 2000) MeV/A et les losanges pour l'abondance isotopique dans le système solaire \cite{RC9}. 
\\On observe une similitude entre les deux compositions chimiques à quelques anomalies notoires pour certains noyaux. Des différences apparaissent cependant pour les éléments immédiatement plus légers que le fer. Le même phénomène touche également les éléments (Li, Be et B) pour lesquels les abondances cosmiques sont beaucoup plus importantes. Ce phénomène peut être interprété comme l'existence de réactions de spallation dans le milieu interstellaire. En effet, la fragmentation d'éléments comme le carbone, l'azote ou l'oxygène (CNO) sur des protons du milieu interstellaire produit une forte quantité d'éléments plus légers comme le Lithium, le Béryllium et le Bore enrichissant leurs portions dans le rayonnement cosmique. La comparaison des abondances entre le CNO et le Lithium, le Béryllium et le Bore nous renseigne sur la quantité de la matière interstellaire traversées.\\
Il existe aussi un déficit de l'hydrogène et de l'Hélium dans le rayonnement cosmique par rapport à l'abondance stellaire. Ceci peut être du à la difficulté d'ioniser ces éléments, les rendant moins disponibles pour l'accélération. On constate aussi la présence de l'effet d'appariement dans les deux compositions. En effet les noyaux pair-pair sont en général stables et plus abondants.\\
Certaines de ces différences peuvent trouver des explications dans les modèles de propagations des RC dans le milieu interstellaire. Avant d'atteindre la terre, certains noyaux subissent la fragmentation et produisent des noyaux plus légers. Ce qu' on observe localement est constitué 
d'une composante primaire provenant directement de la source et d'une composante secondaire résultant de la fragmentation de certains noyaux après interaction avec d'autres particules durant le passage dans le milieu interstellaire. De ce fait, les éléments (Li, Be, B) et (Sc, Ti, V, Cr, Mn) peuvent résulter de la fragmentation du (C, O) et Fe respectivement. Cette caractéristique témoigne du rôle important de la transformation de la composition chimique des RC lors de leur propagation à travers l'espace interstellaire et peut-être au sein des sources (dans la région de la production ou de l'accélération des rayons cosmiques).
\section{Direction d'arrivée}
\begin{figure}[h]
\centering
\includegraphics[height=9cm,width=14cm]{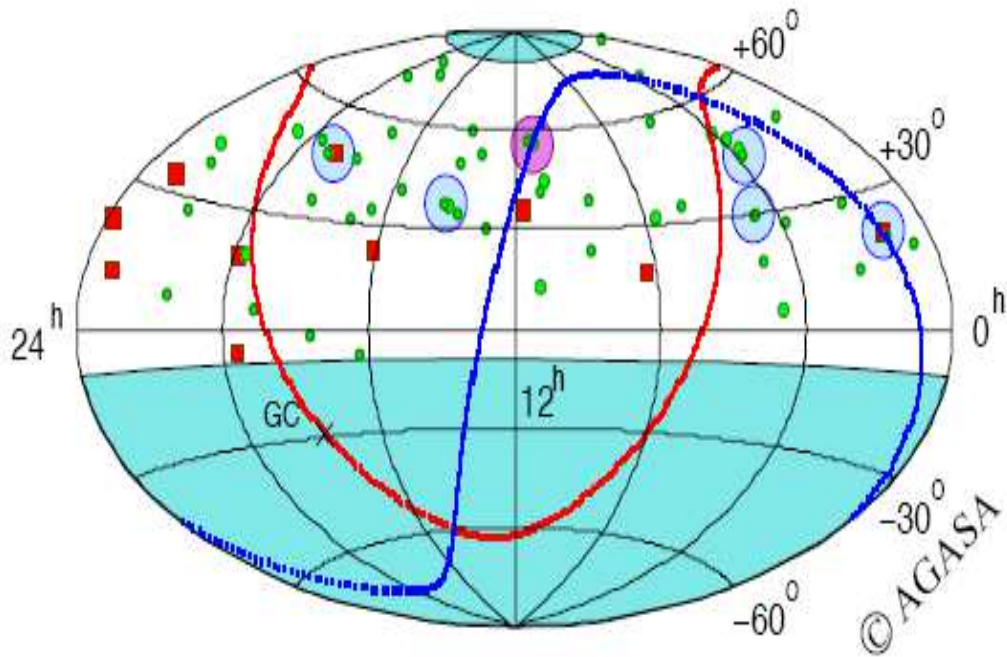}
\caption{\em \textbf{Directions d'arrivée, en coordonnées équatoriales, des rayons cosmiques au-dessus de 4 10$^{19}$eV enregistrés par AGASA. Les carrés rouges et les ronds verts représentent les rayons cosmiques ayant des énergies > 10$^{20}$ et (4-10) 10$^{19}$eV respectivement. Les cercles plus grands indiquent les multiples (triplets en mauve). Ces évènements ont été enregistrés par le réseau d'AGASA de 20 km$^2$ entre 17 février 1990 et le 31 juillet 2001, les angles zénithaux sont inférieurs à 45$^\circ$.}}
\label{fig:RC5}
\end{figure}

A faible énergie, les rayons cosmiques subissent des déflexions de leurs trajectoires causées par les champs magnétiques galactiques et extragalactiques.\\
Au dessus de 10$^{19}$eV les rayons cosmiques sont peu déviés de leur trajectoire par les champs magnétiques galactiques et extragalactiques (de l'ordre de quelques degrés), leur direction d'arrivée devrait donc pointer vers la source.\\
Autour de 10$^{18}$eV, les expériences AGASA et Fly's Eye ont reporté une anisotropie en provenance du centre galactique avec un déficit correspondant dans la direction opposé \cite{RC15}\cite{RC16}. Au delà de 10$^{19}$eV, les résultats des différentes expériences semblent s'accorder sur une répartition isotrope des rayons cosmiques. AGASA a notamment détecté une telle répartition avec la présence de trois doublets et d'un triplets d'évènements dans des cercles de diamètres 2.5$^\circ$. Aucun objet astrophysique n'a été corrélé avec la direction de ces multiplets (figure (\ref{fig:RC5}))\cite{RC17}.
\section{Sources et mécanismes d'accélération des RC}
La densité d'énergie totale des RC mesurés au-dessous de l'atmosphère est dominée par les particules avec des énergies entre 1 et 10 GeV. A des énergies au-dessous de $\sim 1 GeV$ les intensités sont temporellement corrélées avec l'activité  solaire, ce qui est une évidence directe pour une origine solaire. A des hautes énergies le flux observé montre une anti-corrélation temporelle avec une activité solaire, indiquant une origine à l'extérieur du système solaire. Plusieurs arguments tels que la composition, la production des rayons cosmiques secondaires suggèrent que la quantité de RC entre 1GeV et au moins jusqu'à la région du genou est confinées dans la galaxie et qui est probablement produite dans les restes de supernova (SNRs). Entre le genou et la cheville la situation est moins claire. Finalement, au delà de $\sim 10EeV$, les RC sont généralement supposés être d'origine extragalactique, ceci est dû à leur isotropie apparente. \par
\subsection{Mécanismes d'accélération des RC chargés}
\begin{figure}
\centering
\vskip -1cm
\includegraphics[height=8cm,width=8cm]{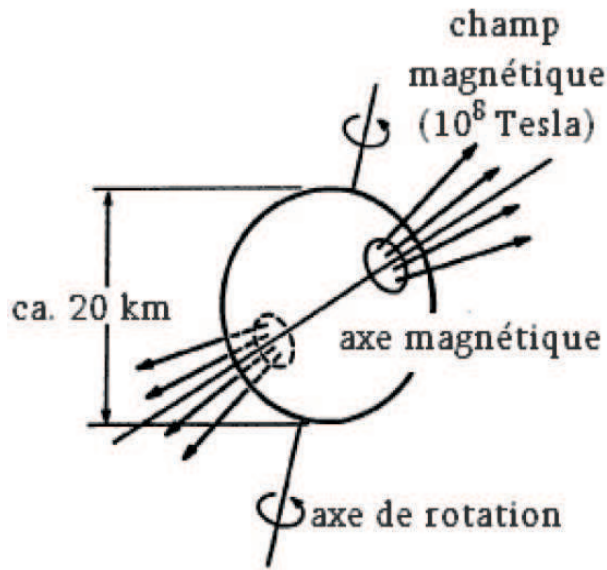}
\caption{\em \textbf{Émission d'un pulsar (modèle du rotateur oblique).}}
\label{RC6}
\end{figure}

De nombreux modèles tentent d'expliquer l'observation des rayons cosmiques au delà de 10$^{15}$ eV en introduisant des mécanismes d'accélération qui permettent aux particules d'atteindre des 
énergies très élevées.\\
La forme lisse du spectre évoquée plus haut laisse penser qu'il existe un nombre réduit de
types de sources. Il existe aujourd'hui un scénario assez communément accepté pour rendre
compte des caractéristiques du rayonnement cosmique jusqu'en dessous du genou. Les supernovae
jouent ici un rôle prépondérant. Celles-ci représentent un moyen simple d'éjecter des
particules relativistes dans le milieu interstellaire. Les abondances relatives dans le rayonnement cosmique correspondraient ainsi de manière assez naturelle aux abondances stellaires.\\
Maintenant que l'on dispose d'une source, il faut accélérer ces particules de manière à obtenir le spectre observé. E. Fermi proposa en 1949 \cite{RC10} un mécanisme où les particules diffusées de manière stochastique sur les irrégularités magnétiques d'un nuage de gaz en mouvement à une vitesse $V$ par rapport à la source gagnent statistiquement une énergie proportionnelle à $(V/c)^2$ à chaque collision \cite{RC11}, où c est la vitesse de la lumière. C'est ce qu'on appelle le mécanisme de Fermi du second ordre. L'intérêt est que cela conduit naturellement à un spectre différentiel en loi de puissance. Étant donné que les vitesses très faibles des nuages de gaz interstellaire et la faible densité de ceux-ci, l'efficacité du mécanisme est insuffisante. Quand on considère le même phénomène, mais dans des ondes de choc, on obtient un gain d'énergie en $V/c$ où $V$ est maintenant la vitesse du front de l'onde de choc. C'est le mécanisme de Fermi du premier ordre. Celui-ci conduit naturellement à une loi de puissance avec un indice spectral de -2. Ces ondes de choc apparaissent naturellement lorsque la matière éjectée par une supernova rencontre le gaz environnant.\\
De plus, les supernovae sont assez fréquentes et énergétiques pour maintenir la densité locale
d'énergie contenue dans le rayonnement cosmique. Cette dernière peut être obtenue à partir
du flux observé et est approximativement $\rho_E = 1eV/cm^3$. Si l'on considère un temps de confinement des particules dans la galaxie $\tau=\sim 10^7$ ans on obtient une puissance d'injection nécessaire: 
\begin{equation}
Q=\int_{V_{galaxie}} \frac{\rho_E}{\tau}dV\sim 5.5^{40} erg.s^{-1}
\label{eq:1}
\end{equation}
\noindent où $V_{galaxie}$ est le volume de la galaxie. Compte tenu de l'estimation de la puissance fournie par les ondes de choc de supernova, cela ne requiert qu'une efficacité de quelques pour-cents pour le mécanisme d'accélération \cite{RC12}. \\
Pour expliquer l'allure du spectre énergétique à très haute énergie, le modèle d'accélération directe a été proposé. Dans ce cas, l'accélération est rapide et due à l'existence d'un champ électromagnétique intense. Ainsi, un pulsar par exemple est le siège, à sa surface, de puissantes inductions magnétiques dues à la rotation rapide de la matière condensée. Dans 
ce cas les particules peuvent être produites et accélérées grâce à la force électromagnétique induite(FEM). L'énergie acquise dépend du rayon de l'objet en rotation (R), du champ 
magnétique (B) à la surface et de la fréquence de révolution (f). D'après le modèle du rotateur oblique \cite{RC13}, les pulsars sont interprétés comme étant des étoiles à neutrons en rotation autour d'un axe qui n'est pas aligné avec l'axe magnétique. Les particules chargées sont produites par 
extraction à la surface puis accélérées par la FEM qui se crée. A travers le rayonnement synchrotron elles émettent des $\gamma$. De cette manière l'émission peut être observée seulement lorsque le cône formé des deux axes est orienté vers la terre (figure (\ref{RC6}). 
Il faut signaler que ce mécanisme peut expliquer l'émission des rayons gamma ayant des énergies de l'ordre du MeV; par contre il ne peut expliquer celle des rayons gamma de haute énergie.\\
Ce mécanisme présente quelques problèmes: d'une part, l'accélération des particules dépend naturellement de la distance à laquelle l'influence du champ magnétique cesse et d'autre part, il ne conduit pas d'une manière claire à la forme du spectre énergétique observé.
\subsubsection{Accélérateurs cosmiques des particules}
\begin{figure}
\hskip 1.2cm
\includegraphics[height=18cm,width=12cm]{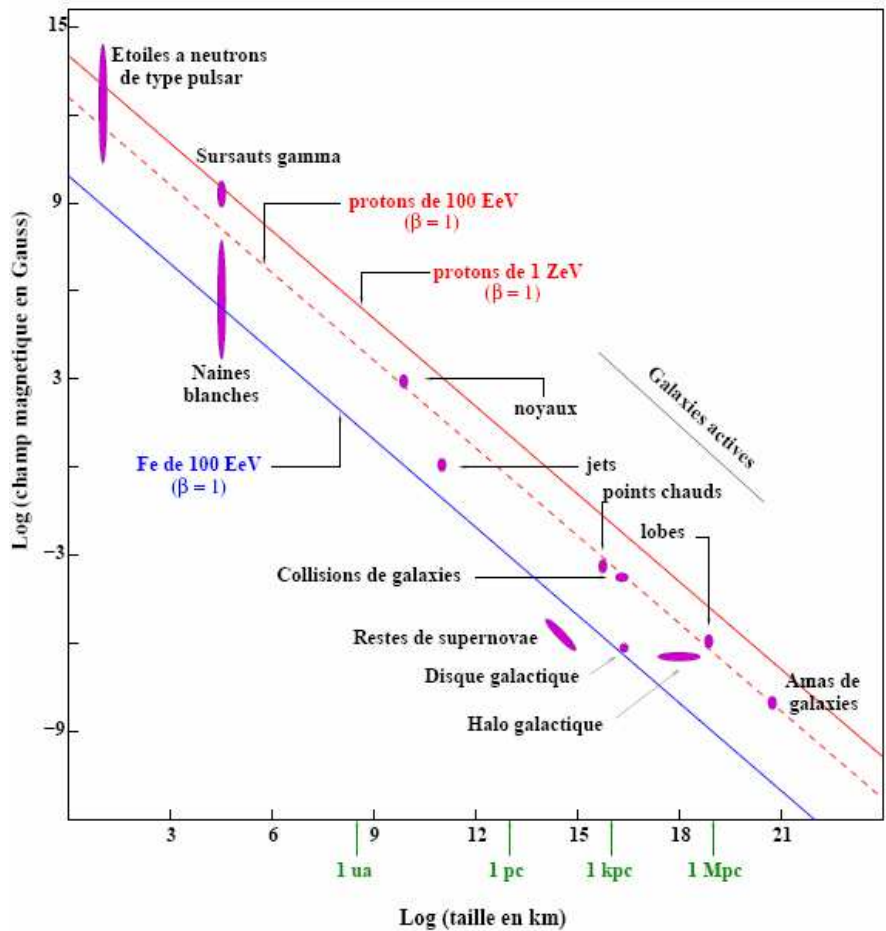}
\vskip -2cm
\caption{\em \textbf{Diagramme de Hillas : les sites sont classés en fonction de leur taille et de leur champ magnétique. Les sources au-dessus des différentes lignes peuvent potentiellement
accélérer les particules indiquées.}}
\label{RC7}
\end{figure}

De manière générale pour être accélérées aux plus hautes énergies, les particules doivent rester confinées dans le site accélérateur pour pouvoir interagir avec celui-ci. 
Le confinement d'une particule dans un certain site d'accélération va dépendre de l'intensité du champ magnétique que celui-ci contient. En comparant le rayon de Larmor $r_{L}$ de cette particule à la taille L de l'objet astrophysique qui contient le champ B, il est possible d'estimer l'énergie limite $E_l$ jusqu'à laquelle la particule de charge $Ze$ va pouvoir être accélérée, avant de pouvoir s'échapper :
\begin{equation}
r_L=\frac{E}{Z\;e\;c\;B}\geq L\Rightarrow E_l>\approx Z\frac{B}{1\mu G}\frac{L}{1kpc}EeV
\end{equation}
\\Cette équation va nous permettre de dessiner un diagramme dit de Hillas \cite{RC14}, qui va représenter les sites supposées capables d'accélérer des particules selon leur champ magnétique et de leur taille. La particule est accélérée tant qu'elle est confinée dans le site, la condition d'accélération est donc :
\begin{equation}
B(\mu G)L(kpc)>E(EeV)/Z
\end{equation}
Cela se traduit sur le diagramme (figure \ref{RC7}) par des sources potentiellement accélératrices qui doivent se trouver au-dessus des lignes représentées.\\
Il faut donc un compromis entre le champ magnétique qui doit être suffisamment grand pour confiner les particules dans le site accélérateur, et la taille de ce dernier qui doit également être suffisamment grand pour que les particules gagnent assez d'énergie avant de s'échapper. Cela restreint déjà les types d'objets candidats. Certains d'entre eux sont déjà écartés à cause de leur petite taille. Une exception à cette règle et un des candidats les plus sérieux à l'accélération des RC d'ultra haute énergie  sont les étoiles à neutrons (pulsars), qui compensent leur petite taille (quelques km) par des champs magnétiques très intenses. Les restes de supernovae
sont également exclus à cause de leur champ magnétique trop faible. Les objets extragalactiques semblent les candidats les plus sérieux, comme les noyaux actifs de galaxies, les lobes de radio galaxies et les sursauts gamma. Dans le cas où les rayons de Larmor sont plus grands que la taille de la Galaxie, c'est une raison qui favorisent l'hypothèse de la nature extragalactique des rayons cosmiques au delà de la cheville.
\subsection {Sources des $\gamma$ de haute énergie}
Les particules neutres telles que les $\gamma$ et $\nu$ de très haute énergie conservent la direction de la source et fournissent des informations précises sur la position de la source émettrice. Les neutrinos peuvent être détectés par des expériences souterraines grâce à la détection des muons produits par les interactions de ce $\nu$ avec la roche de couverture. Les $\gamma$ peuvent être détectés par des détecteurs à la surface EAS {\it"Extensive Air Shower"} en étudiant les gerbes électromagnétiques en cascade.  \\
Les $\gamma$ de haute énergie sont due probablement aux interactions des protons avec le milieu environnant d'une étoile à neutrons. Les protons accélérés, selon les modèles 
cités précédemment, jusqu'aux hautes énergies peuvent interagir avec l'étoile accompagnatrice et le milieu interstellaire et donnent naissance aux $\pi^o$ qui se désintègrent en deux $\gamma$. Des $\pi^\pm$ sont également produits et donnent après désintégration des $\nu$ et $\mu$.\\   
Parmi les sources de $\gamma$ de haute énergie nous citons les pulsars, qui émettent des $\gamma$ dans la bande d'énergie (1-100)TeV, dans la bande des rayons X nous trouvons le pulsar Crab associé à la nébuleuse Crab et Vela \cite{RC15b}.\\
Les systèmes binaires peuvent également être responsables de l'émission des $\gamma$ de haute énergie. Un système binaire est défini comme une étoile à neutrons en rotation autour de l'étoile massive dite "accompagnatrice".
Parmi les sources qui ont été signalées comme émettrices dans les bandes $\gamma$ et X nous citons Cyg-X3 et Hercule X-1.
\section {Interaction des RC avec l'atmosphère}
\begin{figure}[p!]
\begin{center}
\leavevmode
\epsfig{file=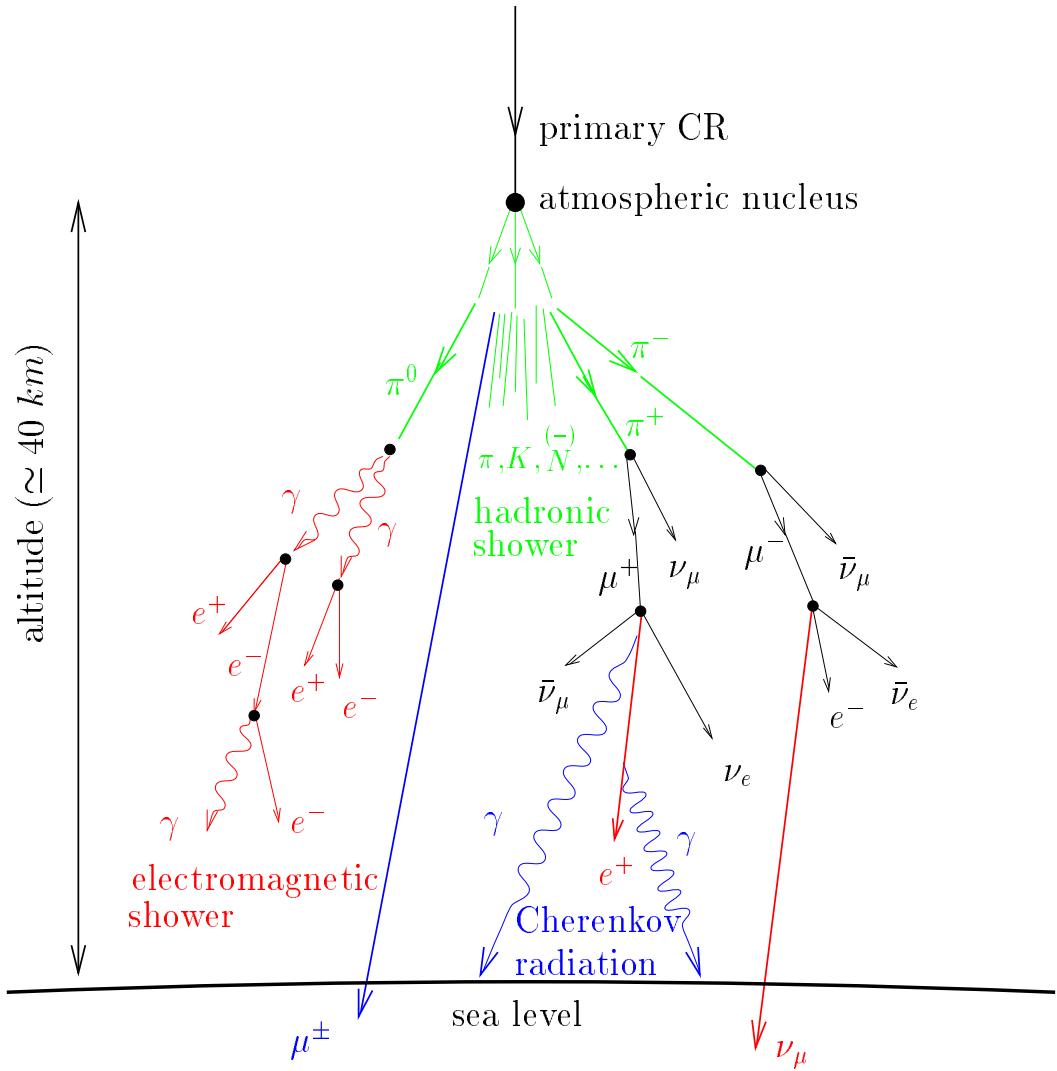,width=24cm}
\vskip -8cm
\caption{\em \textbf{Représentation schématique d'une gerbe produite par collision d'un rayon cosmique avec les noyaux atmosphériques.}}
\label{fig:RC8}
\end{center}
\end{figure}

Dés leur arrivée à la haute atmosphère, les RC interagissent avec les noyaux de ce dernier donnant naissance à un grand nombre de particules secondaires, essentiellement des mésons, qui peuvent ré-interagir ou se désintégrer dans l'atmosphère. En interagissant avec les noyaux de l'atmosphère, ces hadrons produisent d'autre particules en induisant une cascade formée de photons, d'électrons et de hadrons. \\
Sur la figure (\ref{fig:RC8}), nous donnons une représentation schématique des processus d'interaction d'un rayon cosmique primaire avec un noyau atomique dans l'atmosphère. 
Une cascade est formée:
\begin{itemize} 
\item d'une composante électromagnétique provenant de la désintégration 
rapide des $\pi ^{\circ}$ en photons; chaque photon engendre à son tour une cascade électromagnétique secondaire par l'intermédiaire des effets de création de paires et de bremsstrahlung, pour chaque interaction du RC primaire chargé, plus du tiers de l'énergie est absorbé par la composante électromagnétique
\item d'une composante hadronique constituée principalement de pions et kaons chargés qui se désintègrent pour atteindre le sol sous forme de muons et de neutrinos.
\end{itemize}
Dans notre travail on s'intéresse à la composante muonique. Cette dernière joue un rôle important dans l'étude des rayons cosmiques. Les muons ont une petite section efficace d'interaction, ils interagissent faiblement et peuvent passer librement entre les noyaux atmosphériques et finissent par disparaître en se désintégrant. Leur vie moyenne est de $2.2\times 10^{-6}$ s. En effet, les muons sont traditionnellement appelés la composante pénétrante des rayons cosmiques, ils constituent la composante dominante des RC au niveau de la mer et ils peuvent être détectés facilement par des détecteurs souterrains\\
En effet, les protons et les neutrons interagissent fortement et ne peuvent traverser une longue distance dans l'atmosphère, les noyaux lourds fragmentent par collisions avec les noyaux de l'air, les électrons et les photons subissent intensivement les processus électromagnétiques de perte d'énergie et atteignent le niveau de la mer en nombre réduit; enfin toutes les particules ayant une vie moyenne inférieure à $10^{-10}$ s se désintègrent dans la région de la haute atmosphère. Seuls les neutrinos arrivent en nombre élevé, mais à cause de leur faible section efficace d'interaction ils ne peuvent être détectés aussi facilement que les muons par des détecteurs souterrains.\par
\subsection{Production des $\mu$ lors d'une cascade hadronique}
Les muons cosmiques sont produits à la suite de la désintégration des 
pions et des kaons chargés qui ont été générés à leur tour par l'interaction des RC primaires avec les noyaux atmosphériques. La vie moyenne des pions est $2.6\times 10^{-8}$ s, celle des kaons est $1.2\times 10^{-8}$s. Les principaux modes de désintégration et les rapports d' 
embranchement sont les suivants:\vskip .1in
$$\pi^{+}\rightarrow \mu^{+} + {\nu_{\mu}} \ \ \ \ \ \ \         
\pi^{-}\rightarrow \mu^{-} +\bar{\nu_{\mu}} \ \  99.98770 \%$$
$$K^{+}\rightarrow \mu^{+} + \nu \ \ \ \ \ \ \         
K^{-}\rightarrow \mu^{-} +\bar {\nu_{\mu}} \ \  63\%$$ 
$$K^{+}\rightarrow \pi^{+} + \pi^{0} \ \ \ \ \ \ \ 
K^{-}\rightarrow \pi^{-} + \pi^{0} \ \ \ 21.\%$$
\vskip .1in
Les muons de basse énergie sont générés durant la dernière phase de développement de la cascade. Ceux de  haute énergie, sont produits durant la phase initiale de la création de la cascade, sont généralement très pénétrants et contiennent des informations sur les primaires qui les ont générés. \par
Une étude cinématique simple dans le système du centre de masse (CM) puis dans le système du laboratoire (SL) de la désintégration $\pi \rightarrow \mu +\nu $ permet d'obtenir l'expression de l'énergie moyenne du muon dans le (SL) en fonction de l'énergie du $\pi$, $E_{\pi}^{SL}$, et de son angle d'émission par rapport à la direction initiale du pion.
\begin{equation}
E_{\mu }^{SL}=E_{\mu }^{CM}/\sqrt {1-v_{\pi}^{2}/c^{2}}
\end{equation}
\vskip .1in
\noindent avec   $E_{\mu}^{CM}\approx 109.7MeV$.\par
\noindent o\`u encore  
\begin{equation}
E_{\mu}^{SL}=E_{\mu}^{CM}\gamma _{\pi}^{SL}=
E_{\mu}^{CM}.E_{\pi}^{SL}
/(m_{\pi}c^{2})=0.78\cdot E_{\pi}^{SL}
\end{equation}
L'angle moyen par rapport à la direction du $\pi$ avec lequel est émis le muon dans le (SL) est de l'ordre de $10^{-3}$ radian (qui peut être considéré en première approximation égal à 0). Ceci permet d'admettre que le muon conserve la direction du pion initial. Un raisonnement analogue peut être fait également pour la désintégration des K.
\subsection{Production des $\mu$ lors d'une cascade électromagnétique}
Les $\gamma$ de haute énergie interagissent avec l'atmosphère terrestre pour donner une cascade électromagnétique. Le processus de photoproduction peut également avoir lieu avec une faible 
probabilité selon:
\begin{equation}
\gamma + noyau \rightarrow hadrons (\pi,K...)
\end{equation}
Par la suite les $\pi$ et K se désintègrent pour donner des $\mu$ et des $\nu$. La section efficace de ce processus a été mesurée par des expériences utilisant des accélérateurs
jusqu'à des énergies $\approx$ 20 GeV des photons incidents. Au delà de la résonance, (0.1 - 0.8) GeV, cette section efficace est de l'ordre de 100 $\mu b$/nucléon et augmente légèrement pour des énergies supérieures à 10 GeV \cite{RC18}. Cependant il a été remarqué que dans le cas de l'interaction $\gamma$p, l'extrapolation des données des sections efficaces aux faibles énergies peut sous-estimer la probabilité de production des $\mu$.\par 
La probabilité qu'un $\gamma$ produise un hadron au lieu de la création 
d'une paire $e^+e^-$ à une énergie $\sim $ 20 GeV est donnée par le 
rapport des sections efficaces des deux processus \cite{RC9}:
\begin{equation}
P_{\gamma \mu}=\frac {\sigma_{\gamma \rightarrow hadron}}{\sigma_{\gamma 
\rightarrow e^+e^-}}\approx 2.8\times 10^{-3}
\end{equation}
La probabilité de production de hadron au lieu de la création de paires est de l'ordre 1/300. Les $\mu$ sont aussi crées par la production de paires $\mu^\pm$ mais en petite quantité. La section efficace de ce processus est donnée par:\\
\begin{equation}
\sigma_{\gamma \rightarrow \mu^+ \mu^-}=\sigma_{\gamma \rightarrow e^+ e^-}
(\frac{m_\mu^2}{m_e^2})\approx 2\times 10^{-5}\cdot \sigma_{\gamma \rightarrow 
e^+e^-} 
\end{equation}  
Les cascades électromagnétiques riches en muons et qui ont été observées par certaines expériences peuvent trouver leur explication dans une grande section efficace de photoproduction, autrement il est difficile de donner une justification à ces résultats.\par
Il faut signaler enfin que pour avoir un $\mu$ de 1.4 TeV (énergie minimale 
qu'un $\mu$ doit avoir pour atteindre MACRO) (voir ci-après 1.6) le $\gamma$ primaire doit avoir environ 300 TeV, alors que pour une cascade hadronique il faut 20 TeV.\par
\section{Perte d'énergie des muons dans la roche de MACRO.}
Les muons très énergétiques réussissent à traverser la roche de couverture sous laquelle est installé un détecteur comme MACRO. En traversant une épaisseur X, les $\mu$ perdent de l'énergie d'une manière continue ou discrète.\\ 
Cette perte d'énergie peut avoir lieu suivant 4 mécanismes possibles: ionisation, Bremsstrahlung, production directe de paires et interactions électromagnétique avec les noyaux.\par
La perte d'énergie moyenne peut s'écrire:\par
\begin{equation}
\left(\frac{dE}{dx}\right)_{totale}=\left(\frac{dE}{dx}\right)_{ion}+\left(\frac{dE}{dx}\right)_{brem}+\left(\frac{dE}{dx}\right)_p+\left(\frac{dE}{dx}\right)_{nucl}
\end{equation}
La perte d'énergie par ionisation domine aux faibles énergies, alors que les autres mécanismes sont importants à des énergies supérieures au TeV. \par
Dans le cas de l'ionisation, la perte d'énergie est un processus continu qui peut être décrit par l'équation:\par
\begin{equation}
\left(\frac{dE}{dx}\right)_{ion}=-[1.9+0.08\;ln(E_{max}/m_{\mu }c^{2})]
\label{ion}
\end{equation}
$E_{max}$ est l'énergie maximale transférée à un électron du milieu et qui est donnée par:
\begin{equation}
E_{max}=\frac{E_{\mu }^{2}}{E_{\mu }+0.5m_{\mu }^2c^{2}/m_{e}}
\end{equation}
La relation (\ref{ion}) peut être approximée par 
\begin{equation}
\left(\frac{dE}{dx}\right)_{ion}=-\alpha
\end{equation}
 o\`u  $\alpha \approx 2.0$ $ MeV/g.cm^{2}$.\par
\begin{figure}
\centering
\includegraphics[height=14cm,width=14cm]{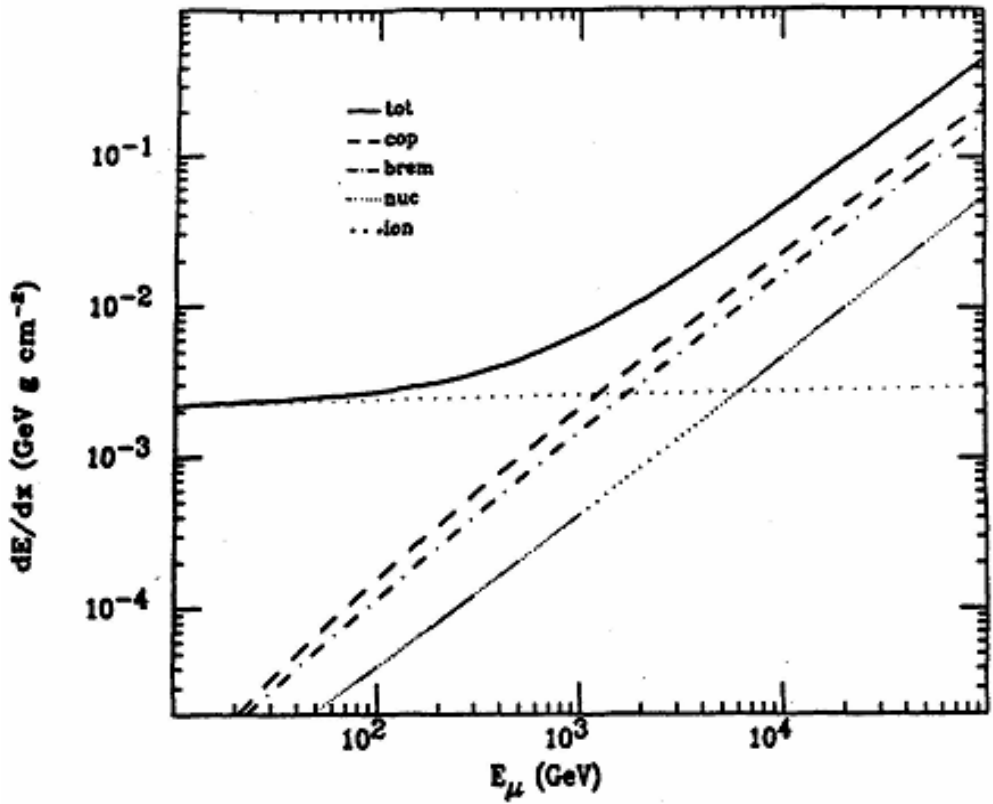}
\caption{\em \textbf{Courbes de perte d'énergie des muons en fonction de l'énergie dans
la roche standard.}}
\label{RC9}
\end{figure}

 
En ce qui concerne la perte d'énergie par Bremsstrahlung, la section efficace varie comme 
$d\sigma $/d$v$ $\approx $1/$v$ o\`u $v$ est la fraction d'énergie cédée par le muon au photon durant le processus radiatif. La section efficace de création de paires se maintient toujours finie et décroît rapidement, après avoir atteint le maximum pour des $v$ voisins de la limite cinématique $v_{min}= 4m_{e}/4E_{\mu }$, comme $ 1/v^{3}$. En négligeant les fluctuations pour ces deux processus discrets de la perte d'énergie des $\mu $, on obtient les courbes de perte d'énergie correspondant aux divers processus et à l'ensemble (voir figure (\ref{RC9}) \cite{RC9}). On constate l'allure constante de la perte d'énergie par ionisation alors que les deux autres mécanismes croissent en fonction de l'énergie selon une forme logarithmique.\par
La perte d'énergie totale est paramétrisée par:
\begin{equation}
(\frac{dE}{dx})_{totale}= -\alpha - \frac{E}{x_0}
\end{equation}
avec $x_0 \approx 2.5\times 10^{5}g/cm^{2}$ dans la roche.\par
Une estimation de l'énergie minimale qu'un $\mu $ doit avoir à la surface 
pour atteindre une profondeur X est \cite{RC9}
\begin{equation}
E_{0}^{min}=E_{c}(e^{\frac{X}{x_0}}-1)
\end{equation}
$E_{c}$ est appelée énergie critique et correspond à la valeur pour laquelle les pertes d'énergie par bremsstrahlung deviennent plus importantes que celles par ionisation. Les muons qui se propagent dans la roche ont $E_{c}\approx 500 GeV$. Dans le cas de MACRO, pour traverser la roche de couverture (3500 m.w.e.)\footnote{1 m.w.e {\it(1 meter of water equivalent)} est une notation utilisée pour exprimer la profondeur des laboratoires souterrains en mètre d'eau équivalent; (1 m.w.e$= 1$ $hectogramme/cm^{2}$).},
les muons doivent avoir une énergie minimale de 1.4 TeV à la surface, ce qui correspond à des primaires d'énergie 20 TeV. A cette profondeur le flux des $\mu $ atmosphériques se réduit d'un facteur 10$^{6}$ par rapport à celui à la surface (c'est à dire 100 $ \mu /cm^{2}s).$
\section{Technique de détection des rayons cosmiques}
\begin{figure}
\centering 
\vskip -4cm
\includegraphics[height=8cm,width=8cm]{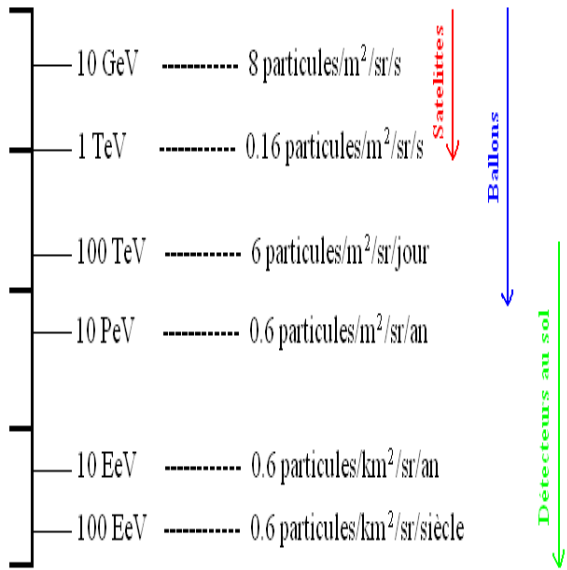}
\caption{\em \textbf{Schéma indiquant le domaine d'application des différentes méthodes de détection des rayons cosmiques. Un ordre de grandeur du flux est indiqué en fonction de l'énergie.}}
\label{RC10}
\end{figure}

Les méthodes expérimentales développées par les physiciens depuis près d'un siècle pour étudier le rayonnement cosmique sont conditionnées par le type de particule que l'on cherche à détecter, mais surtout par la valeur du flux à l'énergie considérée (voir figure (\ref{RC10})). Pour des énergies inférieures à une centaine de TeV, le flux est suffisamment élevé pour qu'un détection directe des particules soit possible ($\sim$ 10 particules/m$^2$/sr/s à 100 GeV). Au dessus de l'atmosphère les détecteurs sont installés dans des ballons atmosphérique, ou sur des satellites. Les détecteurs utilisés sont soit passifs tels que les chambres à émulsion, les films de rayons X et des détecteurs à trace nucléaires, soit actifs tels que : les calorimètres, les compteurs à scintillation et les compteurs Cherenkov.\par
Pour les énergies supérieurs à quelques centaines de TeV le flux des RC est plus faible (inférieur à quelques particules/m$^2$/sr/an). Ceci nécessite une importante surface de détection, ce qui est problématique pour des expériences embarquées et exposées pour de brèves périodes de temps (un à trois jours dans le cas des ballons). La détection directe des particules n'est plus possible et l'étude des RC ne peut se faire qu'à partir du sol en utilisant des détecteurs de grandes surfaces et qui sont exposés pour de longues périodes. Les expériences souterraines, comme MACRO, détectent la composante muonique (voir \ref{chap2}). Sur les sommets des montagnes sont placés des expériences "EAS" qui étudient les cascades électromagnétiques produites par des RC primaires d'énergies variant de 10$^{14}$ à 10$^{20}$eV. Parmi les détecteurs disposés au sol, nous citons l'expérience Auger, c'est un détecteur Hybride utilisant des propriétés de la gerbes pour en analyser les caractéristiques et remonter au rayon cosmique initial.


\chapter{L'expérience MACRO}
\section{Introduction}
\label{chap2}
Actuellement, la physique des accélérateurs nous permet d'atteindre des énergies de l'ordre de 14 TeV dans le centre de masse. Ce seuil est associé aux particules supersymétriques ou à des constituants des quarks et des leptons; ceci pourra être vérifié dans l'avenir avec des accélérateurs puissants tel que le grand collisionneur de hadrons LHC (Large Hadron Collider) au CERN. Un deuxième seuil pourrait être associé à la grande unification qui se propose d'interpréter les trois interactions électromagnétique, nucléaire forte et faible et pouvant se manifester à des énergies de l'ordre de $10^{14}eV$. Parmi les conséquences de ces théories, la désintégration du proton et l'existence de certaines particules exotiques tel que les monopôles magnétiques, les nucléarites, etc.... L'observation de telles particules ne peut se faire auprès des accélérateurs. Dans ce cas, les expériences sans accélérateurs, en particulier les expériences souterraines, jouent un rôle très important pour la recherche des particules traces et la violation de certaines lois de conservation.\\
Les expériences souterraines sont souvent installées dans des sites miniers ou dans des tunnels, la roche couvrant l'expérience joue le rôle de filtre qui absorbe les particules chargées des RC dont le nombre est élevé au niveau du sol et peuvent rendre impossible la détection des évènements peu fréquents.
Les expériences souterraines sont nombreuses et sont destinées à différents sujets de recherche en physique fondamentale. Néanmoins on peut les diviser en deux types, d'une part celles qui se basent sur l'étude des rayons cosmiques et d'autre part celles qui ne dépendent pas de ces derniers.\par
Les expériences souterraines qui ne sont pas destinées à l'étude des rayons cosmiques s'intéressent généralement aux problèmes de violations des nombres leptonique ou baryonique en particulier la désintégration du proton \cite{MAC1}.\par
Les expériences souterraines basées sur les informations fournies par l'étude des rayons cosmiques ont pour objectifs de résoudre les problèmes relatifs :
\begin{enumerate}
	\item à l'astrophysique tel que:
	\begin{itemize}
			\item  l'étude de la composition et du spectre énergétique des RC 
			\item l'étude de la variation temporelle et spatiale des RC, recherche des sources 	
						ponctuelles et détection des neutrinos de l'effondrement gravitationnel.
			\item l'étude des neutrinos atmosphériques et leur oscillation.
	\end{itemize}
	\item aux particules élémentaires et à la recherche des particules exotiques telles que les 					monopôles magnétiques, les nucléarites, les Q-balles, les WIMPs, les particules 			
				supersymétriques, etc...
\end{enumerate}
En général le nombre d'évènements attendus est très petit ce qui nécessite des détecteurs assez massifs et volumineux.\par
Les détecteurs souterrains utilisés sont : les scintillateurs liquides, les détecteurs Cherenkov à eau , les calorimètres avec reconstruction de traces, les tubes à streamer et les détecteurs plastiques. Pour l'étude des RC de haute énergie (astronomie des muons et des neutrinos), on choisit généralement un des trois premiers types de détecteurs ou la combinaison de deux d'entre eux, afin d'augmenter la surface de détection et d'avoir une bonne résolution spatiale. Dans le cas de l'étude de la désintégration du proton, la compétition se fait entre les détecteurs Cherenkov à eau et les calorimètres.\par
Parmi les laboratoires souterrains nous citons: Le laboratoire Baksan en Russie, l'expérience Homestake, l'expérience IMB et Soudan 1-2 aux Etats Unis, l'expérience NUSEX sous le Mont Blac en France, l'expérience Kolar Gold Field en Inde, l'expérience Kamiokande et super-Kamiokande au japan enfin le laboratoire de Gran Sasso en italie.
\section{Généralités sur le Laboratoire National du Gran Sasso (LNGS)}
Le Laboratoire National du Gran Sasso est considéré parmi les plus grands laboratoires souterrains du monde, il est destiné à la recherche en physique nucléaire, en physique des particules élémentaires et en astrophysique. Il comprend plus d'une dizaine d'expériences qui s'intéressent à différents sujets de recherches \cite{MAC11}.\par
\begin{figure}[t]
\begin{center}
\leavevmode
\epsfig{file=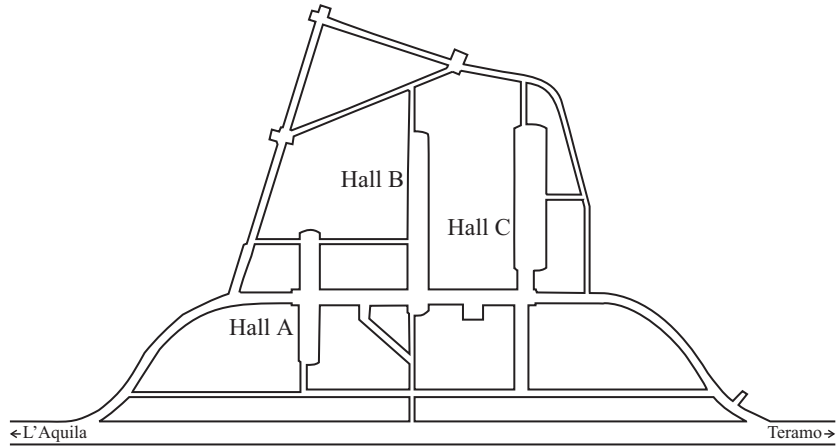,width=11cm}
\caption{\em \textbf{Vue générale du laboratoire souterrain. L'expérience MACRO est localisé dans le Hall B du laboratoires Gran Sasso.}}
\label{labs}
\end{center}
\end{figure}

Le LNGS est situé sous la chaîne montagneuse du Gran Sasso au centre d'Italie dans le tunnel autoroutier qui lie Rome à Teramo (voir figure (\ref{labs})). La latitude et la longitude du LNGS sont respectivement $42^0\;27'\;09"$ Nord et $13^0\;34'\;28"$ Est; l'altitude est d'environ 963m
par rapport au niveau de la  mer.\par
La roche couvrant le laboratoire est constituée de calcaire avec une densité de $2.71 ± 0.05 	g/cm$ , un numéro atomique moyen $11.4 ± 0.2$ et un poids atomique moyen de $22.9 ± 0.4$; l'épaisseur moyenne de la roche est de l'ordre de 1400 m. Un muon arrivant au sommet de la montagne doit avoir au minimum une énergie de 1.4 TeV pour arriver au laboratoire souterrain \cite{MAC13}.\par
Le LNGS contient trois grands Halls ayant chacun une dimension approximative de 100m x 17m x 17m \cite{MAC1}.
\section{Généralités sur le détecteur MACRO}
MACRO (Monopôle Astrophysics and Cosmic Ray Observatory) \cite{MAC15} est situé au Laboratoire National du Gran Sasso (LNGS), il a été considéré comme étant un des plus grands détecteurs souterrains du monde \cite{MAC16}. Il a été réalisé en tenant compte des objectifs exigeants : une acceptance suffisamment grande pour le flux de particules détectées, des techniques de détection variées (Scintillateur liquide, tube à streamer (TS) et le détecteur CR39) assurant une bonne résolution et une grande efficacité pour l'identification des diverses particules, et enfin un système d'acquisition performant.\\
Il a permis d'étudier différents thèmes de recherches. \par
\begin{figure}[t]
\begin{center}
\includegraphics[height=10cm,width=14cm]{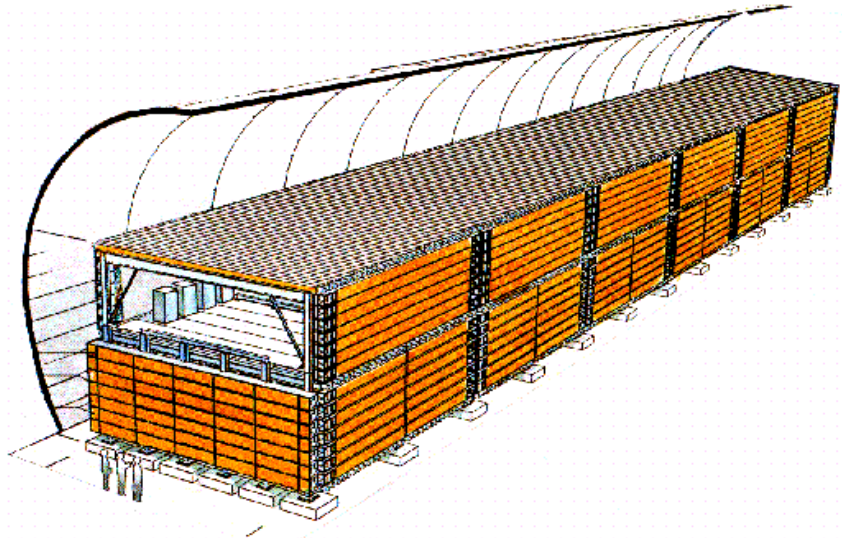}
\caption{\em \textbf{Vue générale de l'expérience MACRO.}}
\label{macro}
\end{center}
\end{figure}
\begin{figure}[p]
\begin{center}
\leavevmode
\epsfig{file=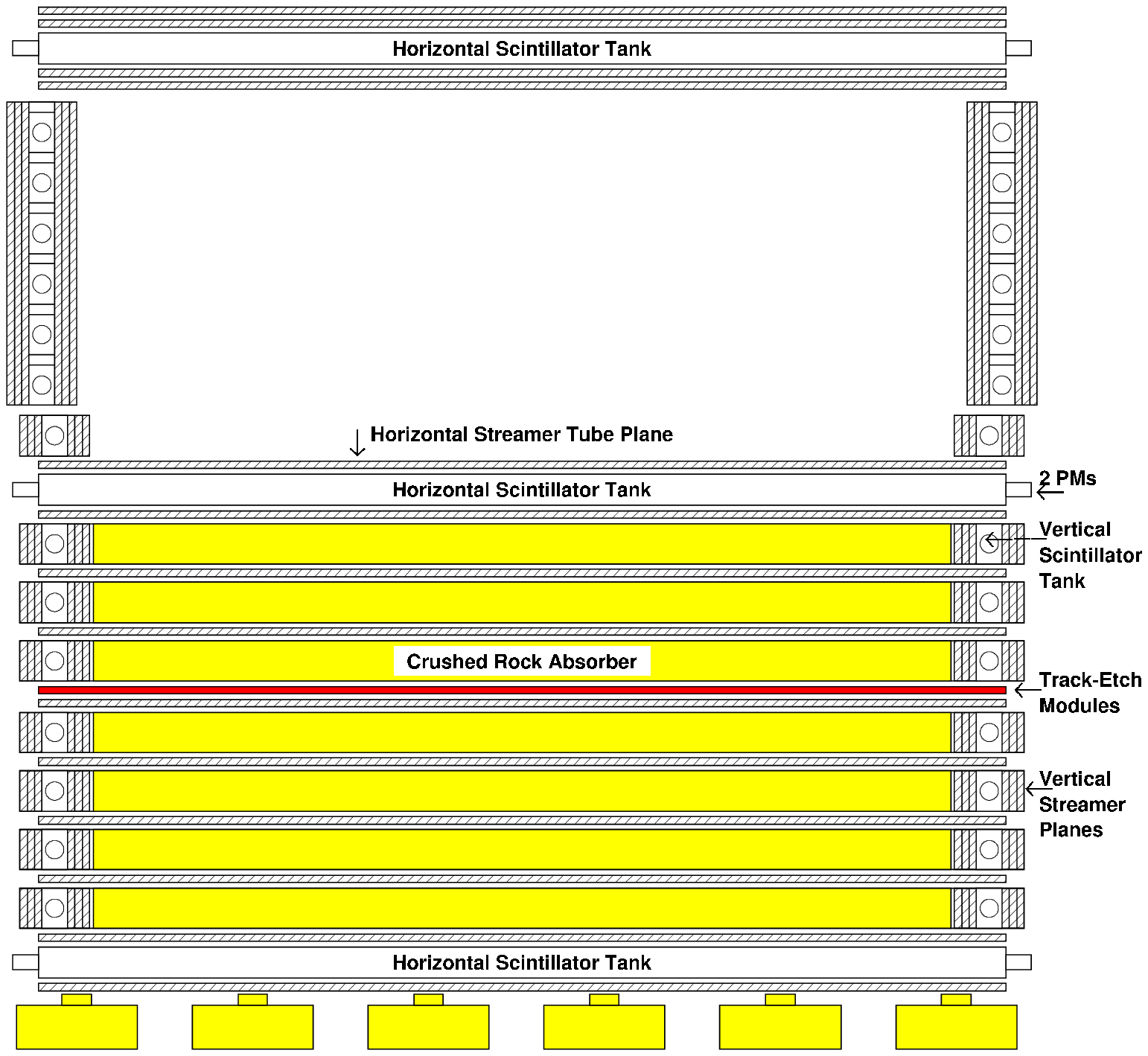,width=13cm,height=14cm}
\caption{\em \textbf{Vue de la section transversale du détecteur MACRO où apparaît tous les constituants du détecteur ainsi que leurs dispositions.}}
\label{MACRO_section}
\end{center}
\end{figure}

MACRO est un détecteur à grande surface d'une longueur de 76.7 m, d'une largeur de 12 m et d'une hauteur de 9.3 m. Il est constitué de six supermodules (SM) placés l'un à côté de l'autre (figures (\ref{macro}) et (\ref{MACRO_section})). Les six (SM) ont une structure identique. Un SM a une dimension de l'ordre de 12 m $\times$ 12 m $\times$ 9.3 m et il est formé de deux parties \cite{{MAC16r}}.\par
Les deux modules sont séparés par une distance de 30 cm qui est occupée en partie par le matériel supportant le détecteur.\par
La partie inférieure est formée de deux plans horizontaux de scintillateurs liquides séparés entre eux de 4.8 m par dix plans de tubes à streamer limités. Entre les plans des tubes à streamer on trouve 60 cm d'absorbant, servant à éliminer le fond dû au rayonnement $\gamma$ mous qui peuvent produire l'effet compton et de reconnaître les muons ayant une énergie > l GeV. Au milieu de chaque SM il y a un plan de détecteurs nucléaires solides à traces (CR39). Ce dernier est placé également sur la partie latérale Est et frontale Nord.\par
Afin d'augmenter l'acceptance du détecteur, les parois latérales et frontales sont constituées de six plans de tubes à streamer et un plan de scintillateurs. Il faut noter que chaque supermodule peut fonctionner indépendamment des autres, permettant la continuité d'acquisition des données même durant la réparation d'une partie du détecteur.\par 
La partie supérieure de MACRO est appelée "attico" placée à une hauteur de 4.8 m par rapport la partie inférieure. Elle est formée d'un plan de scintillateurs horizontaux avec deux plans de tubes à streamer au dessus et deux autres au dessous. Comme pour la partie inférieure, L'attico a également des scintillateurs verticaux et des tubes à streamer latéraux. L'électronique et le système d'acquisition des données sont situés entre la partie inférieure et l'attico comme c'est indiqué sur la figure (\ref{MACRO_section}).\par
A l'intérieur de MACRO est installé un détecteur à rayonnements de transition (TRD), ce module à pour objectif de mesurer l'énergie des particules chargées qui le traversent.\\
\section{Les détecteurs constituant MACRO}
Le détecteur MACRO est constitué de divers subdétecteurs, (scintillateurs, tubes à streamer limités et le détecteur plastique à traces (CR39)). Dans ce paragraphe, on va donner une description de ces détecteurs ainsi que leur mode de fonctionnement et leurs contributions à la détection des différentes particules.
\subsection{Description des scintillateurs liquides}
Le rôle des scintillateurs dans l'expérience MACRO consiste à mesurer la perte d'énergie, le temps de vol, la direction et l'instant d'arrivée des particules en plus de la détection des neutrinos de l'effondrement gravitationnelle.\par
\begin{figure}
\begin{center}
\leavevmode
\hskip -3cm
\epsfig{file=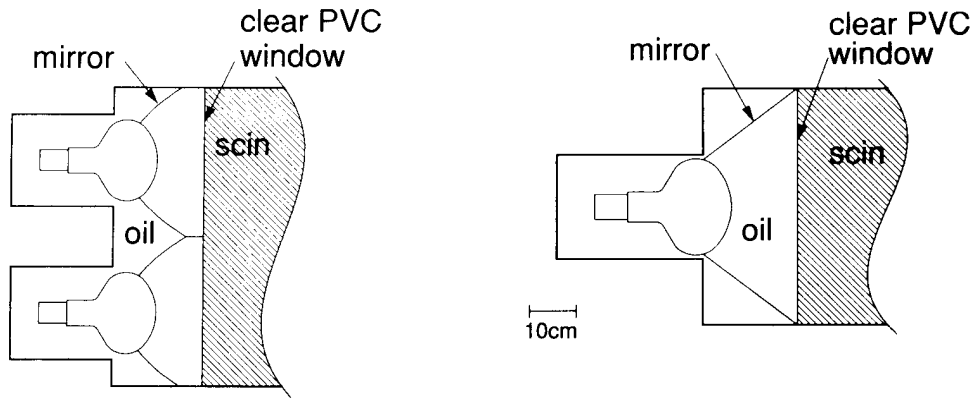,width=12cm}
\caption{\em \textbf{Géométrie de l'une des extrémités des compteurs à scintillation horizontaux (gauche) et verticaux (droite). La partie centrale est remplie de liquide scintillant qui est vue par des PMT.}}
\label{scint}
\end{center}
\end{figure}

MACRO contient 294 compteurs horizontaux et 182 compteurs verticaux. La masse totale du liquide scintillant est 600t. La partie inférieure du système de scintillateurs de MACRO de chaque SM contient 32 compteurs horizontaux et 21 verticaux. Chaque compteur a une longueur de 12m dont 11m rempli d'un liquide scintillant transparent.\\
La figure (\ref{scint}) donne une vue horizontale et verticale des scintillateurs horizontaux.
Le scintillateur est constitué d'un récipient en PVC rempli de liquide scintillant, dont les parois internes sont enveloppées d'une couche de FEP-Teflon afin que ces parois deviennent totalement réfléchissantes. Le liquide scintillant a la composition suivante \cite{MAC17}:
\begin {itemize}
\item 96.4\% d'huile minérale avec peu de parafine.
\item 3.6\% de pseudocumène.
\item 1.44 mg/l de bis-MSB.
\item 1.44 g/l PPO
\end{itemize}
La densité de ce mélange est de 0.85g/cm$^2$ et une longueur d'atténuation de $\sim$12m. L'huile minérale est très claire et a une longueur d'atténuation de 20 m. Le pseudocumène est un composé organique dont la composition chimique est (1,2,4-thriméthyl benzène); le PPO (2,5-diphényl-oxazole) et le Bis-MSB (p-biso-méthylstyrylbenzène).\par
Les compteurs à scintillation horizontaux ont une dimension de 12 m $\times$ 25 cm $\times$ 75 cm chacun et ils sont mis sur trois plans horizontaux. La lumière produite dans un compteur horizontal est recueillie aux extrémités par un couple de photomultiplicateurs (PMT) et les signaux issus de ces derniers sont additionnés.\par
Les compteurs à scintillation verticaux ont une dimension de 12 m$\times$25 cm$\times$50cm et couvrent les deux plans latéraux et le plan frontal de MACRO. Ces compteurs ne contiennent qu'un seul photomultiplicateur à chaque extrémité.\par
MACRO permet de reconstruire l'énergie déposée dans un compteur avec une résolution $\sigma_E/E=0.3/\sqrt{E}$ (E est exprimée en MeV). La résolution temporelle est inférieure à 1ns, ce qui permet de mesurer le temps de vol des particules et de déterminer ensuite leur vitesse. On peut alors distinguer les muons provenant du haut de ceux provenant du bas et qui sont produits lors de l'interaction des neutrinos muonique avec la roche. \par
Les scintillateurs servent également à la détection des $\bar{\nu_e}$ produits durant l'effondrement gravitationnel par l'intermédiaire de la réaction 
\begin{equation}
\bar{\nu_e}+p\rightarrow e^+ + n
\label{grav}
\end{equation}
dans le liquide scintillant. L'énergie moyenne des antineutrinos de supernovae est $\left\langle E\right\rangle\approx 12\;MeV$ car la section efficace de la réaction \ref{grav} croit avec l'énergie de neutrino. Cette réaction sera suivie, après modération du neutron dans le compteur, par la réaction de capture suivante :
\begin{equation}
n+p\rightarrow d+\gamma
\label{capture}
\end{equation}
avec $E_{\gamma}=2.2\;MeV$. Le temps de modération des neutrons est de l'ordre de $10\mu s$ et le temps de capture est $180\mu s$. Le positron donne naissance à un signal de particule rapide et permet à l'électronique de diminuer momentanément le seuil du même compteur de $5 MeV$ à $1 MeV$ pour un temps de $850 \mu s$ afin de détecter le photon de $2.2 MeV$. Les deux signaux provenant du positron et du neutron sont les signatures caractéristique d'un tel événement. 
\subsection{Description des tubes à streamer limités}
\begin{figure}[t!]
\begin{center}
\leavevmode
\vskip 1cm
\epsfig{file=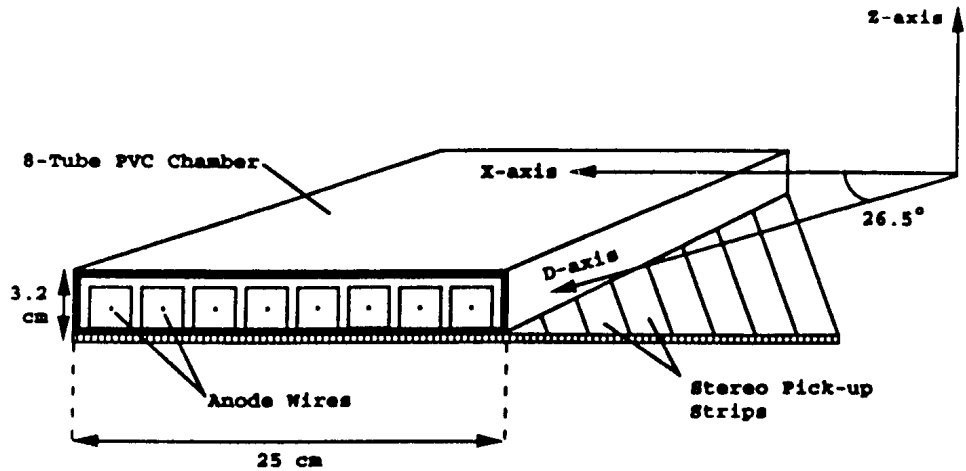,width=8cm}
\vskip 2cm
\caption{\em \textbf{La structure d'une chambre formée de 8 tubes à streamer avec les strips qui forment un angle de 26.6$^\circ$ par rapport aux fils.}}
\label{eight_tube}
\end{center}
\end{figure}

Les compteurs tubes à streamer limités permettent de détecter les particules chargées rapides et lentes. Ces détecteurs sont aussi appelés détecteurs à reconstruction de traces. Ils sont constitués de fils et de strips. Ces derniers sont inclinés de $26.6^0$ par rapport aux fils. Les fils permettent la mesure des coordonnées de la particule entrante dans le détecteur dans une direction appelée X, les strips à leur tour permettent la mesure des coordonnées de la particule dans une direction appelée D \cite{MAC18}(voir figure \ref{eight_tube}).\par
Les tubes contiennent un mélange de gaz constitué de 72\% d'Hélium et de 28\% de n-pentane. Les proportions de cette composition permettent l'exploitation des effets Drell et Penning pour la détection des MMs lents ($\beta=10^{-3}-10^{-4}$).\par
Les tubes à streamer limités sont des cellules ayant chacune une dimension de 3 cm $\times$ 3 cm $\times$ 12 m, 8 cellules sont groupées dans une chambre en PVC ayant une épaisseur de 1.5 mm et une dimension de 25 cm $\times$ 3 cm $\times$ 12 m. L'anode est représentée par un fil de 12 m de longueur et 100 $\mu m$ de diamètre et qui sont liés aux extrémités à un générateur de haute tension. La cathode est sous forme de graphite couvrant trois parois internes de la cellule et elle est liée à la terre. La tension d'alimentation typique des tubes à streamer limités est 4.2 kV.\par
Pour chaque SM il y a 648 unités où sont rangés 5184 fils. Les traces des particules reconstruites sur les dix plans ont une résolution spatiale sur l'axe des X relatif aux fils de l'ordre de 1 cm, pour l'axe D la résolution relative aux strips est de l'ordre de 1.2 cm. On peut donc tracer la trajectoire de la particule qui traverse les tubes à streamer avec une grande précision c'est pourquoi le système des tubes à streamer limité est appelé "système de reconstruction de traces".\par
\subsection{Description du détecteur nucléaire à traces de MACRO}
Le rôle principal du détecteur solide nucléaire à trace de MACRO est la recherche des monopôles magnétiques (MMs). Il est placé verticalement sur le front nord et sur la partie latérale Est et horizontalement au milieu de chaque SM inférieur. La surface totale occupée est de l'ordre de 1263 m$^{2}$ \cite{MAC18}.\par
\begin{figure}[t]
\begin{center}
\leavevmode
\hskip -2cm
\epsfig{file=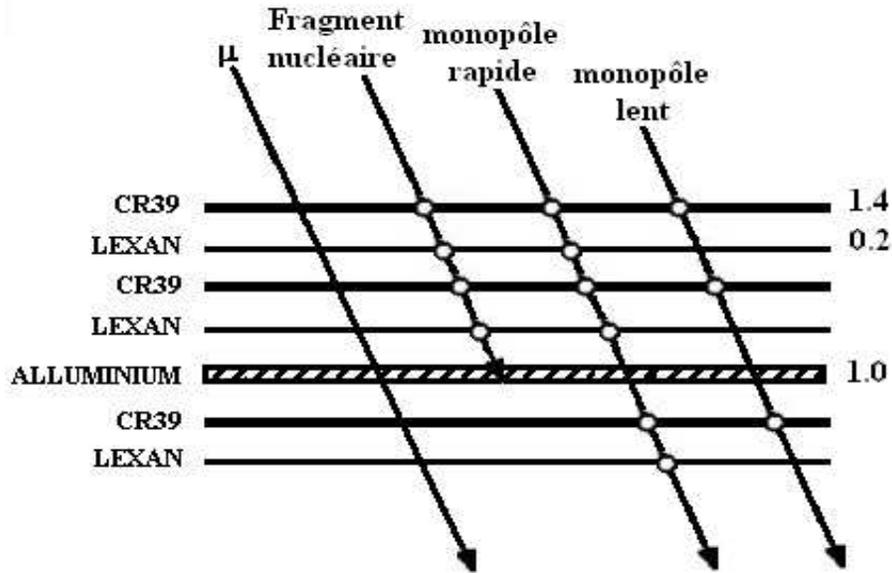,height=8cm, width=12cm}
\caption{\em \textbf{disposition du détecteur track-etch.}}
\label{CR39}
\end{center}
\end{figure}

Ce détecteur se trouve rangé en unités de bases. Une unité de base est  constituée d'un paquet  de dimension 25 $\times$ 25 cm$^{2}$ et formée de 3 feuilles de Lexan d'une épaisseur de 0.2 mm, 3 couches de CR39 d'épaisseur 1.4 mm et d'une feuille d'aluminium d'épaisseur 1 mm (voir figure (\ref{CR39})). 
Les paquets sont mis sur des chariots mobiles en plastique afin de faciliter le retrait si un évènement est candidat à être un MM par les scintillateurs et/ou les tubes à streamer limités. D'autre part le CR39 peut être utilisé comme détecteur passif sans avoir recours aux triggers.\par
C'est un détecteur plastique isolant, ayant pour formule chimique CH$_{12}$H$_{18}$O$_{7}$, qui peut être utilisé dans plusieurs expériences en physique, en particulier la détection des MMs \cite{MAC19}, recherche des particules ayant des charges fractionnelles, l'étude de la section efficace de fragmentation des ions relativistes \cite{MAC20} et la mesure de la concentration du radon \cite{MAC21}.\par
La résolution en charge du détecteur CR39 est de l'ordre de a = 0.16e pour le balayage d'une seule surface. Pour dix mesures sur dix surfaces successives a = 0.05e pour z = 6 et a = 0.07e pour z > 16. Cette résolution peut être considérée comme étant la meilleure comparée à celles des autres détecteurs actifs.\par
Le détecteur CR39 est sensible à la perte d'énergie restreinte (REL) qui représente la fraction d'énergie concentrée dans un cylindre de 0.01 $\mu m$ autour de la trace. La REL représente donc une partie de la perte d'énergie totale, l'autre partie d'énergie est perdue sous forme de rayonnement $\delta$ d'énergie supérieure à une valeur spécifique dépendant du type du matériel utilisé. Pour le détecteur CR39 cette énergie spécifique est de l'ordre de 200 eV. Le rayonnement $\delta$ dépose son énergie loin de la trajectoire de la particule et ne peut donc pas participer à la formation de la trace.\par
Le mécanisme de formation de la trace latente dépend de la particule nucléaire hautement ionisante. Un ion de numéro atomique Z, en traversant un milieu condensé se comporte comme une particule de charge effective positive \cite{MAC22}:
\begin{equation}
Z_{eff}=Z\sqrt{[1-exp(-130\beta/Z^{2/3})]}
\end{equation}
si $v$ est élevée, on a $Z_{eff}\sim Z$.\par
Cela va produire une rupture au niveau de la chaîne moléculaire et à la production de radicaux libres interagissant chimiquement. Les électrons de haute énergie ne contribuent pas à la formation de la trace mais contribuent seulement à la perte d'énergie dans un diamètre de $\sim$100 $\mu m$ de la trajectoire de la particule.
\subsection{Système d'acquisition des données de MACRO}
Le système d'acquisition des données de MACRO est un ensemble de calculateurs permettant la liaison à grande distance, la lecture des données relatives aux évènements ainsi que le contrôle du détecteur.\\
L'acquisition est assurée par un réseau de six microvax et d'un vax central. Les microvax et le vax centrale sont liés par un réseau ETHERNET. Les microvax fonctionnent avec un système opératif VAXELN alors que le vax central qui est un VAX 4000/5000 fonctionne avec le système opératif VAX/VMS.\\
Le système VAXELN optimise des opérations temporelles et permet l'échange rapide et l'écriture des messages entre les différents constituants du réseau. Les microvax assurent le bon fonctionnement du détecteur, lisent et présélectionnent les évènements avant de les transmettre au vax central. Ils sont branchés en parallèle avec les parties électroniques de l'appareillage à travers une interface de modules CAMAC et VME. Ils s'occupent d'autre part des opérations périodiques de calibrations des TS et sont affectés à certains contrôles du fonctionnement de l'appareil. Les trois premiers microvax contrôlent les informations des SM groupés en deux, les trois autres enregistrent les informations relatives aux neutrinos des supernova.\\
Le système opératif VAX/VMS qui gère le VAX central joue le rôle de serveur NETWORK et I/O pour le système opératif VAX/VMS pour le système VAXELN qui gère les microvax et les composants du système VMS. Il constitue ainsi le support des histogrammes et rassemble les alarmes des différents microvax pour les mémoriser sur des disques de haute capacité. Il permet aussi la liaison du système d'acquisition avec l'extérieur par l'intermédiaire de fibre optique.
\subsection{Contrôle spontané d'acquisition}
\begin{figure}[p]
\begin{center}
\leavevmode
\epsfig{file=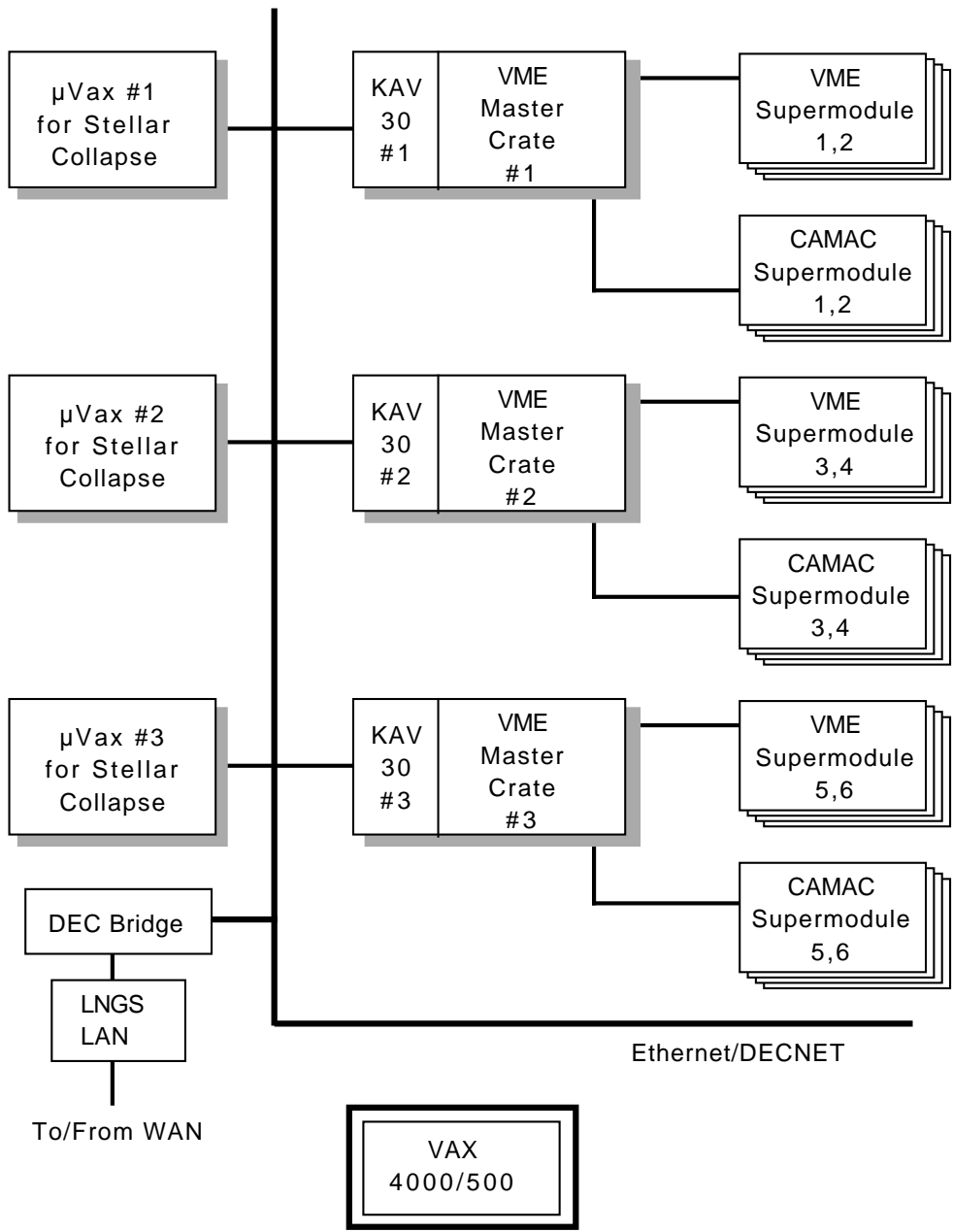,width=13cm}
\caption{\em \textbf{Système d'acquisition des données utilisé par MACRO.}}
\label{das}
\end{center}
\end{figure}

\begin{figure}[p]
\begin{center}
\leavevmode
\vskip -2cm
\epsfig{file=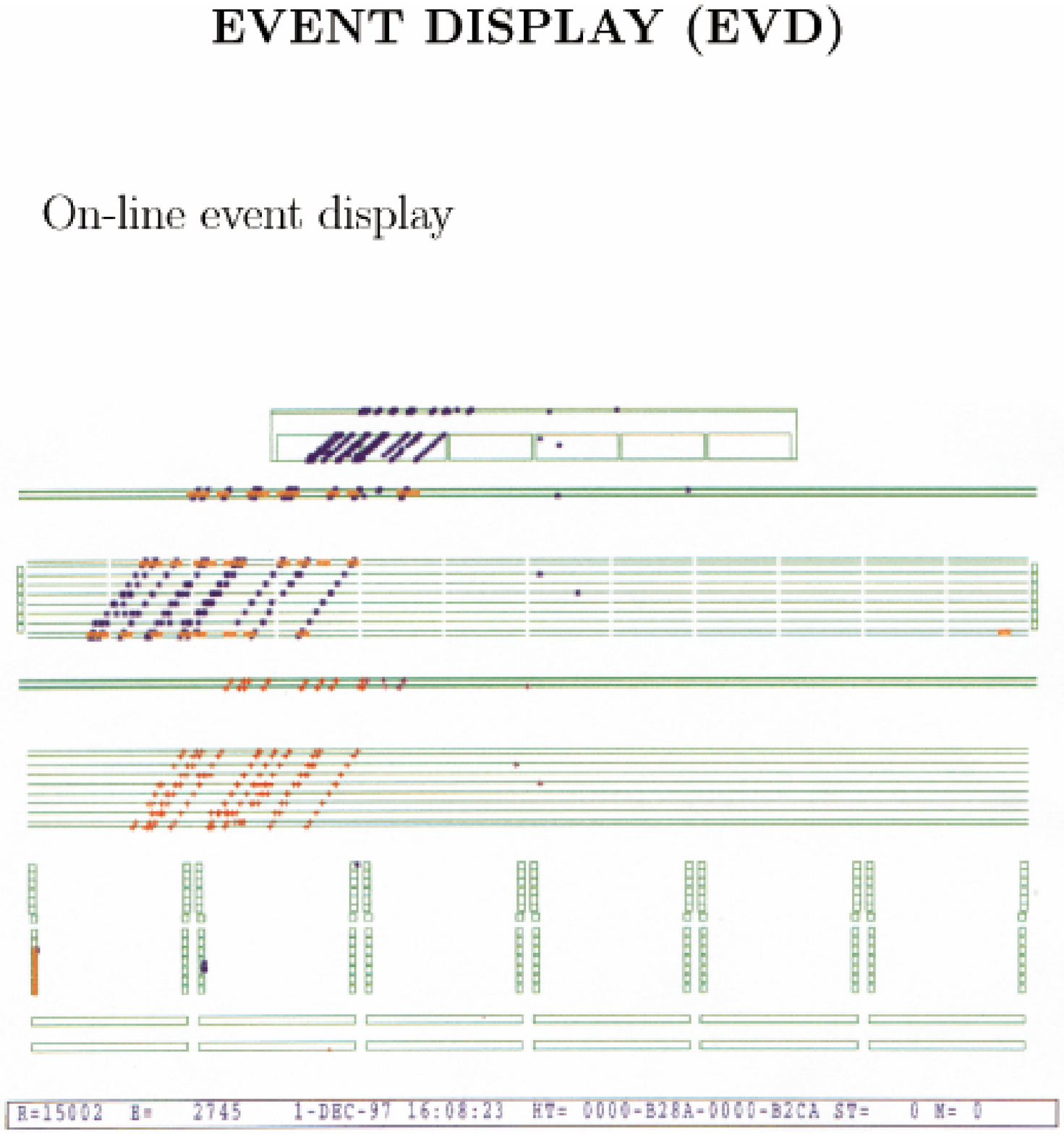,height=20cm, width=14cm}
\vskip -2cm
\caption{\em \textbf{Groupe de 10 muons visualisé par EVD du MACRO.}}
\label{display}
\end{center}
\end{figure}

Le système d'acquisition des données utilisé par MACRO est représenté sur la figure (\ref{das}), il est contrôlé en temps réel par trois principaux processus s'activant sous le système VMS/VAX.
\begin{itemize}
	\item Un programme et un terminal jouant le rôle du console et à partir duquel seront exécutées les commandes de l'initialisation de l'acquisition. Il permet le contrôle des diverses parties de l'appareil en temps réel (la fréquence des triggers, les tensions d'alimentation, le temps mort...). On peut également voir les messages d'erreurs et les alarmes. 
	\item Un programme de présentation des histogrammes: "Histogram Presenter" permet de voir si toute les composantes de l'appareil fonctionnent normalement en visualisant des histogrammes de fonctionnement et comparés à ceux de référence. Parmi les représentations graphiques nous avons le temps de comptage de chaque trigger, le nombre de muons détectés par demi-heure, le temps de vol et le temps mort de chaque microvax par demi-heure.
	\item Un système de visualisation de la trajectoire des évènements dans les différents plans du détecteur "EVent Display" (EVD). Sur la figure (\ref{display}) on présente un groupe de muons visualisé par EVD de MACRO. 
\end{itemize}
\section{Les recherches en physique fondamentale de \\l'expérience MACRO}
MACRO a été destiné à la recherche des monopôles magnétiques et des rayons cosmiques comme son nom l'indique. Cette recherche est tellement variée de telle sorte qu'il avait plus d'une dizaine de sujets qui ont été traités par les physiciens de MACRO. Dans cette section on discute les différents sujets traités par le détecteur.
\subsection{Détection des Monopôles magnétiques avec MACRO}
Dans ce paragraphe, on présente les derniers résultats de MACRO concernant l'acceptance et le flux des MMs \cite{MAC23}\cite{MAC23b}. Pour cela il existe deux méthodes d'analyses: 
\begin{itemize}
	\item l'analyse indépendante qui donne les résultats de recherches concernant chaque détecteur indépendamment de l'autre
	\item l'analyse combinée concernant tout MACRO comme un seul détecteur.
\end{itemize}
Ces études ont été menées pour différentes vitesses des MMs. Sachant qu'après une recherche de plusieurs années aucun candidat MM n'a été trouvé, on présente les limites supérieures sur le flux.\\
A faibles vitesses, la recherche des MMs avec les scintillateurs ou les tubes à streamer est très efficace . Par contre elle est moins efficace pour les MMs de grandes vitesses.\par
Les résultats présentés ici sont relatives au MMs de charge unité. Les résultats sont valables en considérant une section efficace de catalyse inférieure à 10 mb. On suppose que le flux des MMs est isotrope.\\
\textbf{a) Recherches des MMs avec les scintillateurs liquides:\\}
La recherche des MMs avec les scintillateurs liquides se fait sur une large gamme de vitesses; ce qui nécessite des méthodes différentes pour la sélection et l'analyse. On distingue trois domaines de vitesses pour la recherche des MMs:\par
Pour les MMs de faibles vitesses, la perte d'énergie des MMs pour $10^{-4}<\beta<5\times10^{-3}$ est similaire à celle des protons ayant des vitesses analogues dans le liquide scintillant. La limite supérieure sur le flux établie pour des vitesses du MM pour $10^{-4}<\beta<10^{-3}$ est de $3.4\times10^{-16}\;cm^{-2}s^{-1}sr^{-1}$ \cite{MAC24}.\par
Pour les vitesses moyennes des MMs, la lumière produite par unité de longueur et le temps de vol pour parcourir MACRO sont utilisés en même temps. Dans cette procédure, le bruit de fond, principalement dû à la radioactivité naturelle et aux muons atmosphériques arrêtés dans le détecteur, est complètement rejeté. La limite supérieure sur le flux à 90\% CL est $3.2\times10^{-16} cm^{-2}s^{-1}sr^{-1}$ pour des vitesses $1.2\times10^{-3}<\beta<10^{-1}$.\par
Pour les MMs de grandes vitesses, la lumière produite par perte d'énergie des MMs de grande vitesse est collectée par les photomultiplicateurs. Pour ce type d'analyse le bruit de fond est dû essentiellement aux muons de haute énergie et qui peuvent être rejeté puisque l'énergie perdue par les muons dans les scintillateurs est nettement inférieure à celle que perdrait le MM. Cette analyse a permis de fixer une limite supérieure sur le flux et qui est égale à $4\times10^{-15}cm^{-2}s^{-1}sr^{-1}$ pour $\beta>10^{-1}$.
\textbf{b) Recherches des MMs avec les tubes à streamer:\\}
La présence de l'hélium dans les tubes à streamer donne une efficacité de 100\% vis à vis de la détection des MMs en exploitant l'effet Drell et Pening pour les faibles vitesses. L'analyse est basée sur la recherche de traces singulières et la mesure de la vitesse du MM candidat . Seuls les plans horizontaux des tubes à streamer limités de la partie inférieure de MACRO ont été utilisés dans le trigger. Les plans de la partie supérieure et les plans verticaux ont été utilisés pour la reconstruction des traces. La limite sur le flux obtenue par MACRO pour des vitesses $1.1\times10^{-4}<\beta<5\times10^{-3}$ est $8.9\times10^{-15}cm^{-2}s^{-1}sr^{-1}$ \cite{MAC18}.\\
\textbf{c) Recherches des MMs avec le CR39\\}
Le CR39 est utilisé dans MACRO comme un détecteur passif pour la détection des MMs, il peut être également utilisé pour la confirmation du passage d'un MM une fois sa détection est signalée par les scintillateurs et/ou les tubes à streamer limités.\par
Le CR39 a été soumis à des calibrations avec des ions lents et rapides afin de connaître sa réponse lors du passage des particules. La limite supérieure sur le flux trouvée est $6.8\times10^{-16} cm^{-2}s^{-1}sr^{-1}$ pour des vitesses de l'ordre de $\beta\sim 1$ et $10^{-15} cm^{-2}s^{-1}sr^{-1}$ pour des vitesses de l'ordre de $\beta\sim 10^{-4}$.\\
\textbf{d) Analyse combinée\\}
L'analyse combinée est une analyse faite en utilisant les résultats des différents détecteurs constituant MACRO afin d'avoir une limite supérieure combinée sur le flux \cite{MAC25}.\par
\begin{figure}[h]
\begin{center}
\leavevmode
\vskip -3cm
\epsfig{file=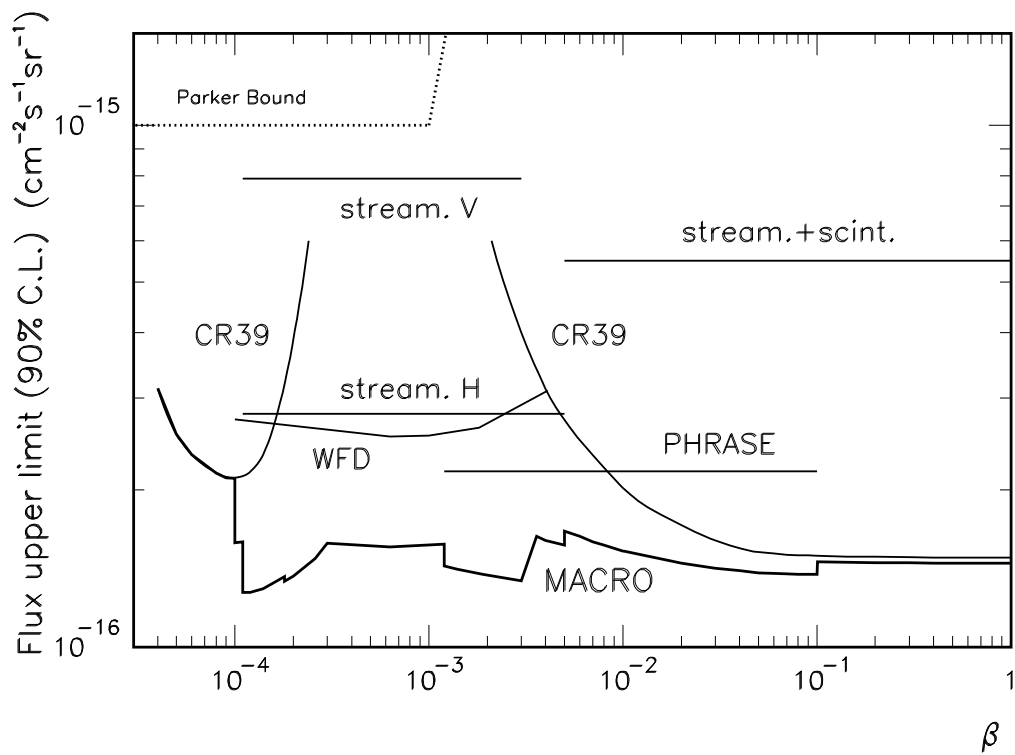,width=14cm}
\caption{\em \textbf{Limite supérieure sur le flux des MMs à 90\%CL avec g = g$_{D}$, obtenue par MACRO et par d'autre expériences.}}
\label{limit}
\end{center}
\end{figure}

Cette analyse a été faite pour 2774 jours avec une efficacité moyenne de 77\% et une acceptance de 3565 $m^2sr$; la limite sur le flux trouvée est $7.6\times 10^{-16} cm^{-2}s^{-1}sr^{-1}$ pour des vitesses $5\times10^{-3}<\beta<0.99$. Sur la figure (\ref{limit}) on présente la limite supérieure sur le flux des MMs du détecteur MACRO confrontée aux autres détecteurs.
\newpage
\subsection{Les Neutrinos des rayons cosmiques}
Les neutrinos sont caractérisés par leur pouvoir de pénétration et leur propagation en ligne droite (sans subir les modifications dues au champ magnétique galactique). Ils peuvent être utilisés comme moyen pour la recherche de sources astrophysiques ponctuelles des rayons cosmiques. Ceci se fait en détectant les muons du bas issus de l'interaction des neutrinos avec la roche se trouvant au dessous de MACRO : 
\begin{equation}
\nu_\mu+X\longrightarrow\mu+Y
\end{equation}
MACRO peut également mesurer le flux des neutrinos du bas et le comparer avec les prédictions des Monte Carlo afin de mettre en évidence l'existence d'un éventuel déficit de ces particules \cite{MAC25}.\par
MACRO, grâce aux scintillateurs, est également sensible aux antineutrinos des supernovae \cite{MAC27} \cite{MAC17} par l'intermédiaire de leur interaction avec les protons libres se trouvant dans le liquide scintillant suivant la réaction suivante:
\begin{equation}
\overline{\nu_\mu}+p\longrightarrow n+e^+
\end{equation}
Le seuil énergétique de détection atteint par MACRO est de l'ordre de 7 MeV.
\subsection{Les muons des rayons cosmiques}
La composante muonique est très importante dans les recherches faites par MACRO. On distingue:
\begin{itemize}
	\item \textbf{Temps d'arrivée des muons :} l'analyse du temps d'arrivée des muons est une technique permettant d'étudier les différents mécanismes aux quels ont été exposés les rayons cosmiques avant leur arrivée sur terre et les modulations introduites dans la distribution du temps d'arrivée de ces derniers \cite{MAC01}\cite{MAC29b}.
	\item \textbf{Astronomie des muons :} cette analyse porte sur la recherche des sources 	astrophysiques ponctuelles des rayons cosmiques, les résultats obtenus concernent la recherche de sources ponctuelles à travers tout le ciel (\textit{All Sky Survey}) et l'étude de sources périodiques connues comme émettrices de rayons X telles que Cyg-X3 et Her -X1 \cite{MAC01}\cite{MAC29b}.
	\item \textbf{Variation saisonnière : }il a été montré que la composante secondaire des rayons cosmiques présente des variations en fonction du temps. Ces variations sont d'origine météorologique \cite{MAC19}.
	\item \textbf{Ombre de la lune : } cette analyse met en évidence un déficit des muons provenant de la direction de la lune. Elle permet de vérifier la précision angulaire du détecteur MACRO. Les résultats obtenus en analysant 30 millions de muons montrent un déficit de l'ordre de $3.7\sigma$ \cite{MAC30}.
	\item \textbf{Décohérence : } c'est la distribution de la distance de séparation de deux muons dans le cas des évènements multiples. Les données expérimentales ont été comparées avec le modèle Monte Carlo HEMAS pour le flux des muons souterrains. Ce qui permet d'examiner les effets systématiques possibles dus aux incertitudes sur la section efficace d'interaction des primaires, la perte d'énergie dans la roche et l'effet du champ géomagnétique \cite{MAC31}.
	\item \textbf{Décorrélation : }la fonction de décorrélation des muons décrit la relation entre la séparation angulaire et spatiale des muons souterrains. Comme pour la fonction de décohérence, elle est sensible aux prévisions du modèle d'interactions des rayons cosmiques de hautes énergies et indépendante de la composition des primaires \cite{MAC32}.
\end{itemize}
\subsection{recherche des nucléarites et des WIMPS}
L'existence de nouvelle forme de matière stable contenant un nombre égal de quarks up (u), down (d) et strange (s) a été proposée par différents auteurs. Cette matière peut avoir des masses allant de quelques GeV à la masse d'une étoile à neutrons. De Rujula et Glashow ont suggéré que cette matière étrange peut exister dans les rayons cosmiques \cite{nuc7}. MACRO avec les détecteurs à scintillation et le CR39, peut étudier l'existence de cette matière sur un large spectre de vitesses qui peut s'étendre de $\beta = v/c = 1$ jusqu'à la zone des $\beta$ des nucléarites qui peuvent être capturés dans notre système solaire.\\
La recherche des nucléarites à partir du détecteur MACRO peut se faire essentiellement par les scintillateurs et le détecteur CR39. Par contre, les tubes à streamer ne sont pas sensibles à ces particules du fait que le gaz se trouvant dans ces tubes a une faible densité. \par
Les WIMPS (Weakly Interacting Massive Particles) sont des particules non baryoniques et représentent un candidat pour la matière obscure \cite{MAC34}. Il font partie de la matière galactique lumineuse qui a une contribution inférieure à $0.1\%$ dans le densité de l'univers.\\
Les WIMPS peuvent être capturés par le soleil ou par la terre et à travers les interactions avec les noyaux ils se concentrent au centre où ils thermaliseront. Une paire de telles particules $(w,\bar{w})$, peut s'annihiler en émettant des paires $\nu\;\bar{\nu}$. Les neutrinos émis peuvent avoir une énergie moyenne $\bar{E_\nu} \approx m_w/3$ (quelques dizaines de GeV). La recherche de ces particules peut être réalisée en détectant les $\mu$ du bas provenant des neutrinos ayant la direction du soleil ou du centre de la terre. 

\chapter{Analyse de la distribution des temps d'arrivée des muons de haute énergie}
\label{chapdt}
\section{Introduction}
\hskip 12pt 
Il est généralement admis que la distribution du temps d'arrivée des rayons cosmiques galactiques est un phénomène aléatoire. Cependant il existe quelques mécanismes introduisant des modulations dans cette distribution. La détection des rayons gamma de haute énergie ($10^{14}-10^{16}\;eV$) provenant des sources galactiques ou extragalactiques a conduit aux développement de différents modèles qui décrivent ces sources dans l'espace environnant (pulsars, systèmes binaires, etc...).\\
Weekes \cite{dtr1} a suggéré que l'analyse des rayons cosmiques (RC) peut révéler la présence de composantes dont l'origine est un pulsar ou un objet céleste similaire qui n'est pas entouré de nébulosité, ayant une courte période et qui n'est pas situé à de grandes distances. De ce fait, une analyse des distributions temporelles des rayons cosmiques est importante. Dans le cas des rayons cosmiques primaires chargés, les modulations des RC sont réduites ou éliminées par l'effet des champs magnétiques galactiques et extragalactiques présents entre la source et la terre. Ceci provoque la perte de la périodicité du signal provenant des sources situées à une distance inférieure à 150~pc. Les mêmes effets sont observés dans le cas des RC arrivant d'une supernovae. Les rayons cosmiques de haute énergie gardent la périodicité du signal, ils peuvent préserver la modulation et donc on ne peut exclure la présence de structures dans la distribution temporelle. Il est donc intéressant d'examiner ces distributions indépendamment de toutes les hypothèses.\par
Plusieurs recherches utilisant des détecteurs souterrains ou en surface exploitant "les gerbes électromagnétiques dans l'air (EAS)", n'ont pas reporté de modulation dans la distribution temporelle des RC de haute énergie et par conséquent confirment l'arrivée aléatoire des rayons cosmiques. Parmi ces études nous citons:
\begin{itemize}
	\item Morello et al. \cite{dtr4}, à l'aide des mesures EAS avec deux différents seuils d'énergie $E>50\;GeV$ et $E>2\;10^{4}\;GeV$.
	\item La collaboration NUSEX \cite{dtr5}, est une expérience souterraine sous le Mont blanc utilisant un détecteur composé de plans de Fer séparés par des plans de tube à streamer. Cette expérience a mesuré le flux des muons à une profondeur de $400\;m.w.e.$ \footnote{$\ 1\  m.w.e (1 meter water equivalent)= 1hg/cm^2$}
	\item La collaboration MACRO \cite{MAC01}, a mesuré le flux des muons singuliers et multiples à l'aide du détecteur MACRO fonctionnant avec 2 et 4 supermodules seulement et en utilisant uniquement la partie inférieure du détecteur. 
\end{itemize}
\par 
Deux expériences ont confirmé la présence d'une composante non aléatoire dans la distribution du temps d'arrivée des rayons cosmiques de haute énergie : 
\begin{enumerate}
	\item Bhat et al. \cite{dtr7} ont analysé les impulsions de la lumière Cherenkov produites dans l'atmosphère par des RC d'énergie primaire $E_{p}\geq 100\;TeV$. Ils ont observé que la corrélation se produit au temps $t<40s$. L'expérience a utilisé deux photomultiplicateurs avec un grand angle d'observation à 2743 m  d'altitude.\\
La figure (\ref{fig:dt1}) donne la distribution des temps d'arrivée de deux évènements successifs et montre une déviation supérieure à $5\;\sigma$ pour $\Delta t<40s$.
\begin{figure}
\centering
\includegraphics{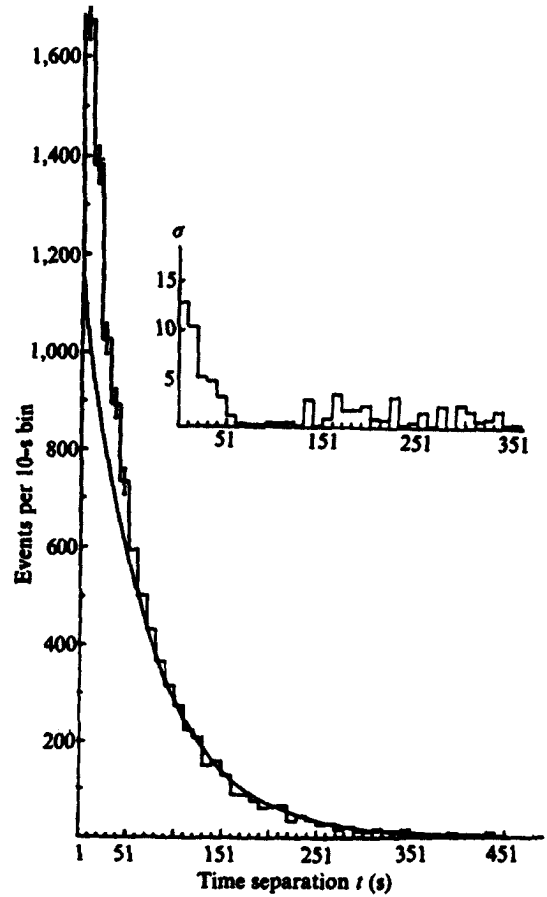}
\caption{\em \textbf{Résultat de l'expérience de Bhat et al. \cite{dtr7}. L'histogramme montre la distribution observée des temps d'arrivée des évènements dans un bin de 10s. La courbe représente l'ajustement par la fonction exponentiel théorique (en considérant que l'arrivée des événements est complètement aléatoire). Une déviation de cette distribution a été observée pour $t<40s$.}}
\label{fig:dt1}
\end{figure}
	\item Badino et al. \cite{dtr8}, en utilisant un détecteur souterrain dans le tunnel du Mont blanc à une profondeur de $5000 hg/cm^2$ de roche, ont observé un excès de muons en étudiant la distribution temporelle des muons. Les muons détectés ont une énergie supérieure à $E_\mu=380GeV$ et sont originaires des RC primaires d'énergie $E_p\approx\;50TeV$. L'excès a été observé pour un intervalle de temps séparant deux évènements successifs d'environ $38s$. Cet excès de muons (1\%) mis en évidence dans un intervalle de temps de quelques dizaines de secondes, révèle la présence d'une composante modulée superposée à la distribution temporelle poissonnienne.
\end{enumerate}
La suite de ce chapitre, est consacré à l'étude des distributions de temps d'arrivée des muons cosmiques, réalisée avec les données du détecteur MACRO fonctionnant avec l'ensemble de ces supermodules (six). Notre objectif est de conclure sur la nature de ces distributions et ainsi se situer par rapport aux résultats obtenus par les expériences citées ci-dessus.\par
Un grand échantillon de muons a été utilisé comparé à celui utilisé dans les travaux ultérieurs de MACRO \cite{MAC1}. La collecte de tel échantillon était possible en utilisant la période durant laquelle le détecteur fonctionnait avec les six supermodules y compris la partie supérieure du détecteur "attico". Il faut noter que cette expérience est en mesure de détecter environ $6\;10^6$ muons/ans grâce à sa grande acceptance. De ce fait, nous nous proposons d'étudier le temps d'arrivée des muons de différentes énergies collectés durant la période allant de 1995 jusqu'à juin 2000. Cette analyse porte sur trois types d'évènements : 
\begin{itemize}
	\item évènements singuliers (multiplicité égale à 1). Un tel évènement correspond à un $\mu$ simple qui est reconstruit aussi bien sur les fils que sur les strips. Ces muons sont générés par des rayons cosmiques primaires ayant des énergies de l'ordre de 20 TeV. Avec une telle énergie, la particule chargée des RC n'est pas influencée par les modulations solaires. Ces muons nous permettent d'obtenir des informations sur les RC primaires ayant une énergie dans le domaine $10^{13}-10^{14}$ eV.
	\item évènements doubles (multiplicité égale à 2). Un tel évènement correspond à deux $\mu$ parallèles et reconstruits sur les fils et sur les strips. De tels muons sont originaires surtout des primaires chargés ayant une énergie de l'ordre de 200 TeV ou des $\gamma$ de haute énergie mais avec une faible section efficace. Cette zone d'énergie est intéressante puisque différents types de sources galactiques peuvent produire des rayons gamma de telles énergies. 
	\item évènements multiples (multiplicité $\geq$3). C'est un groupe d'au moins 3 $\mu$ reconstruits (sur les fils et les strips). Ils correspondent aux primaires ayant une énergie $>$ 200 TeV. 
\end{itemize}
\par
Dans notre étude nous proposons de comparer la distribution temporelle des muons à celle obtenue d'un processus poissonnien. 
\section{Sélection des données}
Vu l'importance de cette étude et sa grande sensibilité à toute anomalie dans la prise de données, ces dernières ont été sélectionnées avec le maximum de soin en respectant des critères de sélection des runs.\\
Le lot de données utilisé correspond à la période allant de Juin 1995 jusqu'à Mai 2000. Durant cette période d'acquisition, les tubes à streamer ont présenté un temps mort faible inférieure à 0.4\%.\\ 
Ce lot d'évènements est sélectionné pour étudier les distributions temporelles des muons d'énergie supérieure à 1,4 TeV au sommet de la montagne, qui arrivent de la partie supérieure du ciel et ceux qui proviennent des régions restreintes définies dans le système de 
laboratoire (cônes).\\
\subsection{Critères de sélection}
\begin{figure}
\centering
\vskip -2cm
\includegraphics[height=18cm,width=15cm]{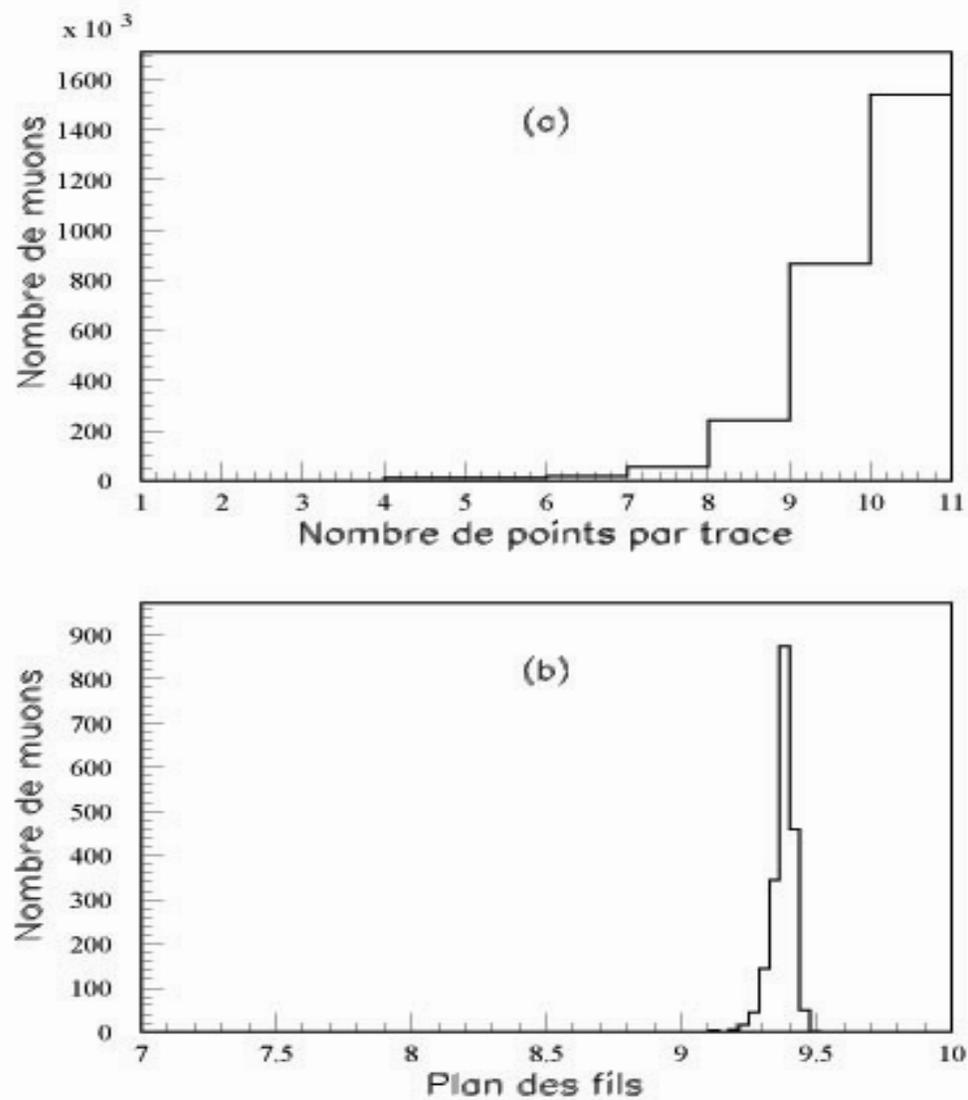}
\vskip -1.5cm
\caption{\em \textbf{(a) Nombre de muons en fonction du nombre de points utilisés pour la reconstruction de la trace; (b) Nombre moyen de plans ayant contribués à la formation de la trace. Seuls les 10 premiers plans ont été utilisés pour cette étude.}}
\label{fig:dt3}
\end{figure}

\begin{figure}
\centering
\includegraphics{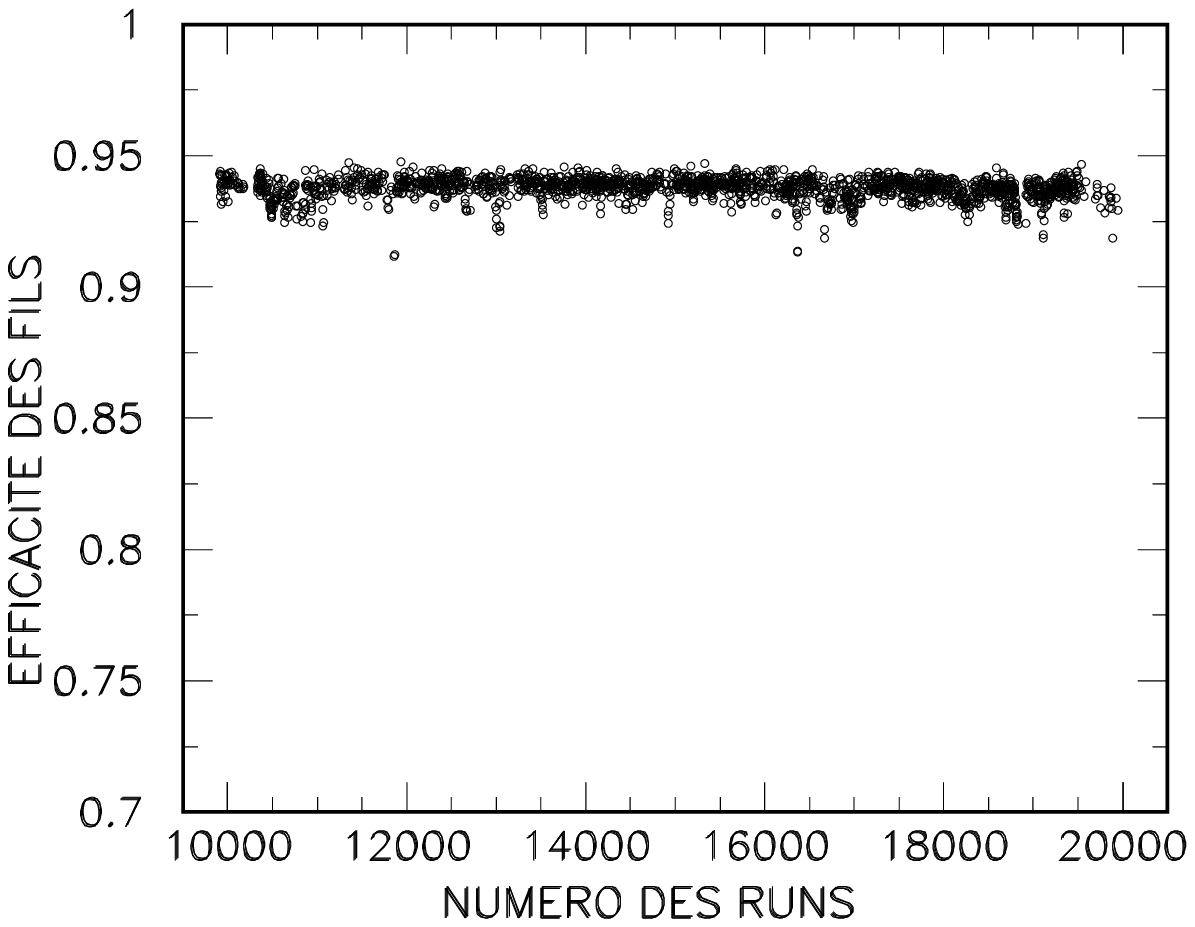}
\caption{\em \textbf{L'efficacité des fils des tubes à streamer pour chaque run sélectionné. Elle est de l'ordre de 94\%.}}
\label{fig:dt4}
\end{figure}

\begin{figure}
\centering
\vskip -4cm
\includegraphics[height=22cm,width=14cm]{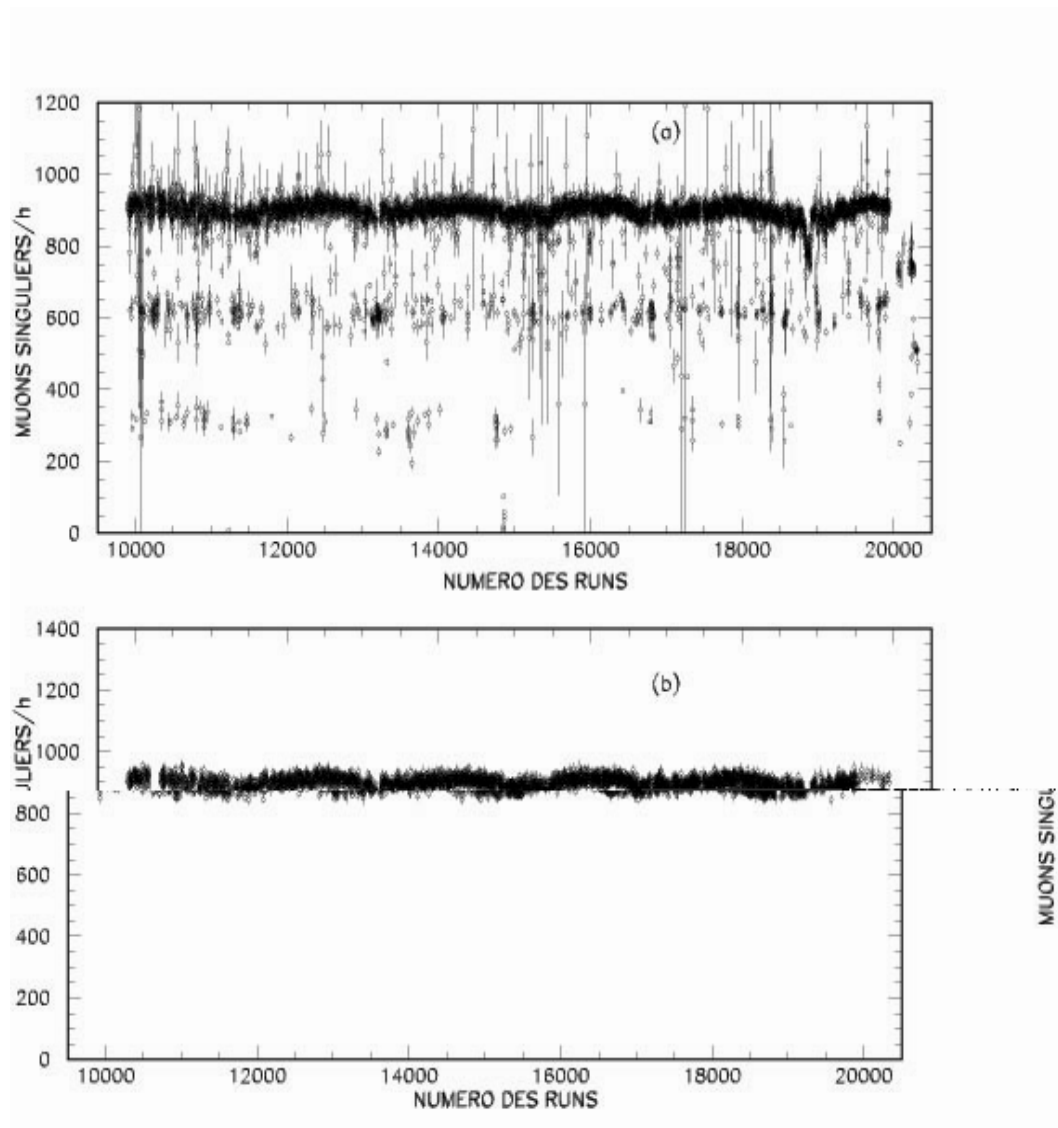}
\vskip -4cm
\caption{\em \textbf{Nombre de muons par heure en fonction du numéro des runs; (a) avant la sélection; (b) après la sélection. On constate bien la présence de la variation saisonnière dans le flux de muons.}}
\label{fig:dt5}
\end{figure}

\begin{figure}
\centering
\vskip -4cm
\includegraphics[height=20cm,width=14cm]{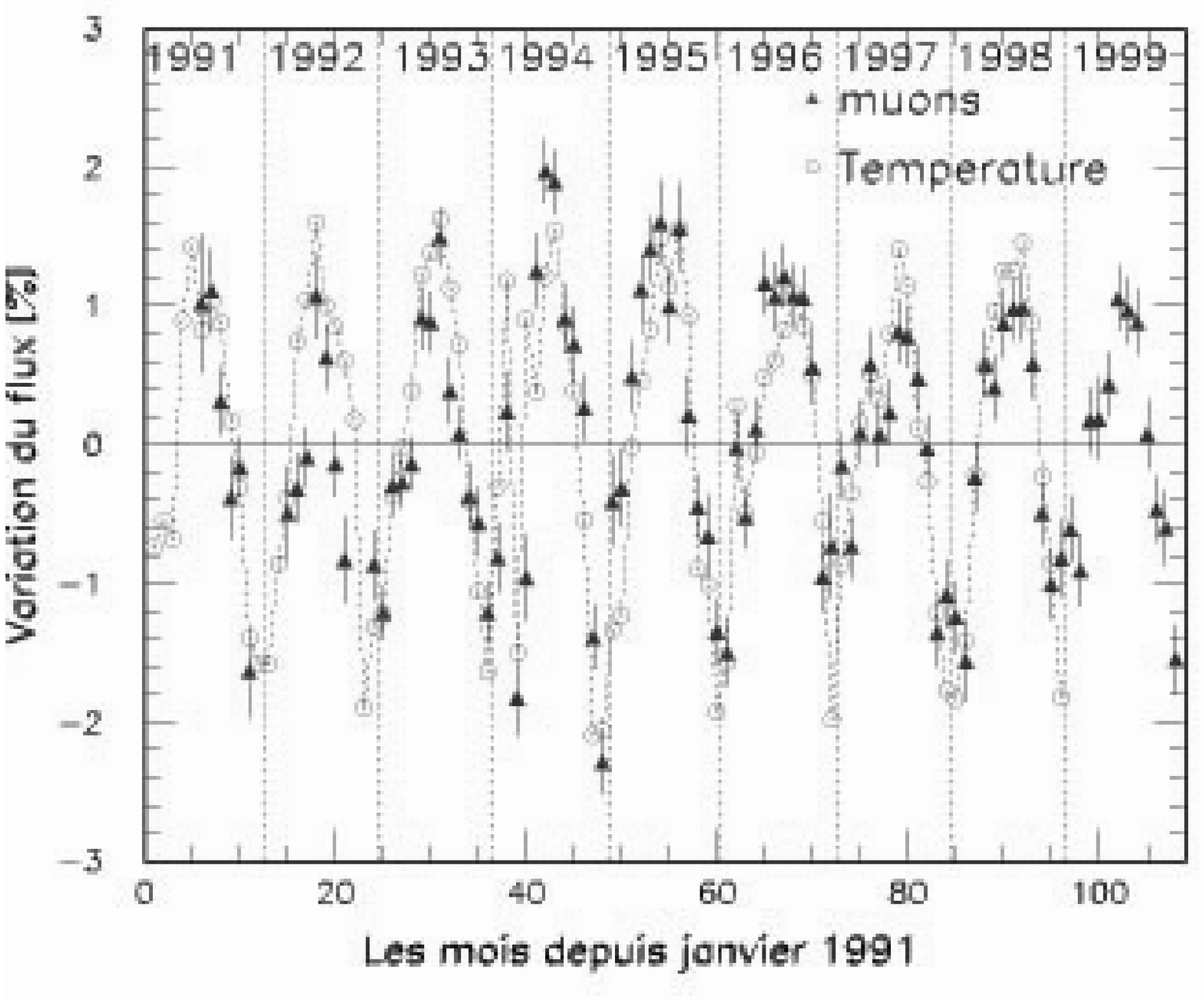}
\vskip -3cm
\caption{\em \textbf{Variation saisonnière du flux de muons \cite{MAC1b}. On observe une diminution du flux de muons en hiver et une augmentation en été. L'amplitude de cette variation est de l'ordre de $\pm 2\%$}}
\label{fig:dt6}
\end{figure}

\begin{figure}
\centering
\includegraphics[height=16cm,width=14cm]{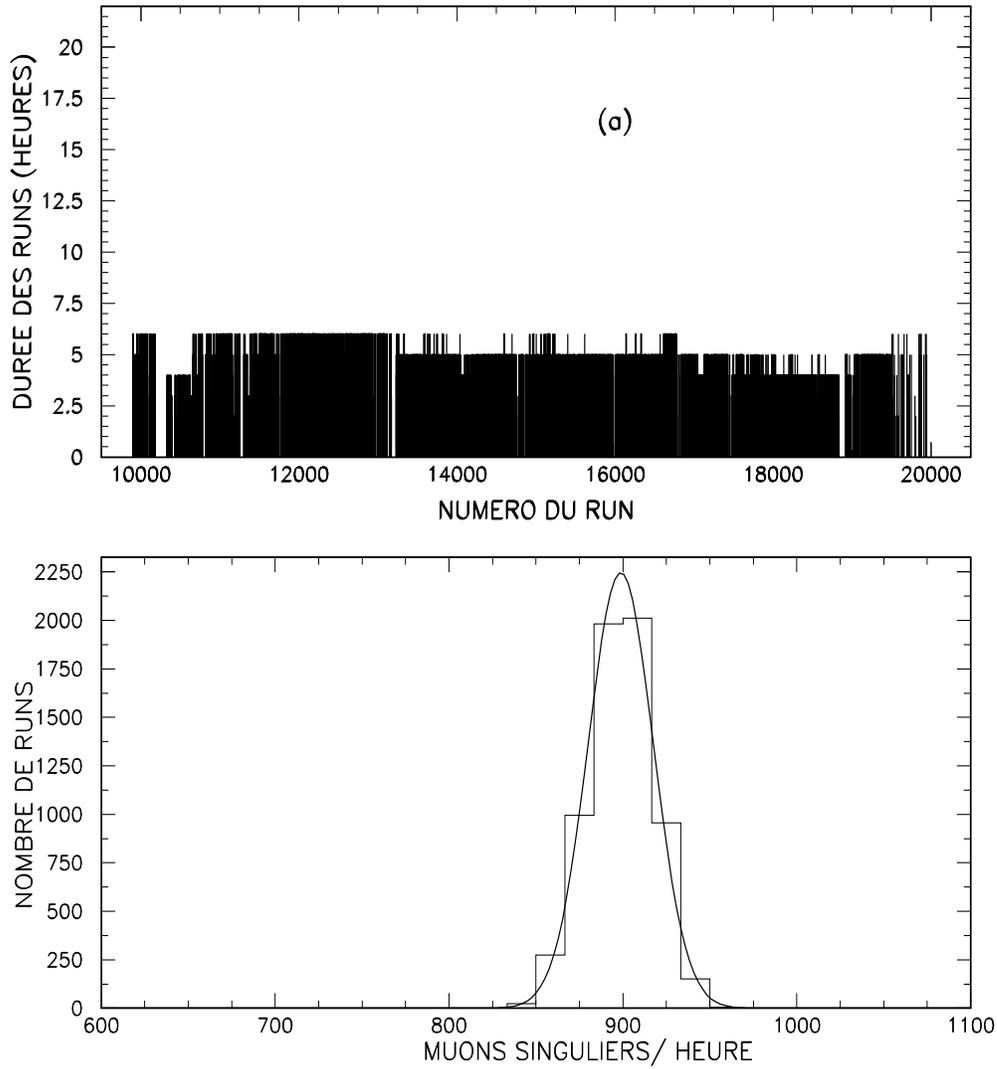}
\caption{\em \textbf{(a) Durée des runs sélectionnés. (b) Nombre de runs en fonction de la fréquence d'arrivée des $\mu$. La fréquence moyenne est 898$\mu$/h.}}
\label{fig:dt7}
\end{figure}

Les critères de sélection auxquels les évènements et les runs doivent satisfaire sont: 
\begin{itemize}
	\item La durée des runs doit être supérieure à 2 heures (la moyenne de la durée des runs est d'environ 5 heures). Ceux qui ont une durée inférieure sont en général des runs de calibration ou  des runs qui présentent des problèmes (électronique, manque de courant, erreurs du système d'acquisition, etc...).  
	\item La fréquence des muons R est choisie dans l'intervalle $840<R<960\;\mu/h$ et située dans un intervalle $\pm 2\sigma$ autour de la valeur moyenne. Les runs qui présentent une fréquence élevée ont été  analysés séparément pour vérifier s'il s'agit d'une fluctuation due au bruit électronique ou d'un éventuel "{\it burst}" de muons. Ce critère nous a permis d'éliminer les runs où l'acquisition n'a pas été réalisée avec tous les 6 SM ou ceux qui présentent des problèmes.
	\item Un temps mort d'acquisition inférieure à 0.4\% pour chaque microvax. Ce qui permet d'éliminer les runs qui présentent un excès de bruits provenant de l'électronique ou des décharges continues des tubes.
	\item Le temps d'arrivée des muons a été mesuré avec une horloge atomique d'une précision d'environ $1\mu s$ absolue et $0.5\mu s$ relative et ne doit présenter aucune anomalie.  
	\item Absence de problème dans le système du gaz dans les tubes à streamer. Un mauvais fonctionnement du système du gaz peut avoir un effet sur l'efficacité des tubes et par conséquent sur la fréquence d'arrivée des $\mu$ détectés. 
	\item L'efficacité de détection des tubes à streamer doit être supérieure à 90\% et 86\% pour les fils et les strips respectivement. 
\end{itemize}
La figure (\ref{fig:dt3}.a) montre le nombre de muons en fonction du nombre de hits utilisés pour la reconstruction de la trace des muons singuliers. Ils faut signaler que seuls les 10 premiers plans ont été utilisés. On constate que la majorité des traces ont été reconstruites avec plus de 8 hits, ce qui donne une idée sur la grande efficacité des tubes à streamer. Les muons reconstruit avec moins de 7 hits, sont généralement les muons qui traversent deux modules juxtaposés ou qui sortent des parois latéraux du détecteur. \\
Sur la figure (\ref{fig:dt3}.b) est présenté le nombre de run en fonction du nombre moyen de plans des tubes à streamer qui ont contribué à la reconstruction de la trace. On constate, là aussi, que le nombre moyen de plan est 9.4 qui est une valeur importante confirmant la grande efficacité de détection des tubes.\\  
La figure (\ref{fig:dt4}), donne l'efficacité des fils de chaque run sélectionné (la valeur moyenne de cette efficacité est de l'ordre de 94\%). Cette efficacité est obtenue en sélectionnant les muons qui traversent uniquement les dix plans des tubes à streamer, sans tenir compte des muons qui sortent latéralement, et en divisant le nombre de points qui contribuent à la reconstruction des traces par le nombre d'évènements par run et par le nombre de plans.\\
D'autres critères de sélection ont été appliqués, un évènement de muon singulier doit avoir une trace avec une multiplicité égale à 1 sur les fils et sur les strips; un évènement de muon double doit avoir deux traces sur le fil et sur le strip et les muons multiples ont des multiplicités entre 3 et 6 et reconstruits sur les fils et sur les strips.\\
Nous présentons dans la figure (\ref{fig:dt5}) la fréquence des muons singuliers en fonction du numéro des runs avant qu'aucune sélection n'a été appliquée (a) et après les sélections (b).\\
Sur la figure (\ref{fig:dt5}.b) on constate la présence d'une variation régulière et qui est la signature d'une variation saisonnière dans le flux de muons due aux changements de la température et de la densité de l'atmosphère. En effet les muons étudiés sont produits par la désintégration des mésons $\pi$ et $k$	dans l'atmosphère. Ainsi, le flux de muons dépend du rapport entre la probabilité de désintégration et celle de l'interaction des mésons parents avec l'atmosphère et qui sont sensibles à la densité et à la température de l'atmosphère. Le flux de muons diminue en hiver, où la température est minimale et l'atmosphère est plus dense, et augmente en été. L'amplitude de cette variation est de l'ordre de $\pm 2\%$ et elle est présentée sur la figure (\ref{fig:dt6}) \cite{dtr10}.\\
L'étude des variations journalière du temps solaire et sidéral du flux des particules montre que le détecteur présente une haute sensibilité pour l'étude des très petites variations du flux de muons (de l'ordre de $10^{-3}$) \cite{dtr11}. Elle montre bien la grande sensibilité de MACRO même aux faibles variations du flux de muons, ce qui est important pour l'étude des variations des temps d'arrivée des muons et pour la recherche des sources astrophysiques des RC.\\
Dans la figure (\ref{fig:dt7}.b) nous avons présenté le nombre de runs en fonction de la fréquence d'arrivée de muons pour chaque run. La fréquence moyenne est égale à $897.7\;\mu/h$.\\
La durée des runs est donnée sur la figure (\ref{fig:dt7}.a). La durée moyenne est de l'ordre de 5 heures.\\
\par 
\section {Effet du temps mort}
\begin{figure}
\centering
\includegraphics{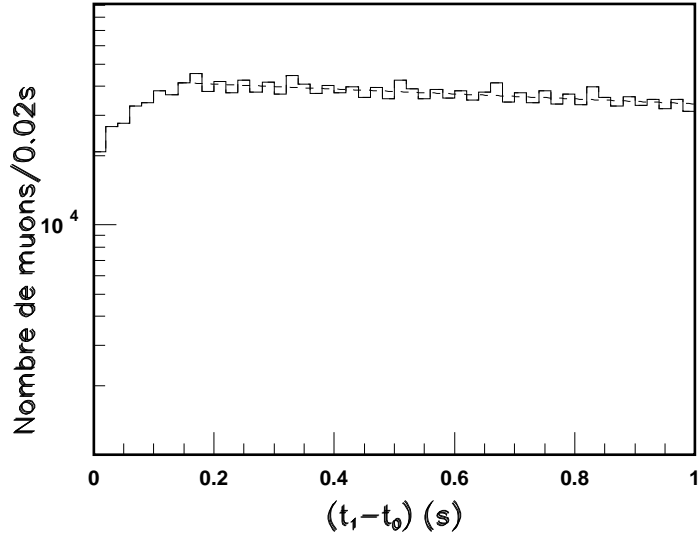}
\caption{\em \textbf{Distribution des temps d'arrivée entre deux muons singuliers consécutifs pour $dt_1<1\;s$, notons l'effet du temps mort sur les premiers bins.}}
\label{fig:dt8}
\end{figure}

Après chaque trigger il existe un temps mort de l'ordre de 100 ms lorsque les scintillateurs étaient en acquisition. L'effet de ce temps mort est apparent dans le cas des distributions $(t_1-t_0)$ pour $(t_1-t_0)<100\;ms$, figure (\ref{fig:dt8}). L'étude de l'effet du temps mort réalisée sur un échantillon de $4\;10^5$ muons détectés par MACRO \cite{MAC01}, montre que l'ajustement avec ou sans correction donne le même résultat.\\
\section{Analyse des corrélations temporelle des muons}
\subsection{Généralités}
Comme il a été déjà mentionné, il est prévu que les rayons cosmiques ont un temps d'arrivée aléatoire à cause des déflexions des particules chargées des rayons cosmiques par les champs magnétiques interstellaires, cependant il  se peut que certains mécanismes introduisent des corrélations dans le temps d'arrivée, par exemple un rayon cosmique originaire d'un pulsar proche peut introduire une modulation aux rayons cosmiques avec une fréquence égale à la fréquence de rotation du pulsar \cite{dtr1}. Des effets similaires peuvent apparaître avec d'autres types de sources de rayons cosmiques. \par
\begin{figure}
\centering
\includegraphics[height=14cm,width=14cm]{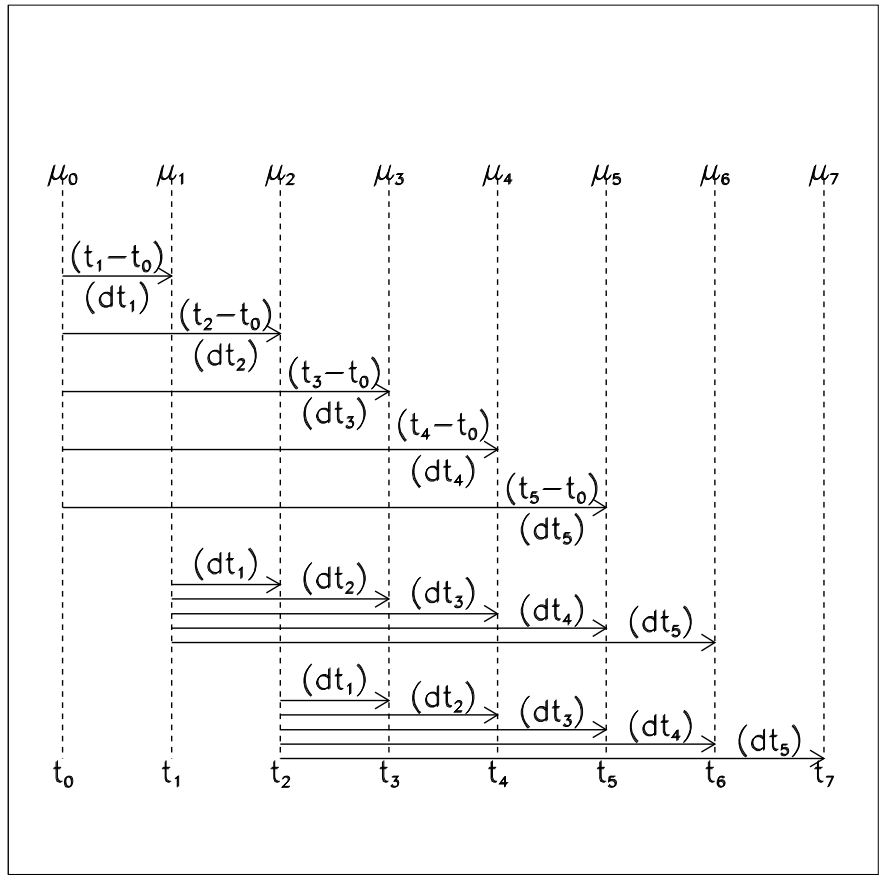}
\caption{\em \textbf{Définition des quantités dt$_{i}$}}
\label{fig:dt2}
\end{figure}

 
Pour chaque muon qui arrive au temps $t_0$, nous analysons la différence de son temps d'arrivée avec le second muon et les quatre muons consécutifs: $t_1-t_0(dt_1)$, $t_2-t_0(dt_2)$, $t_3-t_0(dt_3)$, $t_4-t_0(dt_4)$, $t_5-t_0(dt_5)$ comme c'est définit sur la figure (\ref{fig:dt2}). Cette analyse est effectuée en comparant la distribution temporelle de ces muons à une distribution poissonnienne. On peut prévoir l'observation des clusters dans le temps des évènements générés par des particules chargées ou neutres produites dans une émission périodique d'une source proche.\par
Le flux de rayons cosmiques, en l'absence de modulations dans les mécanismes de production et de propagation, est distribué d'une manière aléatoire dans le temps. Par conséquent nous comparons la distribution des retards entre un $\mu$ et le n$^{ieme}$ $\mu$ (n=1,2,...) successif avec la fonction Gamma qui caractérise les phénomènes aléatoires.\\
La fonction Gamma d'ordre M est définie comme suit \cite{dtr13}: 
\begin{equation}
G(x;N,\lambda,M) = N\lambda\frac{(\lambda x)^{M-1}e^{-\lambda x}}{(M-1)!},\;\;\;\;0<x<\infty 
\label{eq:gamma}
\end{equation}
Où $\lambda$ et M sont deux constantes réelles positives et N un facteur de normalisation. Pour des valeurs entières de M, la fonction Gamma est appelée Erlangienne d'ordre M et elle est équivalente à un ensemble de distributions exponentielles.\par
Pour M=1, l'équation (\ref{eq:gamma}) se réduit à la fonction exponentielle simple: 
\begin{equation}
G(x;N,\lambda,1) = N\lambda e^{-\lambda x}
\label{eq:gamma1}
\end{equation}
et sera appliquée à la distribution de la différence entre les temps d'arrivée d'un $\mu$ et du suivant immédiat ($dt_1$).\\
Une caractéristique de la fonction Gamma est que, au contraire de M, le paramètre $\lambda$ doit rester constant. Dans nos calculs, nous avons déterminé $\lambda$ dans le cas où M=1. Pour les M>1, nous avons laissé le paramètre $\lambda$ libre de varier.\\
Un meilleur ajustement avec la fonction Gamma permet d'obtenir le coefficient $\chi^2/DoF$ voisin de 1. L'ajustement est effectué en utilisant le programme de minimisation MINUIT de la librairie CERN \cite{dtr14}.\par
\subsection{Analyse des Muons arrivant de toutes les directions}
\begin{figure}
\centering
\includegraphics[height=14cm,width=14cm]{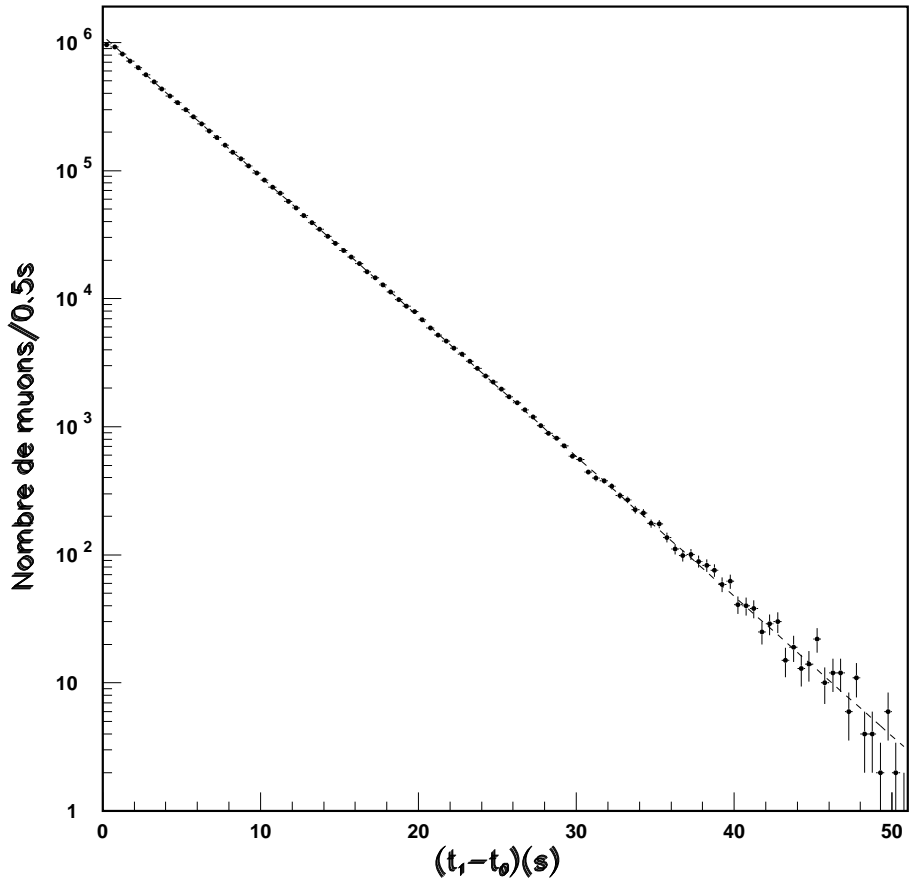}
\caption{\em \textbf{Distribution du temps d'arrivée des $\mu$ singuliers (points). La courbe en pointillés est le résultat de l'ajustement des données par la fonction Gamma d'ordre 1. On constate le bon accord entre les données expérimentales et l'ajustement avec la fonction Gamma d'ordre 1, ce qui montre une arrivée aléatoire dans le temps d'arrivée des muons cosmiques.}}
\label{fig:dt9}
\end{figure}

Nous avons effectué l'analyse sur un large lot de données constitué de $8.6\;10^6$ de muons simples, $0.46\;10^6$ muons doubles et $0.08 \;10^6$ muons multiples qui arrivent de la partie supérieure du ciel. Cette partie est définie en coordonnées célestes par la bande limitée en ascension droite par $0^\circ\leq\alpha\leq 360^\circ$ et en déclinaison par $-20^\circ\leq\delta\leq 90^\circ$ \cite{MAC1}.\\
Nous présentons sur la figure (\ref{fig:dt9}) la distribution temporelle séparants deux muons singuliers consécutifs (dt$_1$). L'ajustement des données expérimentales donne une valeur de $\chi^2/DoF\approx1$, les résultats sont répertoriés dans le tableau (\ref{tab:table}). Ceci indique une adaptation adéquate avec la fonction théorique prévue. Par conséquent la distribution des intervalles des temps d'arrivée entre deux muons consécutifs est compatible avec l'hypothèse qui exclut la présence de toute corrélation temporelle. \par 
\begin{figure}
\centering
\includegraphics[height=14cm,width=14cm]{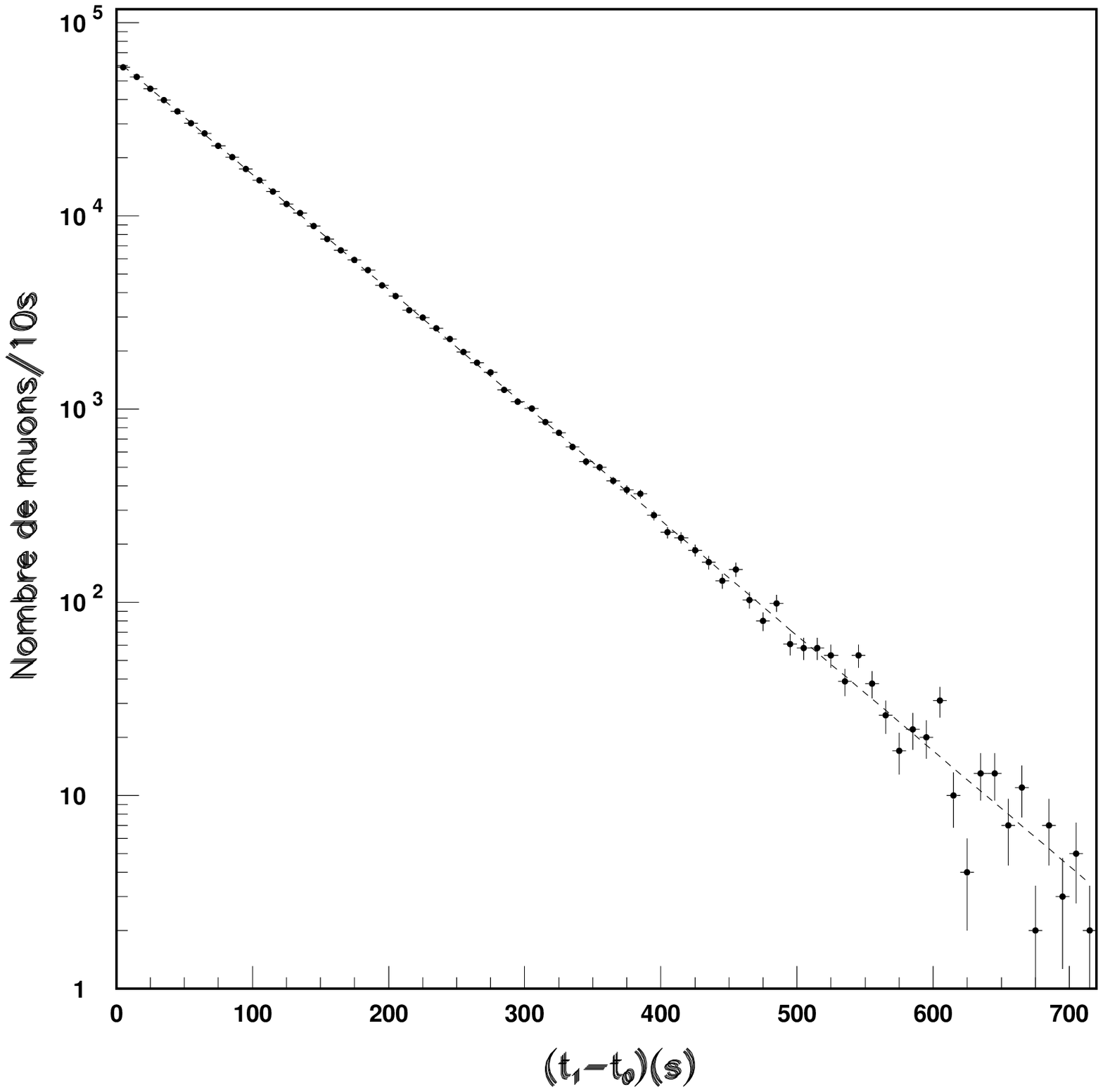}
\caption{\em \textbf{Distribution du temps d'arrivée des $\mu$ doubles. La courbe en pointillés est le résultat de l'ajustement des données par la fonction Gamma d'ordre 1.}}
\label{fig:dt10}
\end{figure}
\begin{figure}
\centering
\includegraphics[height=14cm,width=14cm]{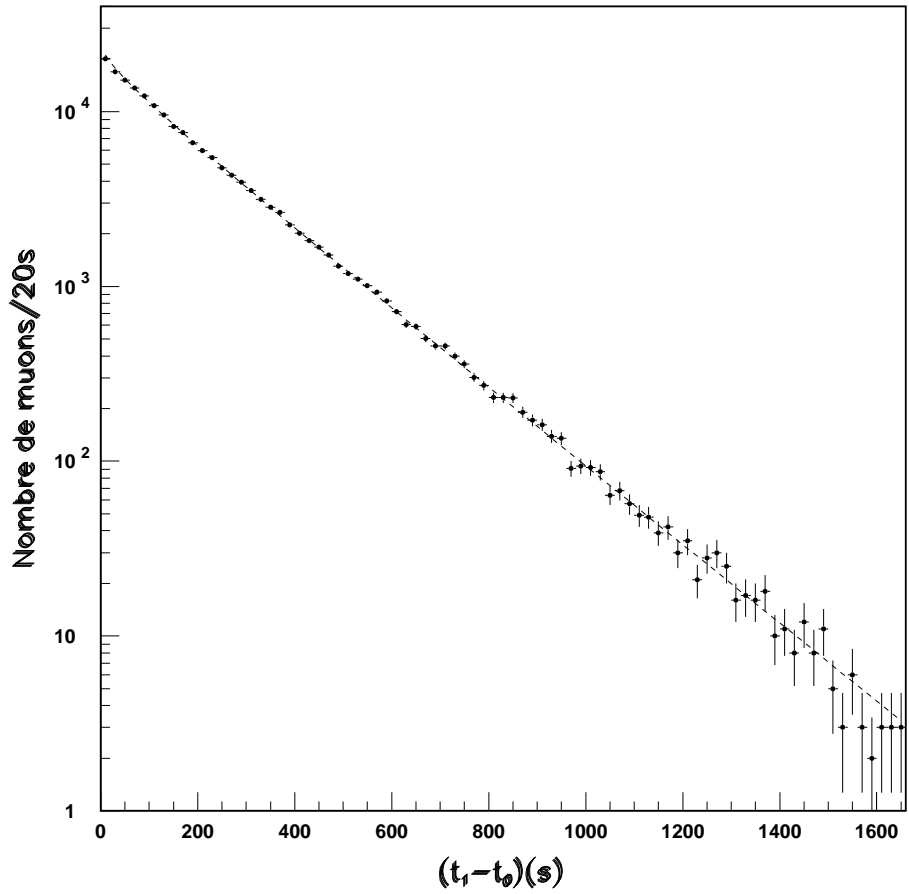}
\caption{\em \textbf{Distribution du temps d'arrivée des $\mu$ multiples. La courbe en pointillés est le résultat de l'ajustement des données par la fonction Gamma d'ordre 1.}}
\label{fig:dt11}
\end{figure}
\begin{figure}
\centering
\includegraphics[height=14cm,width=14cm]{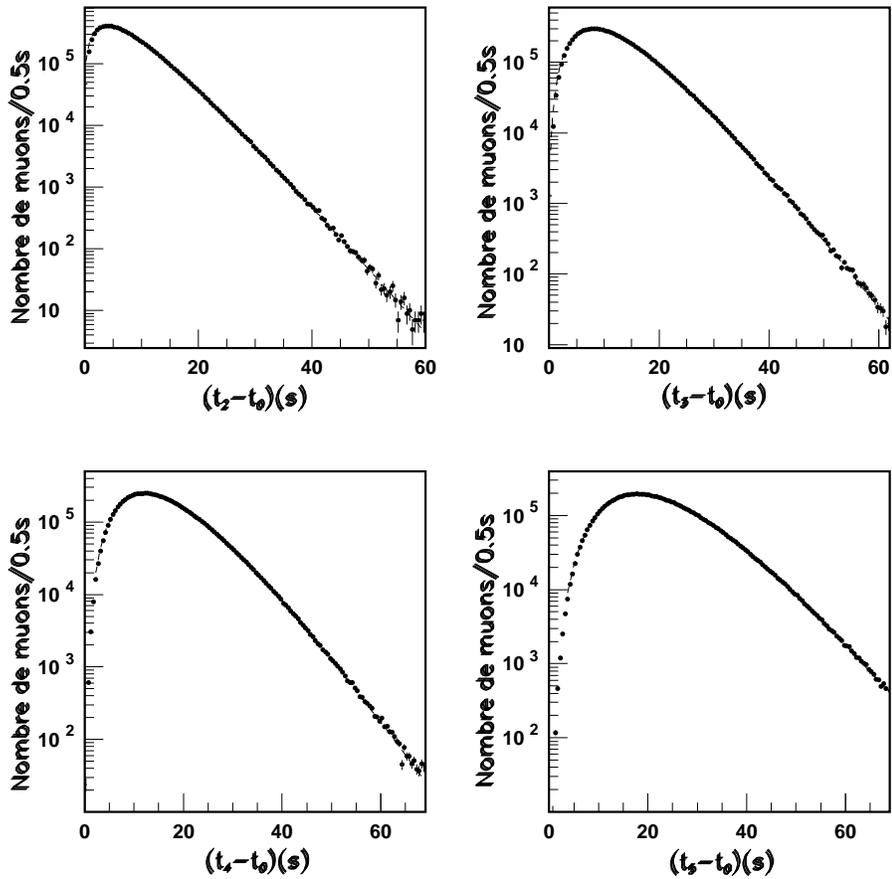}
\caption{\em \textbf{Distribution du temps d'arrivée des $\mu$ singuliers. La courbe en pointillés est le résultat de l'ajustement des données par la fonction Gamma d'ordre 2,3,4,5.}}
\label{fig:dt12}
\end{figure}
\begin{figure}
\centering
\includegraphics[height=14cm,width=14cm]{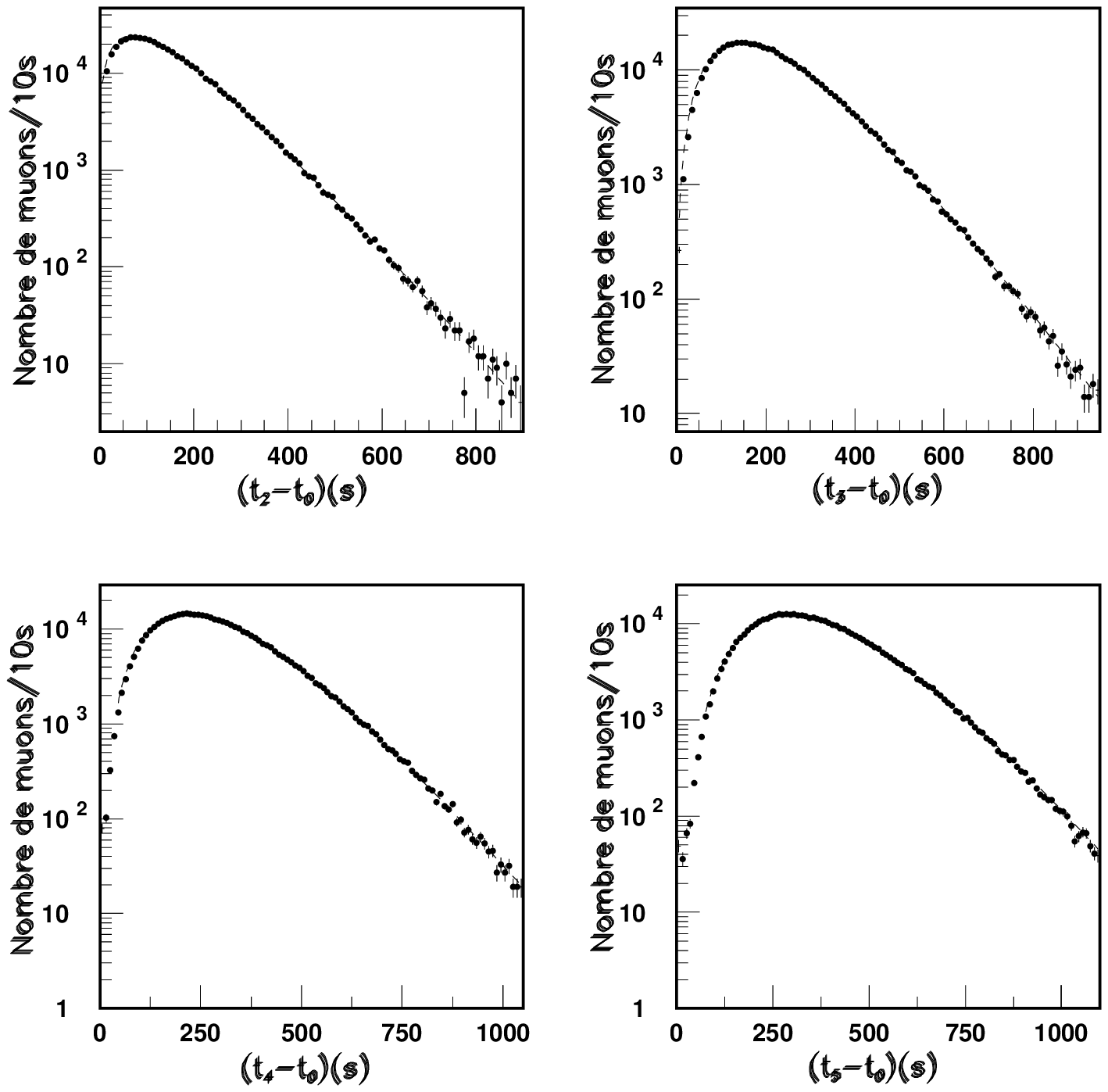}
\caption{\em \textbf{Distribution du temps d'arrivée des $\mu$ doubles. La courbe en pointillés est le résultat de l'ajustement des données par la fonction Gamma d'ordre 2,3,4,5.}}
\label{fig:dt13}
\end{figure}
\begin{figure}
\centering
\includegraphics[height=14cm,width=14cm]{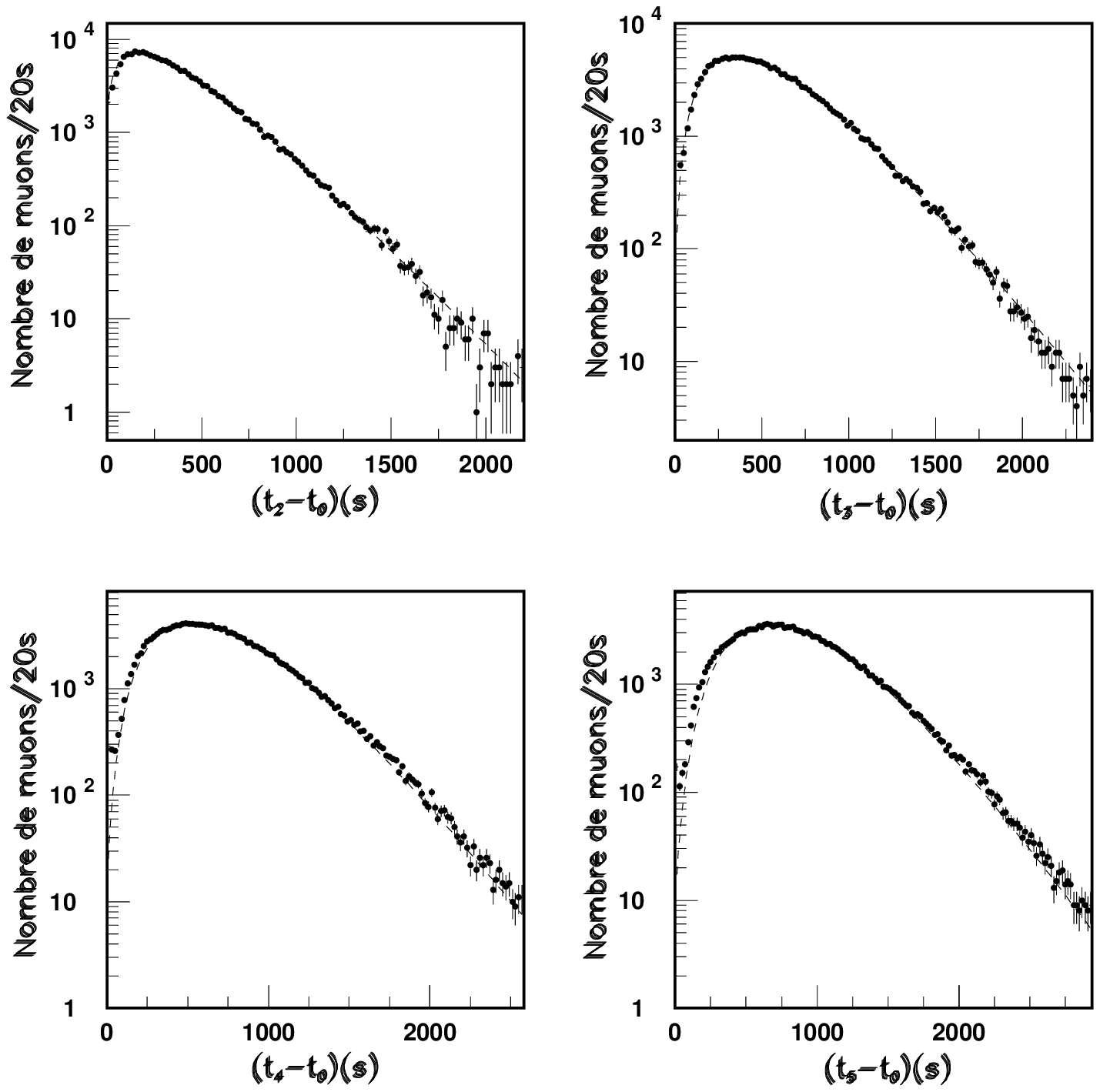}
\caption{\em \textbf{Distribution du temps d'arrivée des $\mu$ multiples. La courbe en pointillés est le résultat de l'ajustement des données par la fonction Gamma d'ordre 2,3,4,5.}}
\label{fig:dt14}
\end{figure}

L'absence d'une éventuelle corrélation temporelle sur $dt_1$, nous a poussé à étudier les ordres supérieurs tels que $dt_2$, $dt_3$, $dt_4$ et $dt_5$.\\
Les résultats relatifs aux muons singuliers, doubles et multiples sont présentés sur les figures (\ref{fig:dt12}), (\ref{fig:dt13}) et (\ref{fig:dt14}) respectivement. Pour chaque ordre, l'ajustement a été effectué avec la fonction gamma d'ordre correspondant (M=2 pour $dt_2$, M=3 pour $dt_3$, etc...). Notons que l'ajustement s'accorde parfaitement avec les données expérimentales, ce qui nous permet de déduire qu'aucune corrélation temporelle n'a été observée et de confirmer les résultats obtenues ultérieurement par la collaboration MACRO \cite{MAC01}.
\subsection{Analyse des muons arrivant des cônes}
\begin{figure}
\centering
\vskip -2cm
\includegraphics[height=22cm,width=16cm]{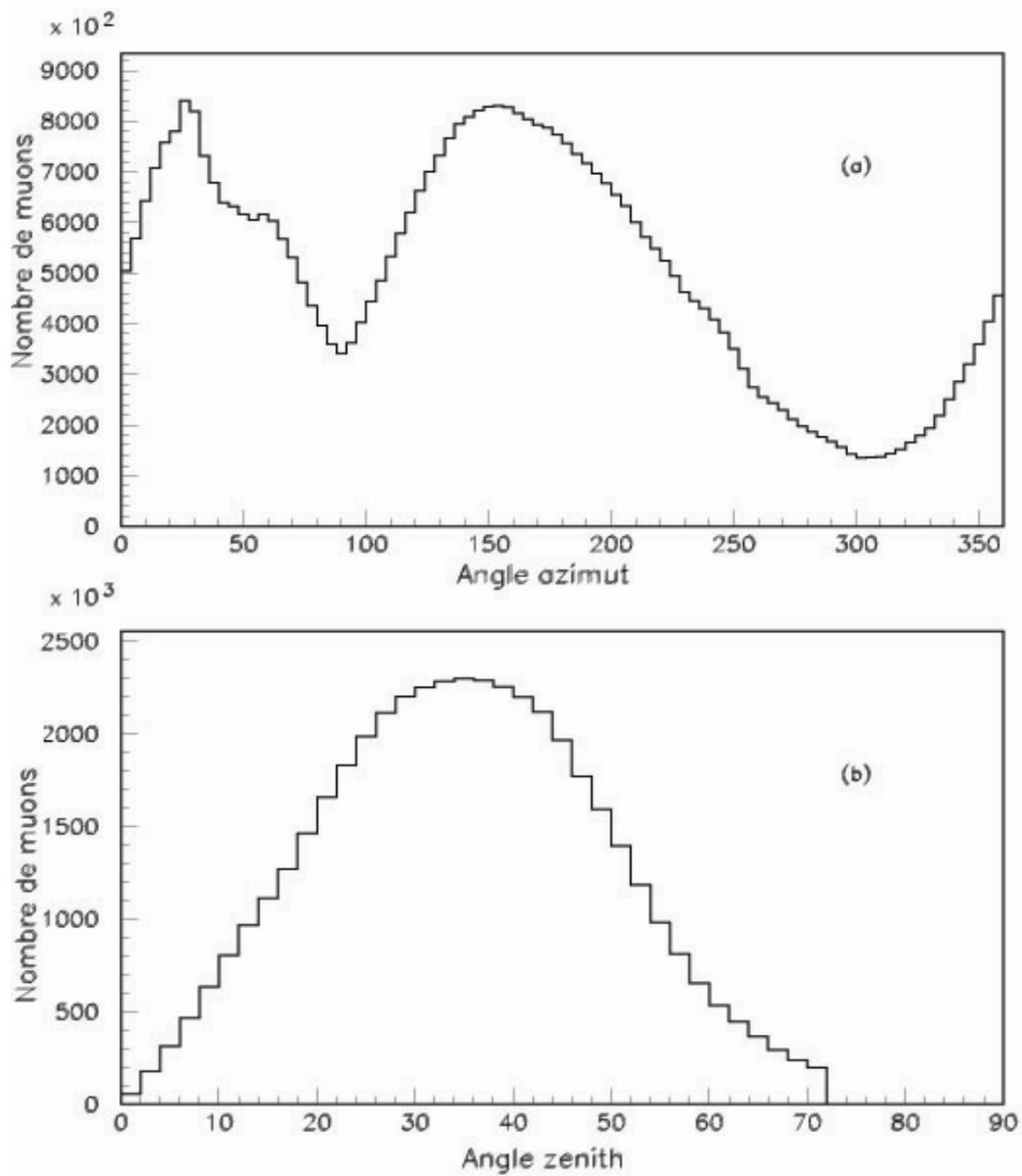}
\vskip -2cm
\caption{\em \textbf{Distributions des angles Zénith (b) et Azimut (a) des muons singuliers.}}
\label{fig:dt15}
\end{figure}

\begin{figure}
\centering
\vskip -5cm
\includegraphics[height=24cm,width=16cm]{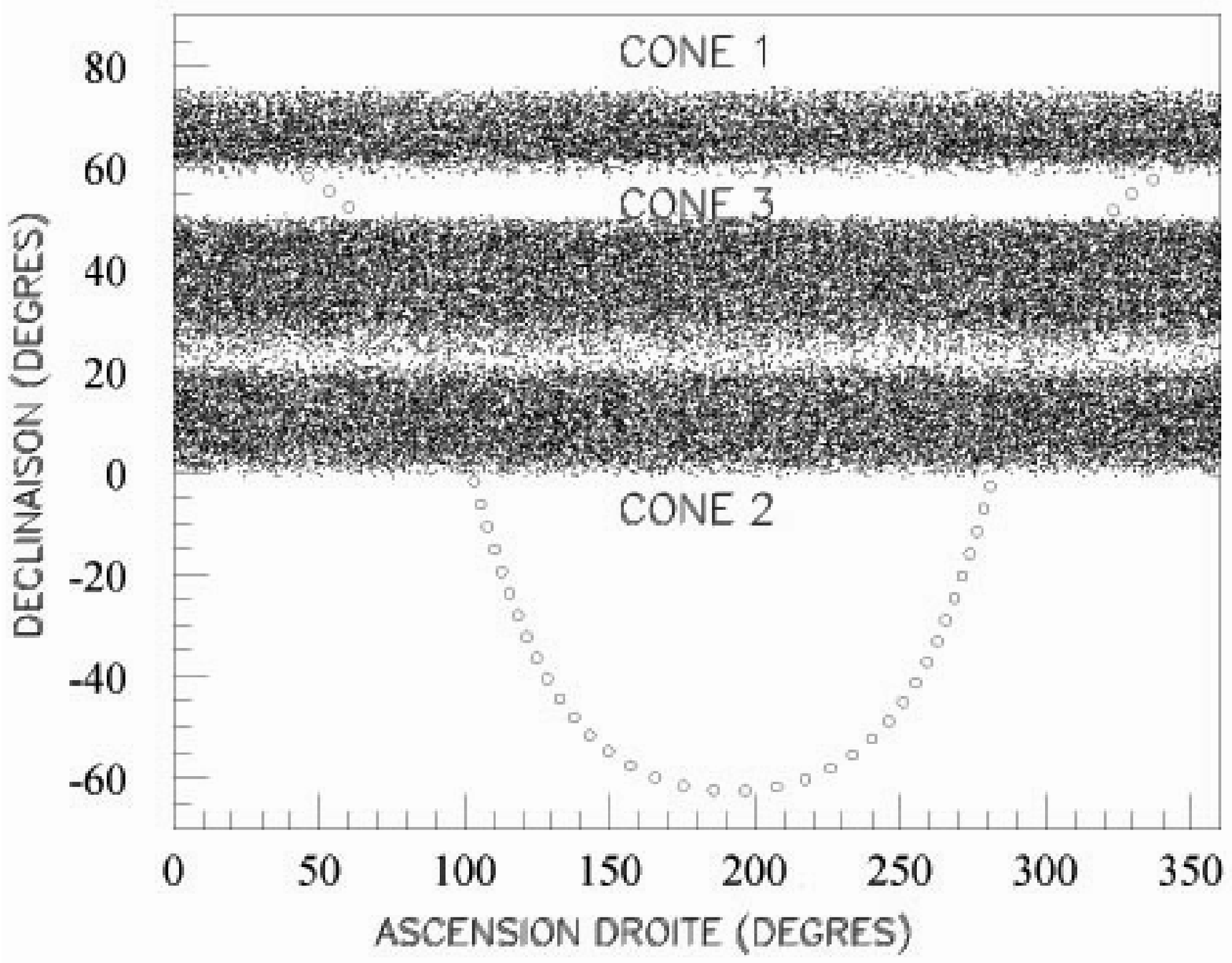}
\vskip -4cm
\caption{\em \textbf{Distribution des muons singuliers venant des cônes sélectionnés en fonction de la déclinaison et de l'ascension droite. Les cercles ouverts indiquent le plan galactique.}}
\label{fig:dt16}
\end{figure}

\begin{figure}
\centering
\includegraphics[height=14cm,width=14cm]{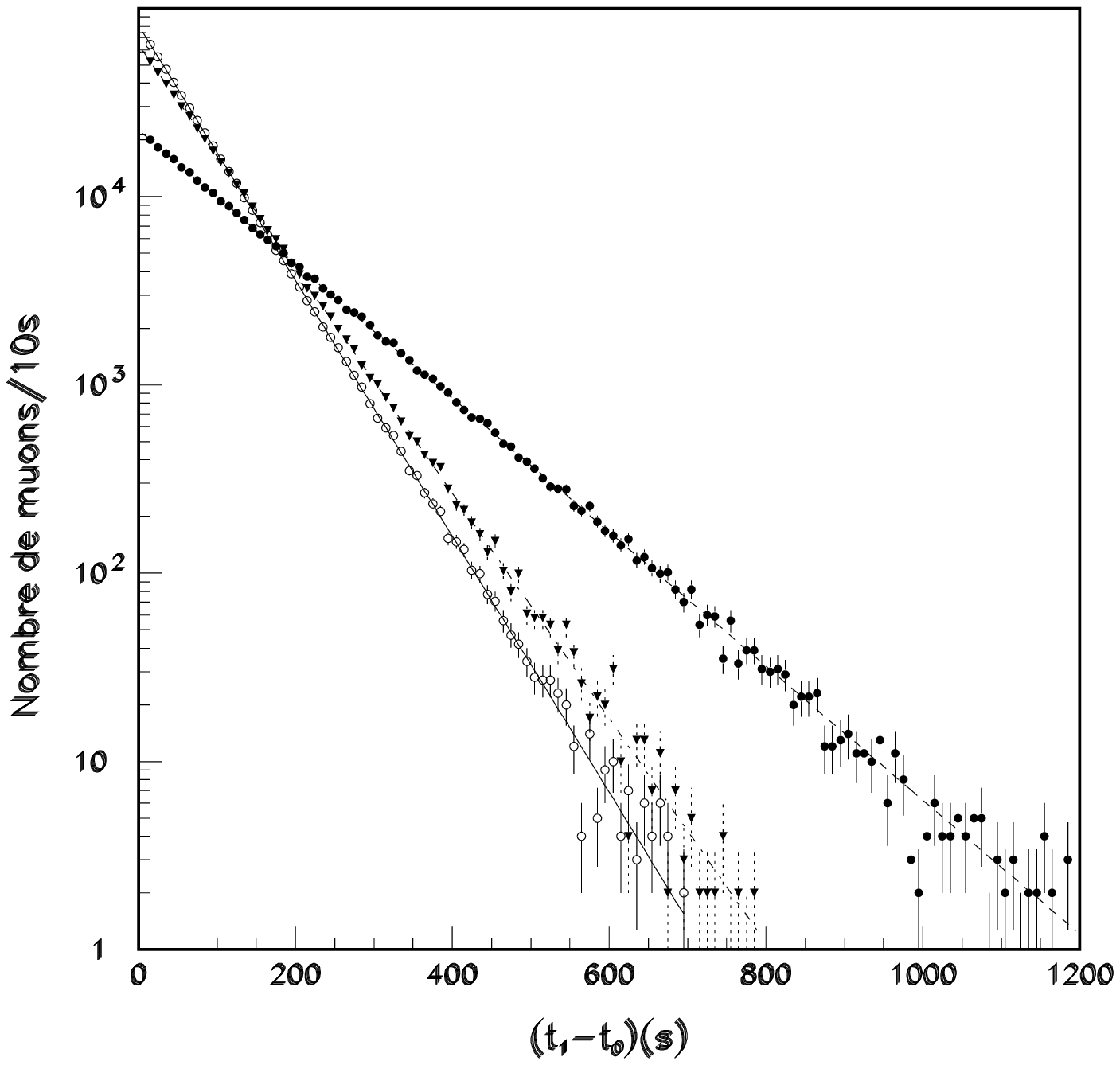}
\caption{\em \textbf{Distribution du temps d'arrivée des $\mu$ singuliers consécutifs venant du cône 1 (cercles pleins), cône 2 (triangles) et du cône 3 (cercles vides). Les courbes représentent le résultat de l'ajustement des données par la fonction Gamma d'ordre 1.}}
\label{fig:dt17}
\end{figure}

Les détecteurs de grande surface fixes par rapport à la terre nous permettent d'observer des petites anisotropies. Avec sa grande acceptance, MACRO nous permet d'observer des directions qui changent avec la rotation de la terre et donc une bande limitée et bien définie dans l'espace sera décrite durant les 24 heures sidérales.\\
Pour chercher l'existence d'une composante non aléatoire, trois zones ayant des déclinaisons séparées dans le temps  ont été choisies. Ainsi nous avons mesuré les distributions des temps qui séparent l'arrivée des $\mu$ dans trois fenêtres spatiales définies en coordonnées locales et centrées sur des valeurs différentes de l'angle azimutale. De ce fait, les trois bandes auront des déclinaisons différentes, non corrélées et se suivent dans le temps.\\
Nous répétons la même analyse pour l'étude des corrélations des muons qui arrivent par des directions définies par les angles zénithal et azimutale. Dans ce cas le détecteur joue le rôle d'un télescope pour observer différentes région du ciel. Les cônes (1) et (2) choisis coïncident avec le maximum d'intensité de la distribution angulaire des muons en zénith et azimut; le cône (3) a été choisi pour couvrir la région de déclinaison centrée sur la direction de Cygnus-X3. Dans la figure (\ref{fig:dt15}) nous présentons les distributions des muons en fonction des angles azimutal (a) et zénithal (b) pour les muons singuliers.\\
Pour le premier cône choisi avec $25\leq\mbox{zénith}\leq 45$ et $20\leq\mbox{azimut}\leq40$, le nombre total des muons singuliers sélectionnés est $280\;10^3$. Pour le deuxième cône, $25\leq \mbox{zénith}\leq45$ et $140\leq\mbox{azimut}\leq160$, nous avons sélectionné $350\;10^3\;\mu$. Pour le troisième cône, nous avons choisi $60\leq\mbox{azimut}\leq100$ et $260\leq \mbox{azimut}\leq300$ et $30\leq \mbox{zénith}\leq50$ et le nombre total des muons singuliers dans ce cas est de $520\;10^3$. Les déclinaisons et les ascensions droites des muons singuliers arrivant des cônes ont été calculées en utilisant les coordonnées locales et le temps d'arrivée des évènements. Dans la figure (\ref{fig:dt16}) nous présentons les bandes de déclinaison des muons singuliers sélectionnés pour les différents cônes, les cercles indiques le plan galactique.\\
Dans la figure (\ref{fig:dt17}) nous montrons les distributions des temps d'arrivée des muons singuliers venant du cône 1 (cercles pleins), cône 2 (triangles) et du cône 3 (cercles vides).\\
Les paramètres d'ajustements des distributions expérimentales $t_1-t_0$ avec la fonction Gamma d'ordre 1 (Eq. \ref{eq:gamma1}) sont présentés sur le tableau (\ref{tab:table}). On constate un bon accord entre les distributions théoriques et expérimentales, ce qui exclut la présence de toute sorte de modulation.
\begin{table}[h]
\begin{center}
\begin{tabular}{|c|c|c|c|c|c|}
\hline
Sélection & N ($10^3$) &  1/$\lambda $ (s) & M & $\chi^{2}$/DoF \\
\hline
$\mu$ singuliers &8638$\pm$4    & 4.03$\pm$0.01 & 1.002$\pm$0.003 & 1.00 \\
$\mu$ doubles    &456.5$\pm$0.9 & 73.3$\pm$0.3  & 1.008$\pm$0.004 & 1.36 \\
$\mu$ multiples  &180.2$\pm$0.3 &196.0$\pm$0.4  & 0.93$\pm$0.07   & 0.98 \\
\hline
\hline
Cône 1  & 276.3 $\pm $0.6 & 121.58$\pm$0.03 & 1.003$\pm$0.004 & 1.02 \\
\hline
\hline
Cône 2  & 353$\pm$1 & 95.9$\pm$0.1 & 1.003$\pm$0.006 & 0.82 \\
\hline
\hline
Cône 3  & 522$\pm$1 & 63.82$\pm$0.3 & 1.004 $\pm$0.004 & 0.91 \\
\hline
\end{tabular}
\end{center}
\caption{\em \textbf{Paramètres d'ajustements des distributions expérimentales ($t_1-t_0$) avec la fonction Gamma d'ordre 1 pour les $\mu$ singuliers, doubles et multiples arrivant de tout le ciel et des 3 cônes.}}
\label{tab:table}
\end{table}
\newpage
\section{Test de Kolomogorov-Smirnov}
Afin de chercher des structures possibles dans les distributions des temps d'arrivée, nous avons aussi utilisé le test de Kolomogorov-Smirnov \cite{{dtr16}}. Ce test nous permet de tester l'hypothèse $H_0$ selon laquelle les données observées sont engendrées par une loi de probabilité théorique considérée comme étant un modèle convenable.\\
Le test compare la distribution cumulative $F(x)$ des données expérimentales avec la distribution théorique aléatoire $H(x)$. \\
La mesure de la déviation est $d=max\left|H(x)-F(x)\right|$, ou $F(x)$ et $H(x)$ sont les distributions cumulatives de $f(x)$ (données) et $h(x)$ (attendue) respectivement. En terme de qualité du test, $F(x)$ s'accorde avec $H(x,\lambda)$, où $\lambda$ est prise des données, avec une probabilité de compatibilité, entre les distributions mesurées et attendues, donnée par:
\begin{equation}
P_k(d>\mbox{observée})=Q_{ks}(\sqrt(N)d)
\label{kolmogorov}
\end{equation}
où:
\begin{equation}
Q_{ks}(x)=2\sum_{j=1}^{+\infty}(-1)^{j-1}e^{-j^2}x^2
\label{kolmo}
\end{equation}
La probabilité des tests pour les distributions $(t_1-t_0)$ est donnée dans le tableau (\ref{tab:table1}). Ces résultats sont en accord avec les distributions aléatoires, bien que dans le cas du cône 3, un désaccord est remarqué (à un niveau de $1\;\sigma$) et peut produire une faible probabilité. Ceci peut être dû à une possible augmentation du rapport d'évènements pour $(t_1-t_0)=500\;s$ (voir figure (\ref{fig:dt17})), mais les faibles statistiques ne peuvent conduire à une conclusion assez claire.

\begin{table}[h]
\begin{center}
\begin{tabular}{|c|c|}
\hline
Sélection & Pr. K-S \\
\hline
$\mu$ singuliers & 0.99\\
$\mu$ doubles    & 0.95\\
$\mu$ multiples  & 0.99\\
\hline
\hline
Cône 1  & 0.99\\
\hline
\hline
Cône 2  & 0.77\\
\hline
\hline
Cône 3  & 0.38\\
\hline
\end{tabular}
\end{center}
\caption{\em \textbf{ Probabilité de Kolmogorov-Smirnov pour les $\mu$ singuliers, doubles et multiples arrivant de tout le ciel et des 3 cônes.}}
\label{tab:table1}
\end{table}
\newpage
\section{Conclusion}
Nous avons présenté les résultats de l'étude des distributions des muons cosmiques avec une énergie plus grande que 1.3 TeV au sommet de la montagne de Gran Sasso. Les données analysées ont été collectées à l'aide du système des tubes à streamer du détecteur MACRO dans sa configuration complète. \\
Les muons singuliers, doubles et multiples arrivant de toutes les directions ainsi que ceux venant des zones sélectionnées formant des cônes ont été considérés \cite{dtr17}.\\
Les résultats de notre étude concernant les différentes distributions temporelles montrent que les données expérimentales sont compatibles avec la fonction Gamma d'ordre M. Ce qui nous a permis d'affirmer que nos résultats sont en accord avec l'arrivée aléatoire des muons dans notre détecteur et permet ainsi d'exclure la présence de toute composante, qui ne soit pas aléatoire dans les temps d'arrivée des $\mu$, similaire à celle observée par Bath et al. \cite{dtr7}, ou par Badino et al. \cite{dtr8}, et ceci pour des énergies de l'ordre de 20 TeV ($\mu$ singuliers) et supérieures à 20TeV ($\mu$ doubles et multiples). L'étude des distributions temporelles des $\mu$ arrivant des cônes, a abouti aux mêmes conclusions.\\
Le test de Kolmogorov-Smirnov, nous a permis de mesurer la déviation entre la distribution théorique et celle expérimentale. Aucune déviation n'a été mise en évidence dans notre échantillon de données, sauf dans le cas dû cône 3 et ceci est dû à la faible statistique.\\  
Dans cette étude nous n'avons mis en évidence l'existence d'aucun signal qui peut être la signature d'une source discrète tels que les pulsars. 

\chapter{Recherche des variations du flux de muons}
\label{chapss}
\subsection{Introduction}
Durant presque dix ans de prise de données à une profondeur de 3800 m.w.e, MACRO a rassemblé un des grands échantillons de muons collectés par les expériences souterraines de ce type. Le flux de muons peut présenter des variations d'origine galactique, solaire ou même terrestre. Lorsque ces variations ne sont pas aléatoires, elles peuvent révéler la présence des modulations au niveau du signal dévoilant ainsi l'existence des objets galactiques émetteurs. Les clusters de muons souterrains peuvent être produites par des événements violents comme les "Gamma Ray Burst" sur l'échelle de temps de quelques secondes, ou par des variations méteorologiques soudaines sur l'échelle de temps de plusieurs heures. Les modulations périodiques sont liés à la variation de la température de la haute atmosphère, qui diminue durant la nuit (hiver) produisant les variations journalières (saisonnières) de la densité de l'atmosphère et cependant du flux de muons.\\
Le temps d'arrivée des muons cosmiques de haute énergie a été analysé précédemment, où nous avons pu confirmer l'absence de modulations dans le flux de muons. \\
Cette étude va être argumentée dans ce chapitre par la recherche de possibles variations de flux. Ainsi deux méthodes sont utilisées : 
\begin{itemize}
	\item 
	la recherche d'éventuel cluster d'évènements (groupement d'évènements) dans le flux de muons.
	\item
	la recherche de variations périodiques.
\end{itemize}
\section{Recherche de cluster d'évènements}
On se propose de chercher la présence des clusters d'évènements dans le flux de muons. L'analyse de ces clusters nous permet d'estimer sa signification statistique, ce qui nous permet  de conclure sur sa nature qui peut être un résultat du hasard où à l'observation d'un excés d'événements originaire d'un objet émetteur des rayons cosmiques. Dans ce dernier cas, qui présente généralement beaucoup d'intérêt, il est important de trouver et d'expliquer les facteurs qui ont conduit à sa réalisation. La présence de groupement d'évènements dans le flux de muons cosmiques, indique l'existence d'une composante émettrice des rayons cosmiques dans une direction privilégiée.\par
Les perturbations locales dans une distribution théorique d'une variable donnée, peuvent signaler la présence d'une modulation, ou d'une composante en désaccord avec le modèle fondamental, utilisé pour décrire les données expérimentales (l'hypothèse nulle). En physique de hautes énergies, ce phénomène est relié généralement à l'apparition d'une résonance inattendue, et/ou à la détermination de la signification statistique de l'excès. Les anomalies globales sont fréquemment traitées à l'aide du test de Kolmogorov-Smirnov et ses extensions. Cependant, le pouvoir de ce test est considérablement réduit dans le cas des perturbations locales. Inversement, le test $\chi^2$ utilise un binage de l'intervalle et compare le contenu de chaque bin avec celui théorique sous l'hypothèse nulle. Ce test est mieux adapté aux perturbations locales. Notons que le test est adapté seulement dans le cas d'un binnage fixe à priori. En général, les perturbations locales partagées entre différents bins, sont moins marquées que ceux où le cluster d'évènements est situé dans un seul bin. De ce fait, les techniques de scan avec une fenêtre de longueur fixe semble les plus appropriées.\\
Ces techniques sont généralement utilisées dans plusieurs domaines de recherche tels que la bio-informatique, la médecine \cite{ssr2}, la physique des particules \cite{ssr7} ainsi que dans des applications en astrophysique \cite{ssr3}.
\subsection{Méthode scan statistics}
La méthode de balayage scan statistics est l'une des plus puissantes méthodes utilisées pour analyser l'apparition de cluster d'évènements. Elle présente un moyen très utile pour signaler la présence de perturbation dans le modèle de probabilité fondamentale qui décrit les données expérimentales. Elle est basée sur l'échantillonnage libre \cite{ssr1}.\\
Soit x une variable continue de l'intervalle $[A,B]$ et qui obéit à un processus de poisson, notons $\lambda$ la valeur moyenne par unité d'intervalle. La probabilité de trouver $Y_x(w)$ évènements dans un intervalle $[x,x+w]$ est 
\begin{equation}
Prob(Y_x(\omega)=k)=e^{-\lambda\omega}\frac{(\lambda\omega)^k}{k!}\;\;\;\;k=0,1,2,...
\label{prob}
\end{equation} 
Le nombre d'évènements dans les intervalles disjoints est indépendamment distribué. Nous appelons scan statistics (SS) le plus grand nombre d'évènements trouvé dans un sous-intervalle de $[A,B]$ de longueur $\omega$: 
\begin{equation}
S(\omega) \equiv \max_{\mathcal{A}\leq x \leq \mathcal{B}-\omega} \left\{ Y_x(w) \right\} 
\end{equation} 
La probabilité pour que le nombre d'évènements dans une fenêtre balayée n'atteint jamais $k$ sera donnée, selon \cite{ssr2}, par: 
\begin{equation}
Q^{*}(k,\lambda\Delta,\omega/\Delta)\equiv 1-Prob(S(\omega)\geq k)
\label{nprob}
\end{equation}
Où $\Delta\equiv B-A$ et le suffixe "*" indique que les probabilités non conditionnelles sont considérées, i.e. que le nombre global d'évènements N dans l'intervalle fluctue selon l'Eq. (\ref{prob}) avec $\omega=\Delta$. La forme exacte de l'Eq. (\ref{nprob}) peut être exprimée sous forme d'une somme des produits de deux déterminants \cite{ssr4}\cite{ssr4r}. La sommation se fait sur un ensemble V constitué par $2H+1$ partitions de N dont les éléments $m_i$ de chaque partition sont des entiers non-négatifs satisfaisants la condition $m_i+m_{i+1}<k$ pour $i=1,...,2H$, où $H$ est le plus grand entier dans $\Delta/\omega$. Les déterminants sont calculés à partir des matrices $(H+1)\times(H+1)$ d'éléments $\left\{h\right\}_{ij}$ et $H \times H$ d'éléments  $\left\{v\right\}_{ij}$ dont:
\begin{eqnarray*}
h_{ij}&=&\sum_{s=2j-1}^{2i-1}m_s-(i-j)k\;\;\;\;\;\;\;1\leq j\leq i\leq H+1\\
&=&-\sum_{s=2i}^{2j-2}m_s+(j-i)k\;\;\;\;\;\;\;1\leq i\leq j\leq H+1\\
v_{ij}&=&\sum_{s=2j}^{2i}m_s-(i-j)k\;\;\;\;\;\;\;\;\;\;1\leq j\leq i\leq H\\
&=&-\sum_{s=2i+1}^{2j-1}m_s+(j-i)k\;\;\;\;1\leq i\leq j\leq H\\
\label{matrice}
\end{eqnarray*}
En utilisant les définitions de V, $h_{ij}$ et $v_{ij}$, nous avons pour $k\geq2$ et $\omega<\Delta$: 
\begin{equation}
Q^{*}(k,\lambda\Delta,\omega/\Delta)= \sum_VR^*det\left|1/h_{ij}!\right| det\left|1/v_{ij}!\right|
\label{nprobb}
\end{equation}
Dans la formule \ref{nprobb}: 
\begin{equation}
R^*=N!\;d^M\;(\frac{\omega}{\Delta}-d)^{N-M}p(N,\lambda\Delta)
\end{equation}
\begin{equation}
M=\sum^{H}_{j=0}m_{2j+1}
\end{equation}
$d\equiv1-\omega H/\Delta$ et $p(N,\lambda\Delta)$ est la probabilité de poisson de N événements et de rapport moyen $\lambda\Delta$.\par
\subsection{Approximation de Naus}
Une approximation très utile de l'équation (\ref{nprobb}) a été introduite par Naus \cite{ssr5}\cite{ssr4r}, basée sur les valeurs exactes des probabilités :
$$Q_2\equiv Q^*(k,2\psi,1/2)$$ 
et 
$$Q_3\equiv Q^*(k,3\psi,1/3)$$
où $ \psi\equiv\lambda\omega$ et $L=\Delta/\omega$. 
\\On obtient alors \cite{ssr1}: 
\begin{equation}
Q^*(k,\psi L,1/L)\cong Q^*_2\left[Q^*_3/Q^*_2\right]^{L-2}
\label{naus}
\end{equation}
\noindent où :
\begin{eqnarray}
Q^*_2 & = & \left[F(k-1,\psi)\right]^{2}-
\left(k-1\right)p(k,\psi)p(k-2,\psi) \nonumber
\\
& & -\left(k-1-\psi\right)p(k,\psi)F(k-3,\psi)
\label{Naus:Q2} \\
Q^*_3 & = & \left[F(k-1,\psi)\right]^{3}-A_{1}+A_{2}+A_{3}-A_{4} 
\label{Naus:Q3} 
\end{eqnarray}
\noindent et :
\begin{eqnarray*}
A_{1} & = & 2\ p(k,\psi)F(k-1,\psi)\left\{\left(k-1\right)F(k-2,\psi)-\psi F(k-3,\psi)\right\}
\\
A_{2} & = & 0.5 \ \left[p(k,\psi)\right]^{2}\left\{\left(k-1\right)\left(k-2\right)F(k-3,\psi) \right. 
\\ 
& & \left.
-2\left(k-2\right)\psi F(k-4,\psi)+\psi^{2}F(k-5,\psi)
\right\}
\\
A_{3} & = & \sum_{r=1}^{k-1}p(2k-r,\psi)\left[F(r-1,\psi)\right]^{2}
\\
A_{4} & = & \sum_{r=2}^{k-1}p(2k-r,\psi)p(r,\psi)
\left\{\left(r-1\right)F(r-2,\psi)-\psi F(r-3,\psi)\right\}
\end{eqnarray*}
\noindent
Dans les formules au dessus $F(k,\psi)$ représente la distribution cumulative: 
\begin{equation}
F(k,\psi) = \sum_{i=0}^k p(i,\psi) \ \ ; \ \ p(i,\psi)=e^{-\psi} 
\frac{\psi^i}{i!}
\end{equation}
\noindent et $ F(k,\psi)=0$ pour $k<0$. \par 
Dans ce qui suit, nous utiliserons la méthode scan statistics basée sur l'approximation (\ref{naus}).
\subsection{Méthode d'analyse}
Pour chercher l'existence d'une composante qui émet des rayons cosmiques dans une direction privilégiée, nous proposons d'appliquer l'approche scan statistics sur le flux de muons détectés par MACRO. Cette méthode nous permet de chercher l'apparition de groupements d'évènements qui peuvent être candidats de sources émettrices des rayons cosmiques dans l'espace environnent \cite{ssr5r}. \\
Les données utilisées pour cette analyse ont été collectées à l'aide des tubes à streamer des 6 supermodules des parties inférieure et supérieure du détecteur MACRO. Ce qui nous a permis de sélectionner les muons reconstruits avec les 14 plans des tubes à streamer.
\subsubsection{Données et critères de sélection}
Vu la nature de cette analyse, nous avons adopté des critères de sélection moins rigoureux par rapport à l'analyse de la distribution des temps d'arrivée des muons :
\begin{itemize}
	\item La durée des runs doit être supérieure à 2 heures (la valeur moyenne de la durée des runs est d'environ 5,5 heures).
	\item La fréquence des muons R est choisie dans l'intervalle $880<R<~1006\;\mu/h$ et située dans un intervalle $\pm 2\sigma$ autour de la valeur moyenne.
	\item Un temps mort d'acquisition inférieure à 2.5\%.
	\item L'efficacité de détection des tubes à streamer doit être supérieure à 90\% et 70\% pour les fils et les strips respectivement.
\end{itemize}
Un lot de 6411 runs a été sélectionné contenant un total d'évènements d'environ $3.29~10^7$ muons.
\subsubsection{Méthode d'analyse :}
\begin{figure}
\begin{center}
\includegraphics[height=6cm,width=12cm]{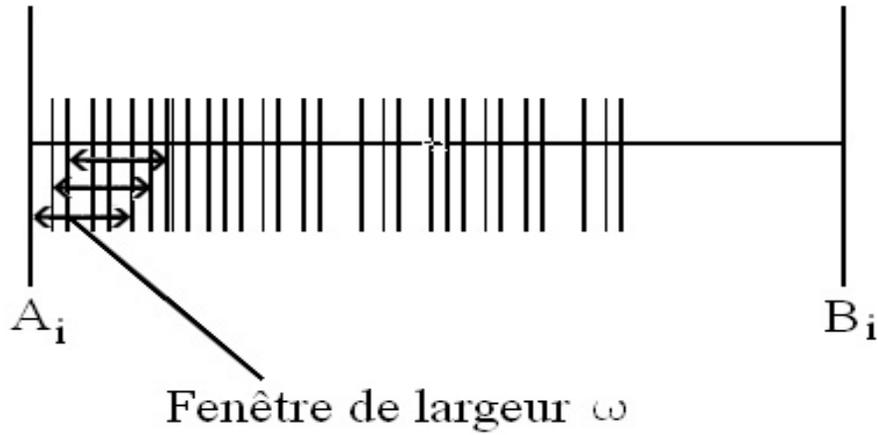}
\caption{\em \textbf{Shématisation du balayage d'un run par une fenêtre de largeur w}}
\label{ss3bis}
\end{center}
\end{figure}

\begin{figure}
\begin{center}
\includegraphics[height=16cm,width=16cm]{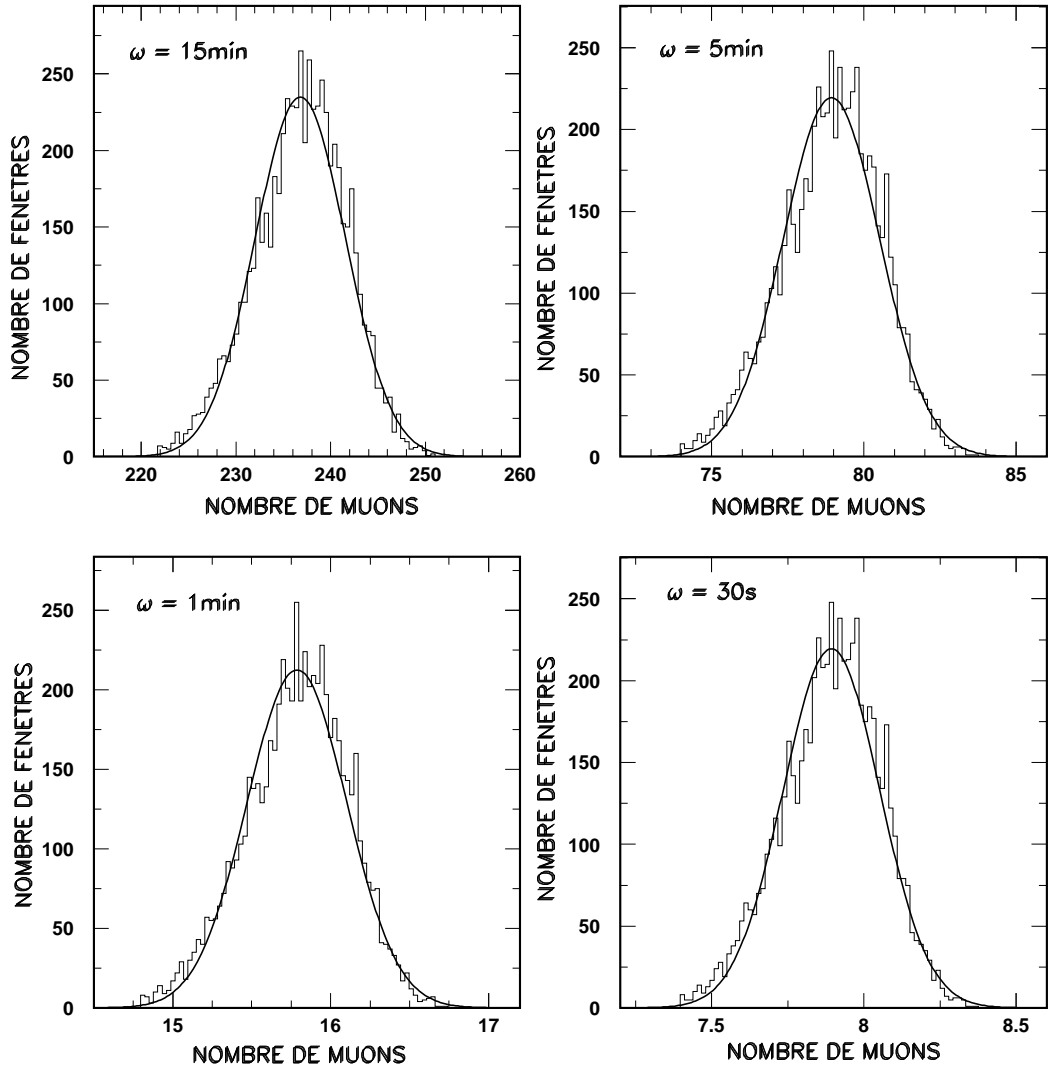}
\caption{\em \textbf{Distribution du nombre moyen d'évènements dans une fenêtre de largeur $\omega$ dans chaque run pour $\omega =15$ min, $\omega=5$ min, $\omega=1$ min et $\omega=30$ s.}}
\label{ss4}
\end{center}
\end{figure}

\begin{figure}
\begin{center}
\includegraphics[height=14cm,width=14cm]{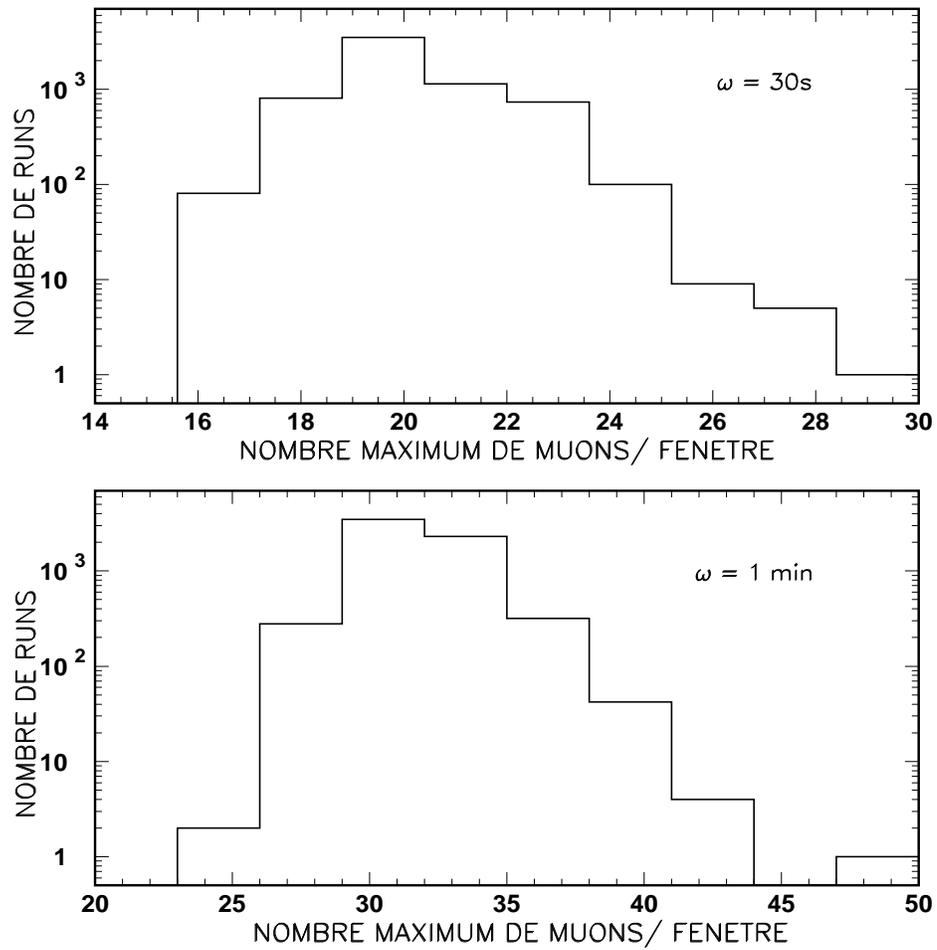}
\caption{\em \textbf{Distributions du nombre maximum d'événements $S(\omega)$ dans les fenêtres de temps $w=30$ s et $w=1$ min.}}
\label{ss5}
\end{center}
\end{figure}

\begin{figure}
\begin{center}
\includegraphics[height=14cm,width=14cm]{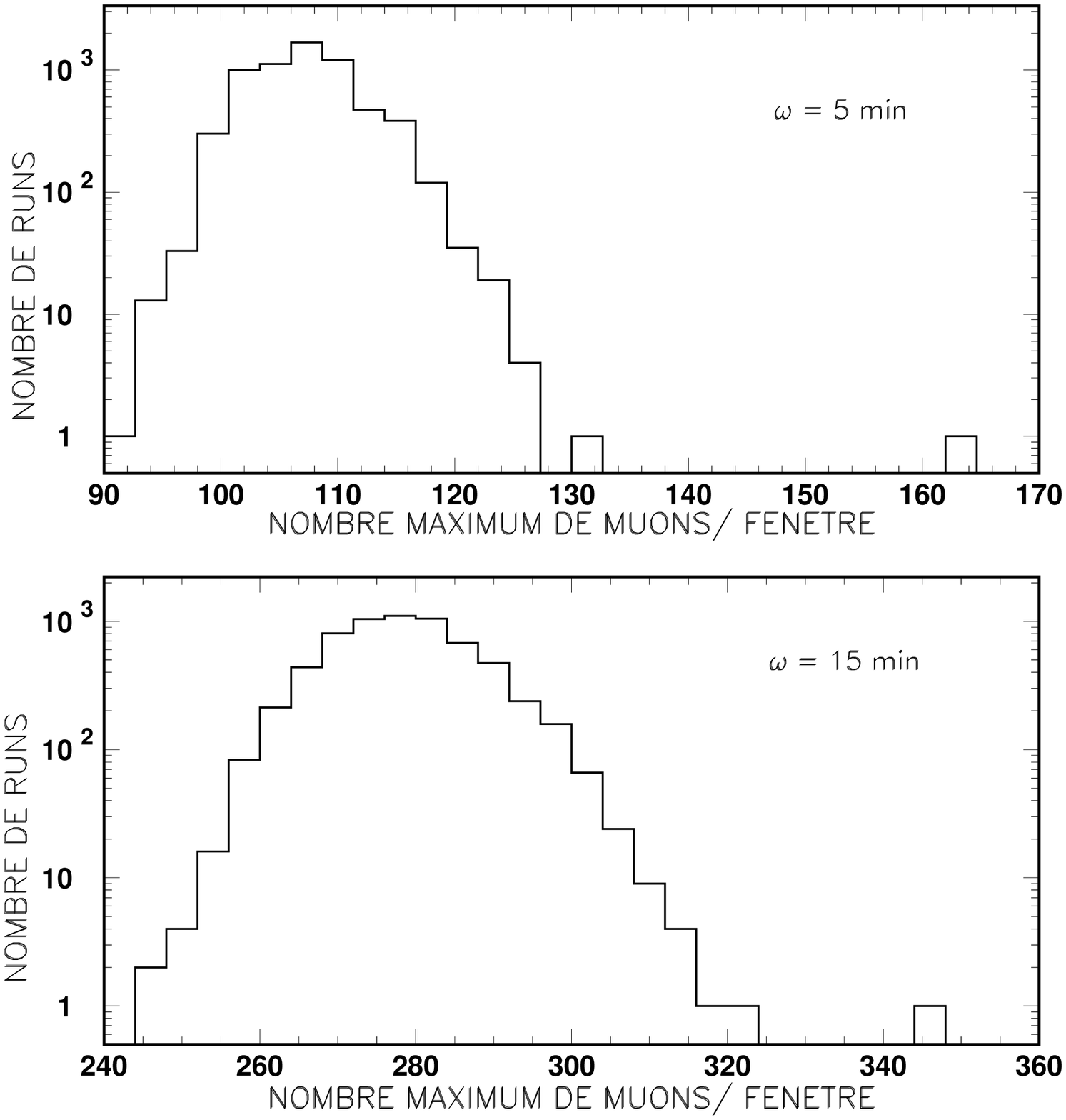}
\caption{\em \textbf{Distributions du nombre maximum d'évènements $S(\omega)$ dans les fenêtres de temps $w=5$ min et $w=15$ min. On constate la présence des fenêtres avec un excès d'évènements (k=162 pour $w$= 5 min et k=346 pour $w$= 15 min) par rapport à la moyenne (<k>$_{5min}$=107 et <k>$_{15min}$=277).}}
\label{ss6}
\end{center}
\end{figure}

\begin{figure}
\begin{center}
\includegraphics[height=14cm,width=14cm]{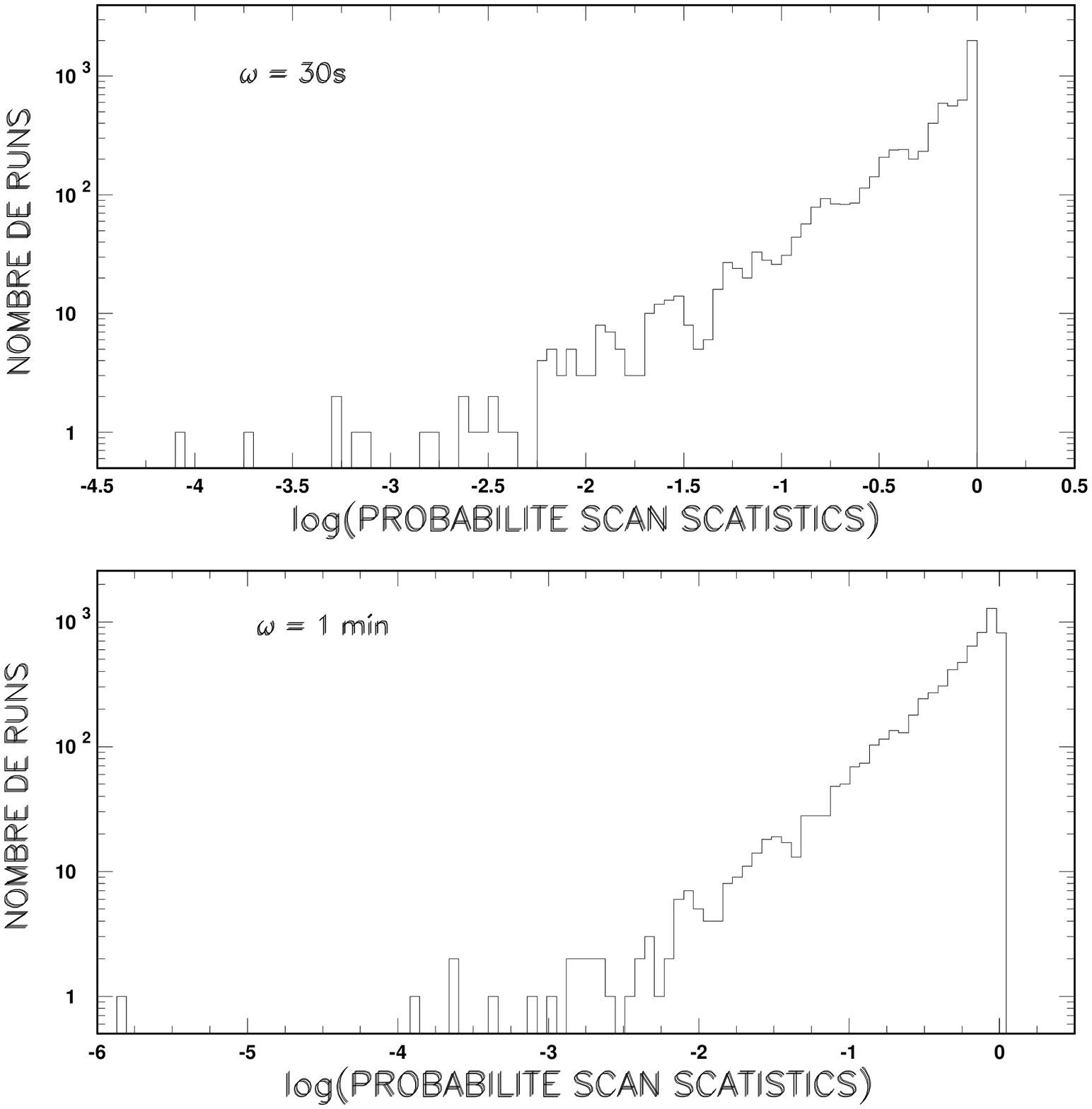}
\caption{\em \textbf{Distributions de la probabilité scan statistics des runs sélectionnés pour les valeurs des fenêtres de temps $w=30$ s et $w=1$ min. Pour $w$=1min, on note l'apparition d'une fenêtre avec une faible valeur de probabilité scan statistics inférieure à $10^{-4}$. Cette fenêtre correspond au run 11056.}}
\label{ss7}
\end{center}
\end{figure}

\begin{figure}
\begin{center}
\includegraphics[height=14cm,width=14cm]{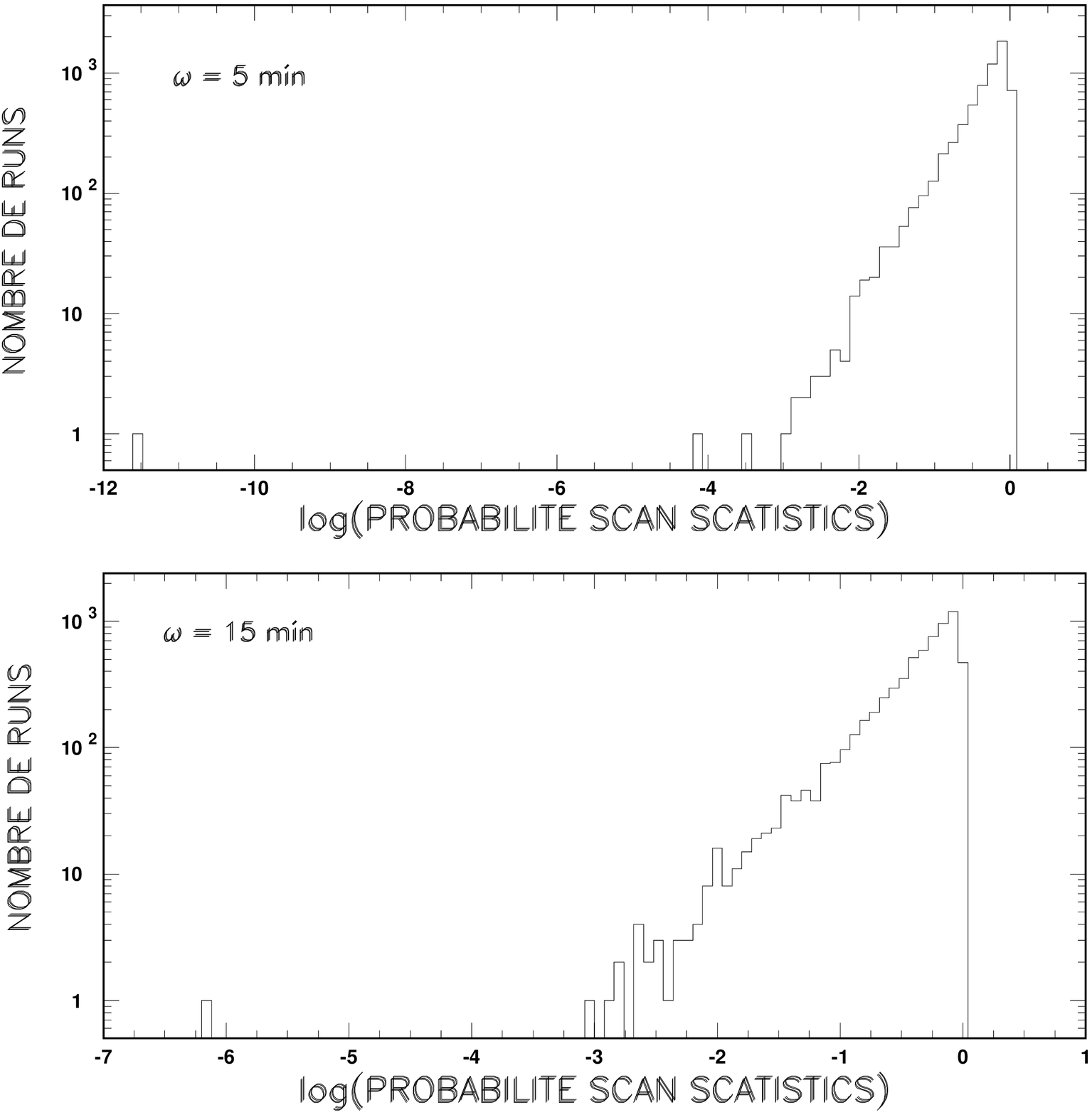}
\caption{\em \textbf{Distributions de la probabilité scan statistics des runs sélectionnés pour les valeurs des fenêtres de temps $w=5$ min et $w=15$ min. Pour $w$= 5 min et $w$= 15 min, on note l'apparition des fenêtres avec de faibles valeurs de probabilité scan statistics inférieure à $10^{-4}$. Ces fenêtres correspondent aux run 11056 et 11079.}}
\label{ss8}
\end{center}
\end{figure}

Soit $[A_i,B_i]$ l'intervalle de temps séparant le temps du début et de fin du run pour chaque run $i$. Une fenêtre de temps de largeur $\omega$ balaye l'intervalle $[A_i,B_i]$ (voir figure \ref{ss3bis}).
La Distribution du nombre moyen d'évènements dans une fenêtre de largeur $\omega$ dans chaque run est représenté sur la figure (\ref{ss4}) pour $\omega =15$ min, $\omega=5$ min, $\omega=1$ min et $\omega=30$ s. \\
Soit $k_i$ le nombre maximum d'évènements comptés durant le balayage. Sur les figures (\ref{ss5}) et (\ref{ss6}) est représenté le nombre maximum d'évènements comptés durant le balayage des runs $S(\omega)$.
\\ Finalement, pour chaque run, on a calculé la probabilité pour que les fluctuations \mbox{statistiques} puissent produire un cluster d'évènements aussi large que $k_i$. Dans cette analyse le paramètre $\omega$ est libre; ici il est fixé au valeurs (15 min, 5 min, 1 min et 30 s). Pour chaque run analysé, la distribution des probabilités scan statistics est représentée sur les figures (\ref{ss7}) et (\ref{ss8}) pour les différentes valeurs des fenêtres de temps ($\omega=30s$ , $\omega=1min$) et ($\omega=5min$ , $\omega=15min$) respectivement.\\

\begin{figure}
\begin{center}
\includegraphics[height=14cm,width=14cm]{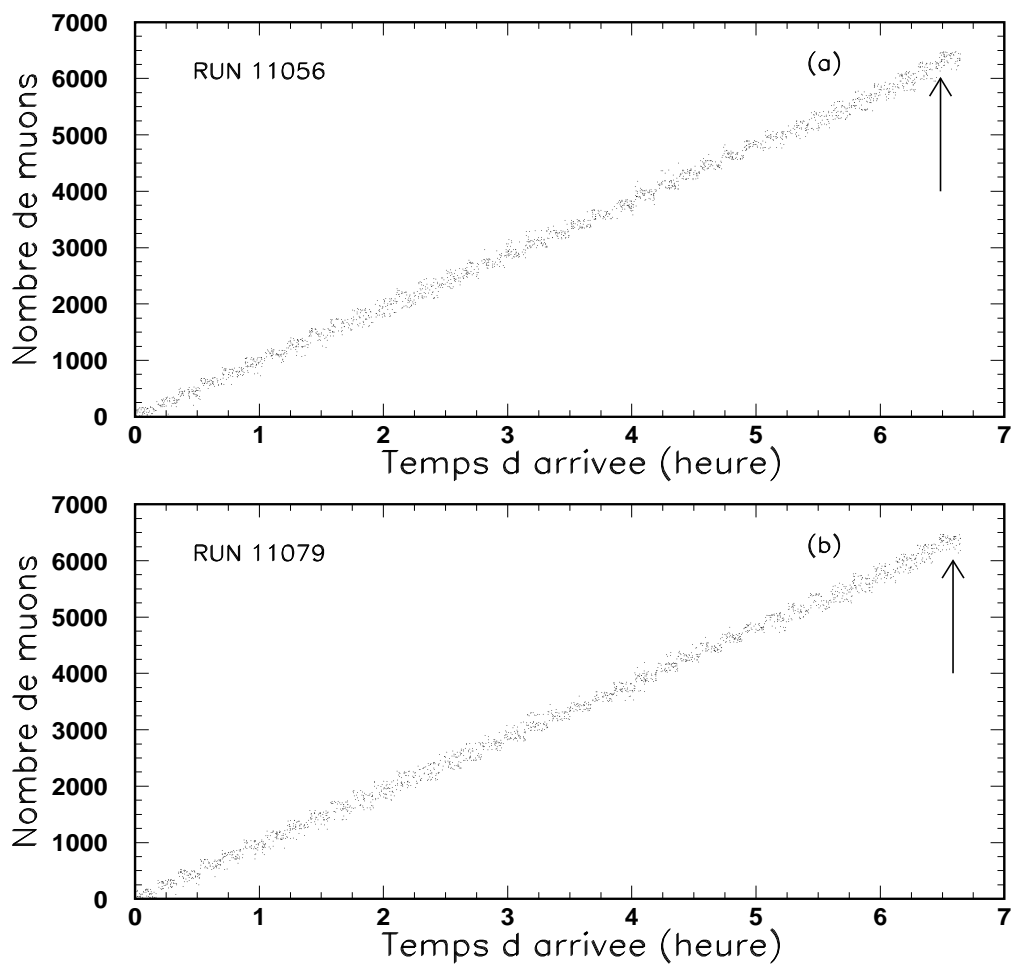}
\caption{\em \textbf{Le nombre d'évènements en fonction du temps d'arrivée des muons pour les runs 11056 (a) et 11079 (b). Les flèches indiquent la position des groupements d'événements repérés dans ces runs.}}
\label{ss9}
\end{center}
\end{figure}

Pour les fenêtres ($w=1min$, $5min$, et $15min$), les figures (\ref{ss7}) et (\ref{ss8}) montrent la présence d'un groupement d'évènements avec une faible valeur de probabilité scan statistics (inférieure à $10^{-4}$) identifiés sur les runs 11056 et 11079. La position de ces groupements d'évènements est repérée à la fin de ces runs, ce qui ne permet pas de tirer une conclusion définitive sur la nature de l'excès. L'inspection de l'allure de la variation du nombre d'évènements en fonction du temps d'arrivée ne montre aucune divergence apparente ( voir figure (\ref{ss9})). Les flèches sur la figure (\ref{ss9}) indiquent la position des groupements d'événements.
\newpage
\subsection{Conclusion}
L'analyse scan statistics portée sur 6411 runs sélectionnées ne présente aucune déviation significative de l'hypothèse nulle. Seules 2 runs ont été identifiés avec une faible probabilité dont l'excès à été repéré vers la fin du run, ceci peut être attribué aux problèmes d'acquisition des données qui peuvent parfois apparaître vers la fermeture des runs. Ceci nous permet de confirmer l'absence d'éventuel cluster d'événements dans le flux de muons et donc aucune variation du flux n'a été identifiée.
\section{Recherche de variations périodiques dans le flux de muons de MACRO}
\subsection{Introduction}
Afin de chercher l'existence de variations périodiques dans le flux de muons détecté par MACRO, qui peuvent être la signature d'une composante émettrice de rayons cosmiques. On se propose, dans ce paragraphe, d'analyser les signaux périodiques dans les séries de temps d'arrivée des muons. \\
La méthode de la transformée de Fourrier rapide (FFT) est la plus adaptée pour ce genre d'étude. Cette technique représente l'une des plus puissantes méthodes qui permet d'identifier les fluctuations régulières dans les séries de temps. Cependant, une telle technique s'applique à des données échantionnées sur des intervalles uniformément distribués et ne tient pas compte des bins vides. \\
Dans le cas d'un échantillon de données non réguliers où les sauts de données apparaissent, la FFT est appliquée après un peuplement des sauts par les techniques d'interpolation, ce qui dissimule généralement les informations originales du signal. Les méthodes d'interpolations ont été étudiées dans \cite{ssr7bis},  il a été recommandé l'utilisation de la méthode spectrale Lomb-Scargle \cite{ssr8}\cite{ssr8bis} qui permet de palier l'effet introduit par la présence de bins vides. 
\subsection{Méthode de Lomb-Scargle}
Les pannes ou les périodes de maintenance du détecteur induisent des interruptions irrégulières dans la prise des données. La méthode spectrale de Lomb-Scargle \cite{ssr8}\cite{ssr8bis} a été développée pour surpasser cet effet et conduire ainsi à la détermination de périodicités temporelles dans le flux de muons.\\
Étant donnée N points mesurés, la moyenne $\bar{h}$ et la variance $\sigma$ des données sont données par : 
\begin{equation}
\bar{h}\equiv\frac{1}{N}\sum_{i=1}^{N}h_{i} \ \ \ \ \ \ \ \ \ \ \ \ \sigma^2\equiv\frac{1}{N-1}\sum_{i=1}^{N}(h_i-\bar{h})^2
\label{lomb1}
\end{equation}
Le périodogramme de Lomb normalisé est donné par :
\begin{equation}
P_N(w)\equiv \frac{1}{2\sigma^2}\left\{\frac{\left[\sum_j(h_j-\bar{h})\ cos\ \omega(t_j-\tau)\right]^2}{\sum_j cos^2\ \omega(t_j-\tau)}
+
\frac{\left[\sum_j(h_j-\bar{h})\ sin\ \omega(t_j-\tau)\right]^2}{\sum_j sin^2\ \omega(t_j-\tau)}\right\}
\label{lomb2}
\end{equation}
$\omega\equiv 2\pi f>0$ et $\tau$ est défini par :
\begin{equation}
tan(2\omega\tau)=\frac{\sum_j sin(2\omega t_j)}{\sum_j cos(2\omega t_j)}
\label{lomb3}
\end{equation}
La constante $\tau$ est un type d'excentrage qui rend $P_N(w)$ invariant par translation de tous les $t_i$ par une constante. Lomb montre que ce type d'excentrage rend l'équation (\ref{lomb2}) identique à l'équation que nous obtiendrons en estimant un ensemble de données harmoniques, à une fréquence donnée $\omega$, par un fit de moindre carrée au modèle:
\begin{equation}
h(t)\equiv A\ cos\omega t + B\ sin\omega t
\label{lomb4}
\end{equation} 
La méthode de Lomb exploite les informations de chaque point contrairement à la FFT qui traite les intervalles. 
Un point très fréquent est que les points mesurés $h_i$ sont la somme d'un signal périodique et le bruit gaussien indépendant.\\
La présence d'un pic dans le spectre de $P_N(\omega)$ peut révéler la présence d'une modulation au niveau du signal étudié. Dans notre cas, l'hypothèse 'non nulle' considère que les données sont les valeurs indépendantes d'une gaussienne. \\
Scargle montre \cite{ssr8bis} qu'à chaque $\omega$ particulier et dans le cas de l'hypothèse 'non nulle', $P_N(\omega)$ a une distribution de probabilité exponentielle avec une moyenne égale à 1. Autrement dit, la probabilité que $P_N(\omega)$ sera entre les positifs $z$ et $z+dz$ est $exp(-z)dz$. \\
Il suit que, si nous balayons quelques M fréquences indépendantes, la probabilité qu'il ne donne pas de valeurs plus grandes que $z$ est $(1-e^{z})^M$, ainsi : 
\begin{equation}
	p(>z)\equiv 1-(1-e^{-z})^M
\end{equation}
est la 'fausse alarme' de l'hypothèse 'non nulle', qui est, le niveau de signification de n'importe quel pic observé dans $P_N(w)$. Une petite valeur de la probabilité 'fausse alarme' indique une grande signification d'un signal périodique. 
\subsection{Recherche de modulations périodiques dans le flux de muons}
\begin{figure}
\begin{center}
\vskip -1cm
\includegraphics[height=15cm,width=16cm]{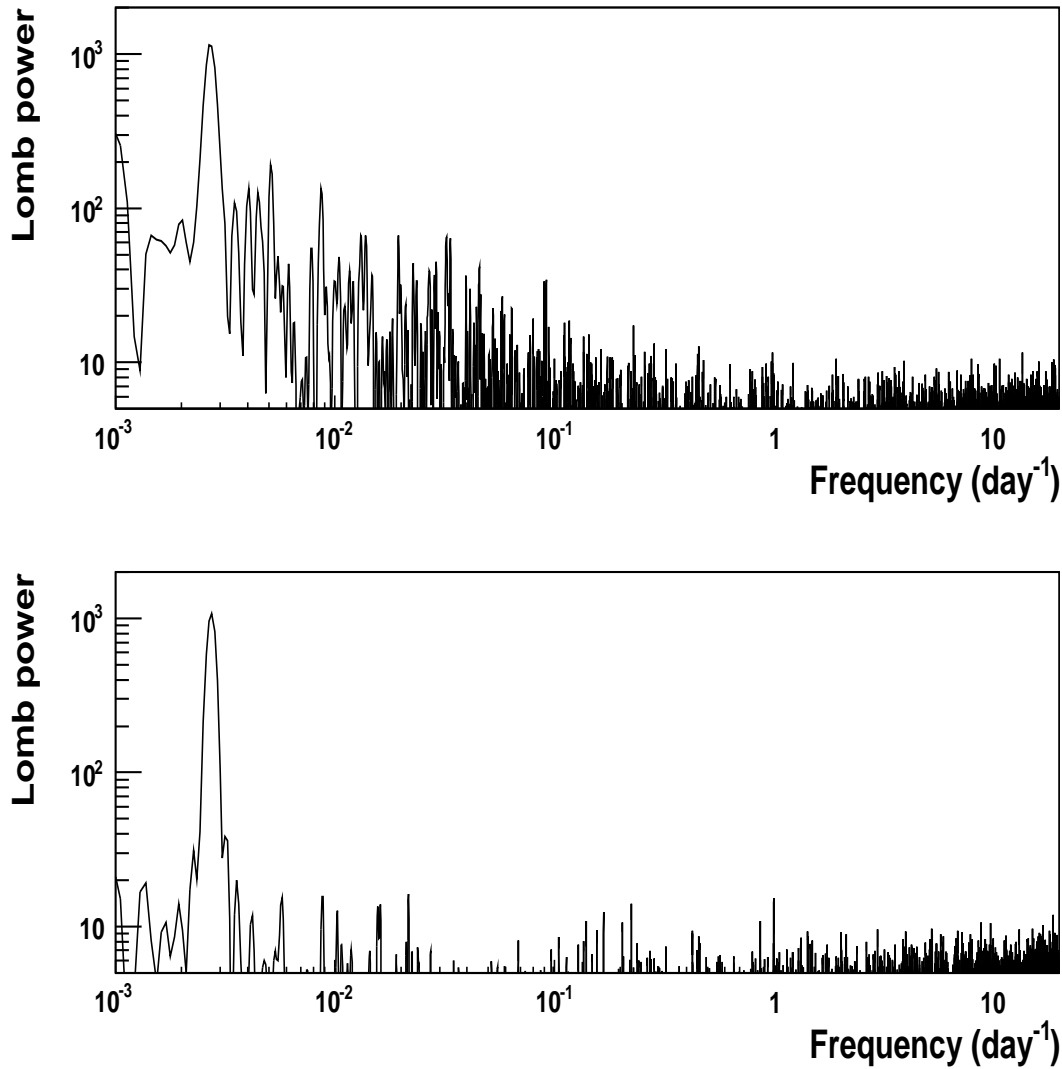}
\vskip -.5cm
\caption{\em \textbf{En haut, la puissance spectrale de Lomb en fonction de la fréquence [jour$^{-1}$] pour les données expérimentales. Le pic à $\sim 0.0027$ correspond à la variation saisonnière. En bas, nous présentons les résultats de la simulation en considérant un échantillon qui contient des périodicités de mêmes ordre de grandeur que les données réelles (saisonnière, journalière et sidérale)\cite{dtr10}.}}
\label{lomb_period}
\end{center}
\end{figure}

\begin{figure}
\begin{center}
\includegraphics[height=14cm,width=14cm]{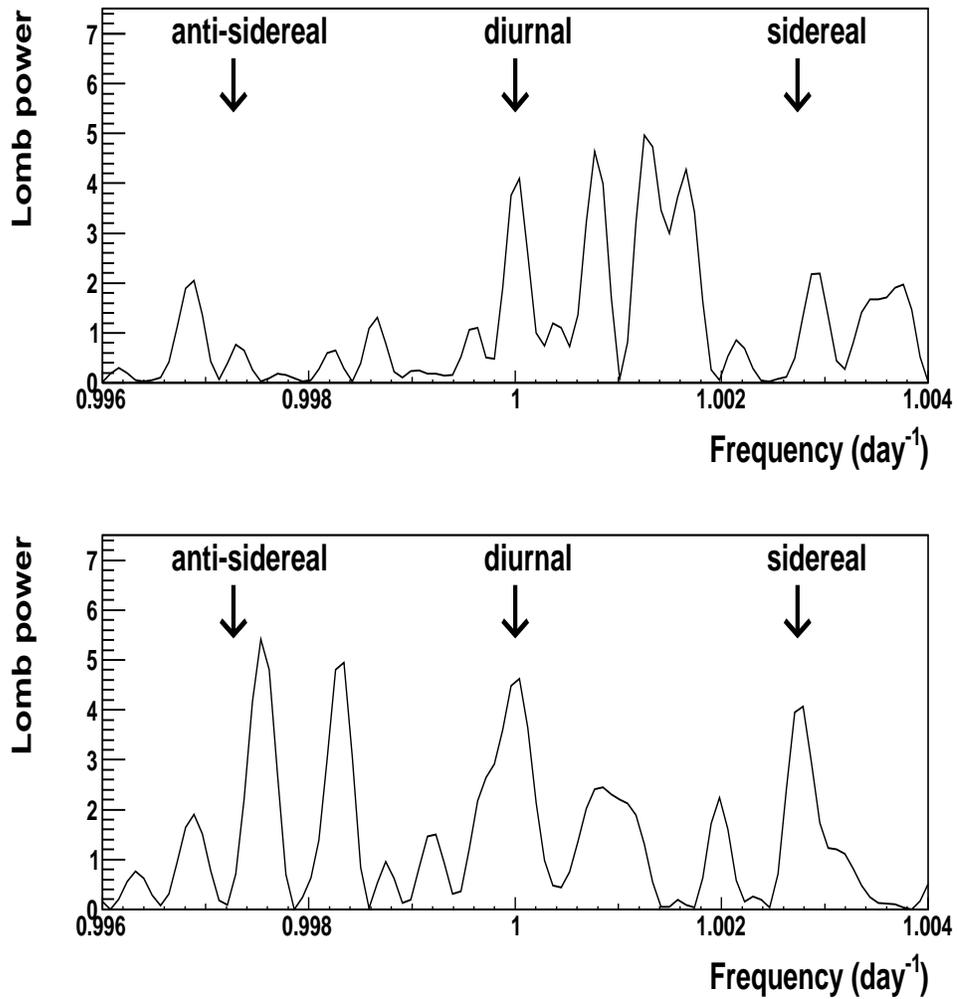}
\caption{\em \textbf{La région de fréquence autour de l'onde journalière pour les données réelles en (haut) et pour la simulation Monte Carlo (en bas). Les flèches montrent la position des pics journalier solaire, sidéral et anti-sedéral.}}
\label{Lomb_zoom}
\end{center}
\end{figure}

En l'absence de modulations dans les mécanismes de production et de propagation, le flux de rayons cosmiques, est distribué aléatoirement dans le temps.\\
L'étude des variations journalière du temps solaire et sidéral du flux des particules montre que le détecteur présente une haute sensibilité pour l'étude des très petites variations du flux de muons (de l'ordre de $10^{-3}$)\cite{dtr11}. Elle montre bien la grande sensibilité de MACRO même aux faibles variations du flux de muons. Sur la figure (\ref{fig:dt5}.b) on observe bien la présence d'une variation périodique due à la variation saisonnière du flux \cite{dtr10}.\\ 
Notre objectif est d'analyser par la méthode Lomb-Scargle les données de muons afin de chercher les modulations périodiques qui peuvent apparaitre.\\
Nous avons échantillonné les muons sur des intervalles de temps de 15 min. Les bins déviant par plus de $3\sigma$ par rapport au taux moyen mensuel ont été éliminés. Le nombre total des bins utilisé est 160242 correspondant à $58\%$ de la totalité de notre échantillon.\\
Les résultats de l'analyse Lomb-Scargle appliquée sur notre échantillon de données sont représentés sur la figure (\ref{lomb_period}). \\
Nous comparons le spectre obtenu par les données expérimentales à celui de la simulation Monte Carlo. Dans cette dernière,  nous avons ajouter des perturbations de même niveau que les données réelles ( à savoir les ondes saisonnière, journalière et sidérale) et des intervalles de temps choisit ont été distribués selon la séquence des séries originales \cite{dtr10}.\\
Sur la figure (\ref{lomb_period}) on remarque le grand pic à $\sim 0.0027$, il correspond à la variation saisonnière du flux.
La figure (\ref{Lomb_zoom}) montre la région de fréquence autour de la fréquence journalière solaire où nous avons indiqué les fréquences qui correspondent aux ondes sidérale et anti-sidérale, pour les données réelles (figure en haut) et pour la simulation Monte Carlo (figure en bas). Afin d'éliminer le bruit dans le spectre des puissances, nous avons considéré un échantillonnage, aussi bien pour les données réelles que pour les données simulées, selon la formule suivante :\\
\begin{equation}
N_i^\prime=\frac{N_i-\bar{N}(\Delta \tau)}{\bar{N}(\Delta \tau)}
\end{equation}
où $N_i$ est le contenu des bins originales et $\bar{N}(\Delta \tau)$ est le contenu moyen dans l'intervalle de temps $\pm \Delta \tau$ ($\Delta \tau$ a été choisit égale 1 jour). Le pic à la fréquence jour$^{-1}$ a une signification statistique de l'ordre de $\sim 2.3\ \sigma$. Le signal correspondant à la variation sidérale est observé, mais des pics avec une taille similaire (ou encore plus large) sont aussi présents dans le spectre. \\
Conclure que les ondes journalières solaire et sidérale sont réelles dépend essentiellement de la stabilité de leur amplitudes et leur phases avec le temps.\\
Après l'analyse des séries de temps avec la méthode de Lomb-Scargle, nous avons trouvé à nouveau la modulation saisonnière (figure \ref{lomb_period}), et des signaux aux positions des ondes journalières solaire et sidérale (figure \ref{Lomb_zoom}). Aucune déviation de l'hypothèse nulle n'a été signalé.
\section{Conclusion}
Pour l'analyse des séries de temps des muons, détectés par MACRO, deux approches complémentaires ont été considérées : la recherche de périodicité en utilisant la méthode Lomb-Scargle et la recherche de cluster d'évènement en utilisant la méthode scan statistics. Les deux méthodes complètent l'analyse de la distribution des temps d'arrivée des muons de haute énergie menée dans le chapitre \ref{chapdt}. Pour les deux approches aucune déviation de "l'hypothèse nulle" n'a été mise en évidence. Ce résultat confirme les conclusions du chapitre \ref{chapdt} et permet à nouveau de conclure sur la nature aléatoire des temps d'arrivée des muons.

\chapter{Étude de la perte d'énergie des nucléarites}
\section{introduction}
\label{chapnuc}
Proposée par Witten \cite{nuc1}, la matière nucléaire étrange "Strange Quark Matter" (SQM) présente un état de la chromodynamique quantique. C'est un agrégat de quarks u, d et s en nombre égale, entouré par un nuage d'électrons assurant la stabilité électrique et formant ainsi la dite "nucléarite". \par
Un grand intérêt a été apporté à l'étude des nucléarites, Bodmer \cite{nuc2} a suggéré que les nucléarites peuvent être plus stables que la matière nucléaire ordinaire. Farhi et Jaffe \cite{nuc3} ont montré que les nucléarites peuvent être stables sur un large domaine de masses, allant de celles des noyaux les plus légers (quelques quarks) jusqu'à celles des étoiles à neutrons ($\sim 10^{57}$ quarks), à conditions qu'ils se composent de quarks u, d et s en quantités approximativement égales. En effet, la taille des nucléarites peut dépasser largement celle des noyaux les plus massifs, car contrairement	à ces derniers, il n'y pas de limitation due à la répulsion coulombienne.\par
En partant de ces conditions, la matière étrange doit être extrêmement stable. Elle peut former des systèmes liés avec n'importe quel nombre de quarks, et peut être considérée comme un candidat à la matière obscure de l'univers \cite{nuc4}. Il est supposé que les nucléarites ont été produits aux premiers instants de l'univers, ($T > 200 MeV$ et $t < 10^6 s$) \cite{nuc5}.
La production des nucléarites, peut avoir lieu aussi dans les phénomènes astrophysiques violents se produisant au sein de l'univers dont on peut citer les collisions entre étoiles à neutrons et les explosions des étoiles \cite{nuc6}.\par 
La possibilité que les nucléarites bombardent la terre a été introduite par de Rùjula et Glashow \cite{nuc7}.\par
Dans ce qui suit, nous discuterons la perte d'énergie de ces particules et leur possibilité pour arriver aux différents niveaux de détection.
\section{Propriétés des nucléarites}
On peut approximer un nucléarite par une sphère de rayon R, dans laquelle les quarks u, d et s sont confinés dans un volume V avec une pression B dite pression de "bag". Les nucléarites sont décrits par le "bag model" (Modèle du Sac) \cite{nuc5} dans lequel les quarks représentent un gaz de Fermi dégénéré. La matière nucléaire (MN), peut être vue comme une association de plusieurs sacs maintenus ensemble par une énergie de liaison (à titre d'exemple le deuton se compose d'un sac uud et d'un sac udd). Cette idée peut être étendue à des sacs contenant plus de trois quarks dans un même sac, c'est la matière de quarks \cite{nuc1}. Lorsque la matière de quarks contient des quarks "strange" c'est la matière nucléaire étrange "SQM".\\
Il est supposé que la charge est distribuée de manière uniforme à l'intérieur du nucléarite et que l'équilibre "chimique" entre les quarks u, d, s et les électrons, est maintenu par les interactions faibles \cite{nuc5}
\begin{equation*}
d \rightarrow u + e + \bar{\nu_e}
\end{equation*} 
\begin{equation*}
u+e \rightarrow d + \nu_e
\end{equation*} 
\begin{equation*}
s \rightarrow u + e + \bar{\nu_e}
\end{equation*} 
\begin{equation*}
u+e \rightarrow s + \nu_e
\end{equation*} 
\begin{equation*}
s+u \longleftrightarrow u + d
\end{equation*} 
L'état du nucléarite est déterminé par le potentiel thermodynamique $\Omega_i (i = u,d,s)$. Ce dernier est fonction du potentiel chimique $\mu_i$, de la masse du quark $s (ms)$, de la pression de bag $B$ et du volume $V$.\par
\begin{figure}[ht]
\begin{center}
\mbox{\epsfig{file=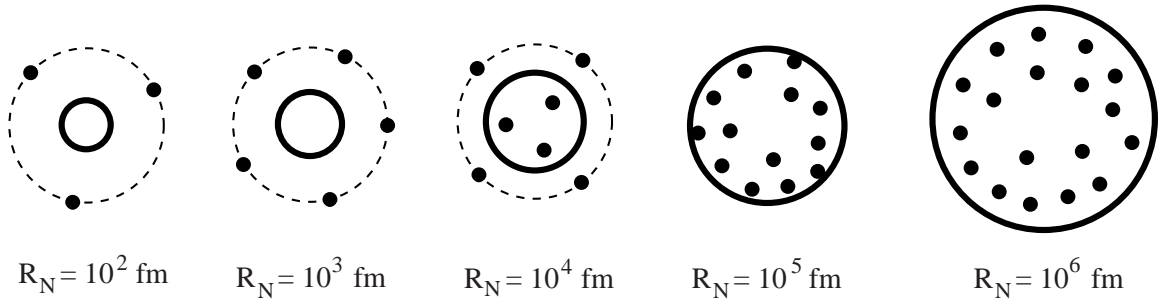,height=4cm}}
\end{center}
\vspace{-0.5cm}
\caption{\em \textbf{Dimensions du sac de quarks ($R_N$) et le système noyau+électrons (nuclearite). Les points indiquent les électrons, le bord du sac de quarks est indiqué par la ligne solide; le bord du système noyau+électron pour les faibles masses est indiqué par la ligne en pointillés.}}
\label{nucleariti}
\end{figure}

Les densités des espèces sont données par:\\
\begin{equation}
n_i = -\frac{\partial\Omega_i}{\partial\mu_i}
\end{equation}
et la densité baryonique
\begin{equation}
n = \frac{n_u+n_d+n_s}{3}
\end{equation}
Afin d'assurer l'équilibre du potentiel chimique des espèces de quarks, SQM doit avoir un nombre de quarks s un peu moins que le nombre des quarks u et d \cite{nuc5}. Cependant le noyau du nucléarite \footnote{On appelle noyau du nucléarite le système quarks sans cortège électronique} doit avoir une charge électrique positive compensée par le nombre d'électron du cortège électronique $N_e\cong (N_d+N_s)/3$, où $N_d$, $N_s$ et $N_e$ sont les nombres de quarks d, s et les électrons respectivement ( il a été considéré que $N_d=N_u$) \cite{nuc9}.\\
Soit $R_n$ le rayon du noyau du nucléarite, le système nucléarite aura un rayon constant $\sim 1$ \AA~ pour $R_N< 1$ \AA~ $=10^5$fm.\\ 
Pour $R_N \geq 10^{5}$ fm, tous les électrons sont à l'intérieure du sac de quark.\\
Pour $10^{4} < R_N< 10^{5}$ fm, une fraction des électrons est à l'intérieure du sac et une autre à l'extérieure, il donne au nucléarite dans ce cas une dimension globale de $\sim 10^{-8}$ cm $= 10^{5}$ fm. Dans ces conditions le nucléarite est similaire à l'atome de Bohr.\\
La figure (\ref{nucleariti}) illustre une distribution spatiale du système noyau+électrons. 
Notons que pour $R_N> 10^{5}$ fm, le nucléarite est considéré comme un atome de Thomson \cite{nuc9}.\\
\newpage 
Pour les nucléarites de masses supérieures à $1.5 \cdot 10^{-9}$ g $\simeq 10^{15}$ GeV/c$^2$, la relation entre la masse et le rayon prend la forme suivante \mbox{(masse $\propto$ volume, $M_N \propto V_N \propto R_N^3$):} 
\begin{equation}
R_N= \left( \frac {3M_N}{4 \pi \rho_N} \right)^{1/3} 
\label{eq:111}
\end{equation}
Pour $M_N=1.5 \cdot 10^{-9}$ g, le rayon du nucléarite devient :  
{\normalsize
\begin{equation}
R_N= \left( \frac {3}{4\pi} \frac{M_N}{3.5 \cdot 10^{14}} 
\right) ^{1/3} \simeq 10^{-8}cm =1~\mbox{\AA} 
\label{eq:2}
\end{equation}
}
Pour $M_N=10^{18}$ GeV/c$^2$, $R_N \simeq 10^5$ fm $(1000)^{1/3}=10^6$ fm. \par
\begin{figure}[h]
\vspace{-0.5cm}
\begin{center}
\mbox{\epsfysize=6.5cm \epsffile{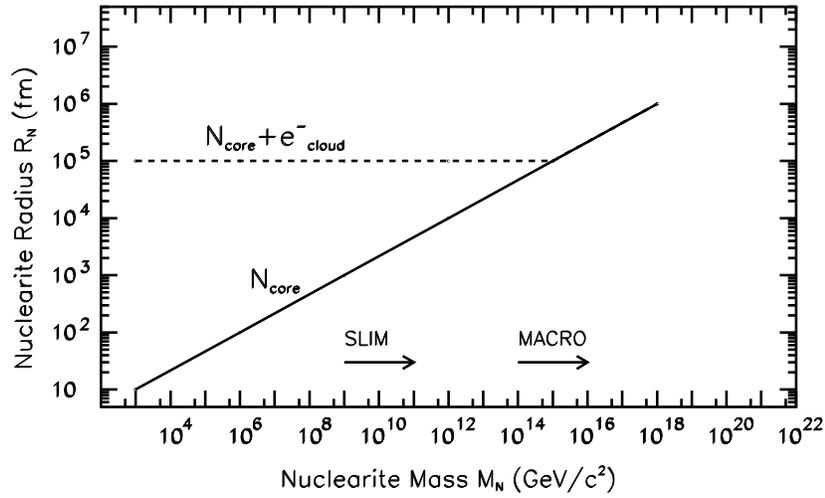}}
\end{center}
\vspace{-0.5cm}
\caption{\em \textbf{Rayon du nucléarite en fonction de la masse (modèle du sac de quarks) : ligne Solide. La ligne en pointillés donne le rayon du système noyau+électrons (nucléarite). Nous indiquons aussi les régions d'accessibilité en terme de masse par rapport aux expériences MACRO et SLIM.}}
\label{fig:r-vs-mass-nuc}
\end{figure}

Le rayon du noyau du nucléarite $R_N$ en fonction de sa masse $M_N$ est représenté sur la figure (\ref{fig:r-vs-mass-nuc}).\\
\section{Perte d'énergie des nucléarites}
Dans la plupart des travaux qui calculent la perte d'énergie des nucléarites dans divers milieux détecteurs, on ne tient pas compte de la contribution de l'atmosphère. Cette dernière peut être considérable dans le cas des expériences souterraines ou celles situées au niveau de la mer vu que la densité de l'atmosphère augmente tout en se rapprochant de la mer.\par
Dans notre travail, on a introduit la contribution de la perte d'énergie dans l'atmosphère afin de déterminer la vitesse minimale avec laquelle les nucléarites peuvent atteindre différents niveaux où sont installés les détecteurs, ainsi que la vitesse d'arrivée des nucléarites en considérant qu'ils atteignent la haute atmosphère avec une vitesse $\beta=10^{-3}$.  \par
La perte d'énergie des nucléarites traversant la matière est due principalement aux collisions élastiques ou quasi-élastiques avec les atomes du milieu \cite{nuc7}. Aux faibles vitesses, l'interaction des nucléarites se fait par des collisions élastiques avec les molécules du milieu.\par
Une partie des objets galactiques massifs non lumineux peut exister sous forme de nucléarites \cite{nuc7}.\par
La densité locale de la matière obscure est de l'ordre de $10^{-24} g/cm^3$ avec des vitesses d'échappement de la galaxie avoisinant 250 km/s. Le rapport de perte d'énergie des nucléarites ayant une telle vitesse est \cite{nuc7}: 
\begin{equation}
\frac{dE}{dx}=-A\rho v^2
\label{derujula}
\end{equation}
Où A est la section efficace effective du nucléarite, v sa vitesse et $\rho$ la densité du milieu traversé.\\ 
La section efficace géométrique des nucléarites est donnée par :
\begin{equation}
A=\pi R^2
\end{equation}
Comme il a été discuté au dessus, l'aire effective des nucléarites de faibles dimensions est contrôlée par son nuage électronique qui ne peut jamais être inférieur à $\sim 1~\mbox{\AA}$.\\
A faibles masses $(A\leq 10^7)$, le nuage électronique n'est pas à l'intérieur du nucléarite  \cite{nuc3}. Pour des masses $(10^7 \leq A \leq 10^{14})$ une partie du nuage électronique va se trouver à l'intérieur.\\
Pour des masses $A \geq 10^{14}$, tout le nuage électronique sera contenu à l'intérieur du nucléarite, ce dernier devient neutre (le nuage électronique fait écran aux charges positives internes de la particule).\\
La section effective d'interaction est donnée par \cite{nuc7} : 
\begin{equation}
A(cm^2)
\left\{
\begin{array}{rl}
\pi \times 10^{-16} & ~~~~~~\mbox{pour}~~M<~1.5 ng \\
\pi(\frac{3M}{4\pi \rho_N})^{2/3} & ~~~~~~ \mbox{pour}~~ M\geq~ 1.5 ng
\end{array}
\right.
\label{sect}
\end{equation}
Où $\rho_N$ est la densité de la matière nucléaire étrange, estimée être égale à $\approx 3.6\times10^{14}~g/cm^3$ \cite{nuc10}.\par
\begin{figure}
\centering
\includegraphics[height=10cm,width=10cm]{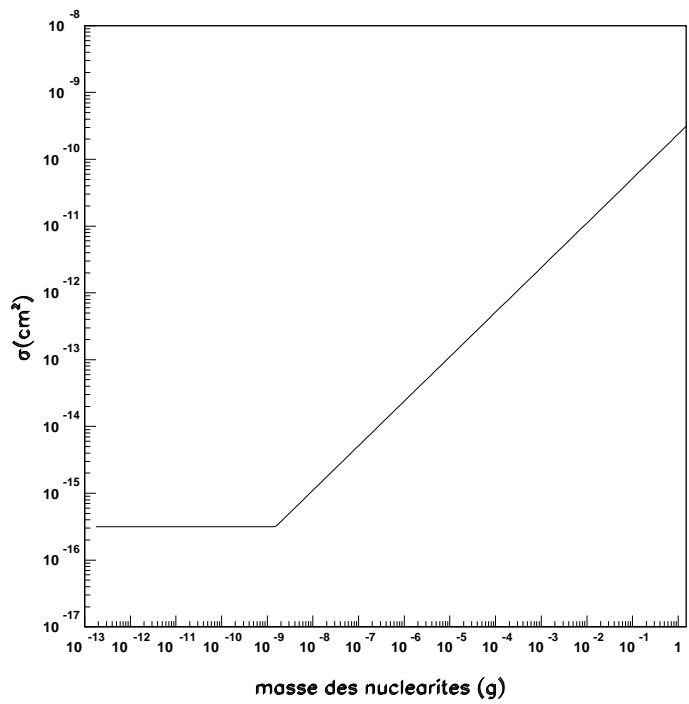}
\caption{\em \textbf{Section efficace d'interaction des nucléarites en fonction de leur masse.}}
\label{section}
\end{figure}

La section efficace d'interaction des nucléarites en fonction de leur masse est représentée sur la figure (\ref{section}).\par 
\begin{figure}
\centering
\includegraphics[height=10cm,width=10cm]{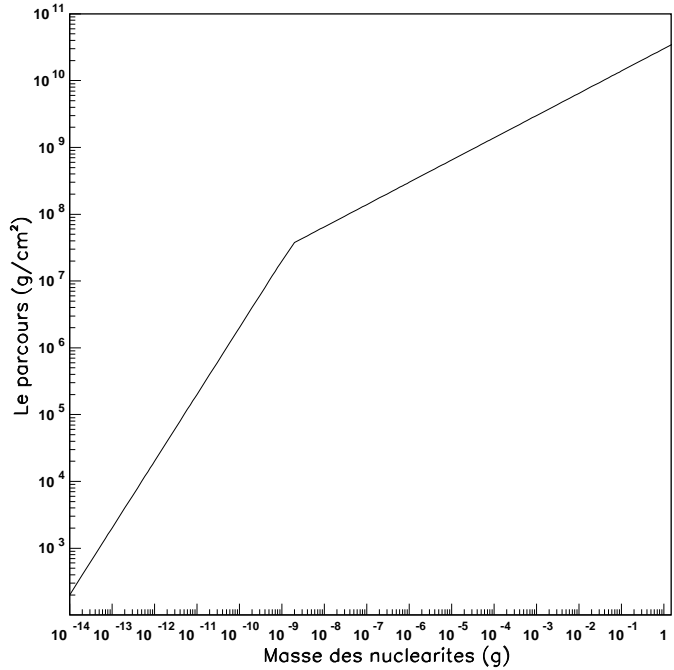}
\caption{\em \textbf{Parcours des nucléarites en fonction de leur masse.}}
\label{range}
\end{figure}

Selon l'équation (\ref{derujula}), la vitesse des nucléarites diminue de façon exponentielle, lors de la traversée de la matière selon l'équation suivante : 
\begin{equation}
v(L)=v_0~exp\left[-\frac{A}{M}\int_0^L\rho~dx\right]
\label{beta}
\end{equation}
Où L est la longueur traversée et $v_0$ la vitesse d'entrée dans l'atmosphère terrestre.\\
Ainsi le parcours R des nucléarites peut s'écrire sous la forme : 
\begin{equation}
R=\int_0^L\rho~dx=\left(\frac{M}{A}\right)Log\left(\frac{v_0}{v_c}\right)
\label{parcour}
\end{equation}
Où $v_c$ est la vitesse minimale pour traverser la longueur L.\\
Pour une vitesse $\beta\sim 10^{-3}$, le parcours est donnée par :  
\begin{equation}
R(g/cm^2)=
\left\{
\begin{array}{rl}
3\times 10^7~[M/1ng]^{1/3} & ~~~~~~ \mbox{pour}~~ M\geq~ 1.5 ng\\
2.3\times 10^7~[M/1ng]^{1/3} &~~~~~~\mbox{pour}~~M<~1.5 ng 
\end{array}
\right.
\label{portee}
\end{equation}
Sur la figure (\ref{range}) on présente le parcours en fonction de la masse des nucléarites. 
Il est à noter que les nucléarites ayant des masses $M\geq4.5\times10^{-14}g$ traversent l'atmosphère et arrivent à la surface de la terre, alors que ceux ayant des masses $M\geq0.1~g$ peuvent traverser la terre sans être arrêtés. \par
\section{Perte d'énergie des nucléarites dans la terre}
\subsection{Le modèle interne de la terre}
La composition exacte de la matière qui constitue la terre présente de grandes controverses \cite{MAC22}, par conséquent la perte d'énergie dans la terre ne peut être qu'approximative.\par
\begin{figure}
\centering
\includegraphics[height=10cm,width=10cm]{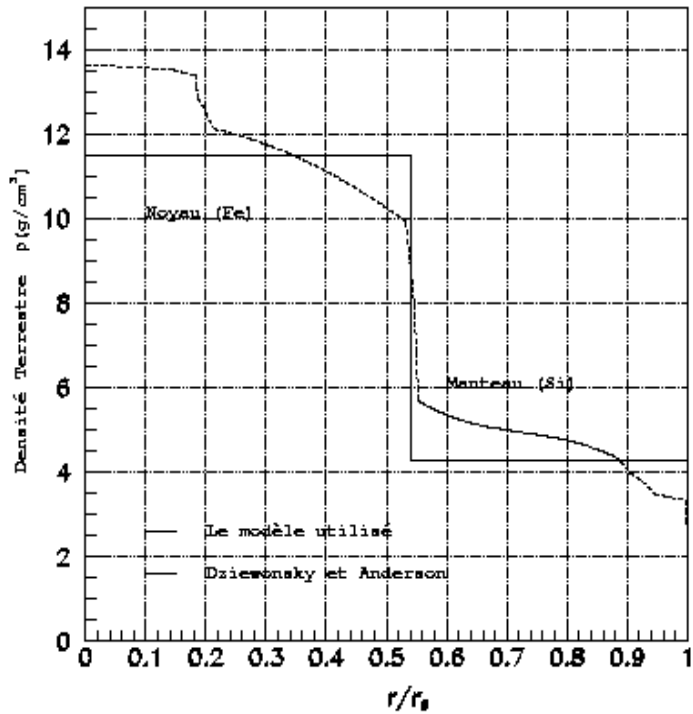}
\caption{\em \textbf{Profil de la densité terrestre est représenté par la courbe discontinue (voir Réf. \cite{MAC22}); La courbe en pointillés représente une approximation de la densité terrestre ($r_0=$ rayon de la terre).}}
\label{terre}
\end{figure}
Le manteau terrestre devrait être formé de dioxyde de silicium SiO$_2$. Le profil de la densité 
terrestre, que nous avons adopté pour notre étude, est représenté en pointillés sur la figure (\ref{terre}), où l'on peut observer trois parties : le noyau, le manteau et la croûte. Un modèle simple a été appliqué, dans lequel la densité et la composition de chaque partie est uniforme et elle est présentée en lignes continues sur la fig. (\ref{terre}) \cite{nuc12}. Dans ce modèle le noyau est constitué de fer, avec une densité $\rho_{fer}= 11.5 g/cm^3$  et une conductivité de $1.6\times 10^{16} s^{-1}$; le manteau est constitué de silicium avec une densité $\rho_{si} = 4.3 g/cm$. Le rayon du noyau constitue 0.54 fois le rayon terrestre.\par
\begin{figure}[ht]
\begin{center}
\mbox{\epsfig{file=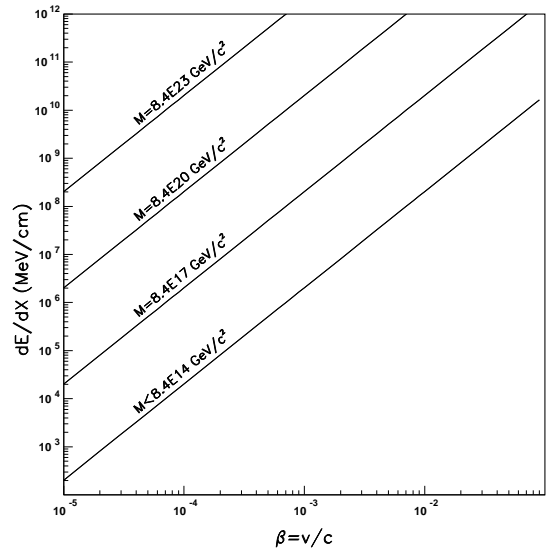,height=8cm}}
\end{center}
\vspace{-0.5cm}
\caption{\em \textbf{Perte d'énergie des nucléarites de différentes masses dans le noyau de la terre.}}
\label{nucleus}
\end{figure}
\begin{figure}[ht]
\begin{center}
\mbox{\epsfig{file=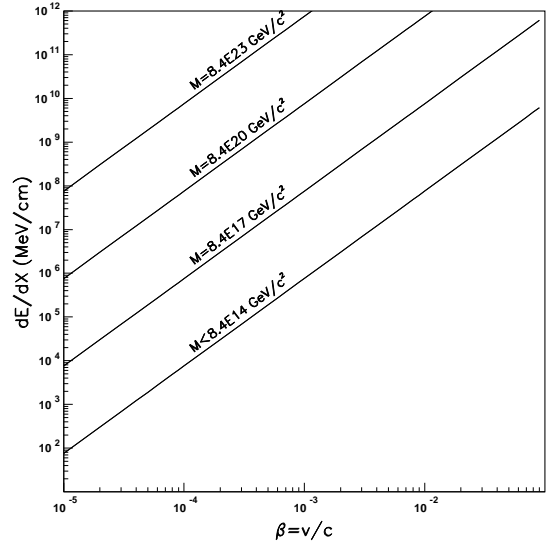,height=8cm}}
\end{center}
\vspace{-0.5cm}
\caption{\em \textbf{Perte d'énergie des nucléarites de différentes masses dans le manteau terrestre.}}
\label{mant}
\end{figure}

Les nucléarites perdent leur énergie dans la terre par des collisions élastiques avec les atomes du milieu. On représente la perte d'énergie des nuléarites de différentes masses dans le noyau de la terre sur les figures (\ref{nucleus}) et dans le manteau terrestre (\ref{mant}). 
\section{Perte d'énergie des nucléarites dans l'atmosphère}
La complexité du calcul de la perte d'énergie des nulcéarites dans l'atmosphère vient de la variation de la densité de l'atmosphère avec la profondeur. Ainsi la perte d'énergie dépend de cette dernière. Le calcul de la perte d'énergie se fait en se basant sur le fait que l'atmosphère est divisée en succession de couches de différentes épaisseurs et de densités.\par
\begin{figure}
\begin{center}
\mbox{\epsfig{file=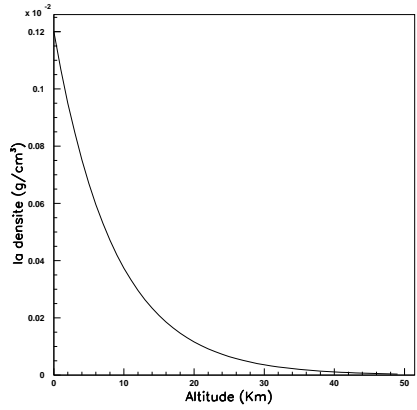,height=6cm}}
\end{center}
\vspace{-0.5cm}
\caption{\em \textbf{Variation de la densité de l'atmosphère en fonction de l'altitude.}}
\label{densite}
\end{figure}

\begin{figure}
\begin{center}
\mbox{\epsfig{file=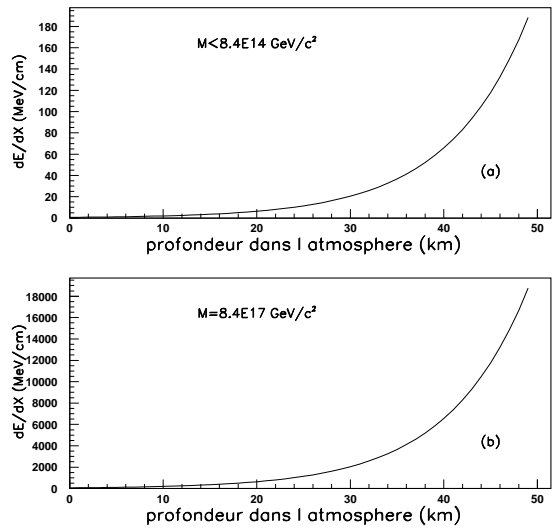,height=8cm}}
\end{center}
\vspace{-0.5cm}
\caption{\em \textbf{Perte d'énergie dans l'atmosphère des nucléarites de masse $M<8.4\ 10^{14} GeV/c^2$ (a) et $M=8.4\ 10^{17} GeV/c^2$ (b).}}
\label{loss_atm}
\end{figure}

 
Dans ce travail la densité de l'atmosphère a été calculée en utilisant la paramétrisation de Shibata \cite{nuc13}:
\begin{equation}
\rho(h)=ae^{-\frac{h}{b}}=ae^{-\frac{H-L}{b}}
\end{equation}
où les constantes $a=1.2\times 10^{-3}g.cm^{-3}$  et $b\simeq 8.57\times 10^5cm$, H est l'altitude totale de l'atmosphère ($\simeq 50~km$) et L la profondeur de pénétration dans l'atmosphère.\\
La variation de la densité de l'atmosphère en fonction de l'altitude est représentée sur la figure (\ref{densite}).\\
La perte d'énergie des nucléarites dans l'atmosphère est montrée sur la figure (\ref{loss_atm}) 
pour les masses $M=8.4\times10^{17}\;GeV/c^2$ et $M<8.4\times10^{14}\;GeV/c^2$.
\subsection{La région d'accessibilité des nucléarites}
\begin{figure}[ht]
\begin{center}
\mbox{\epsfig{file=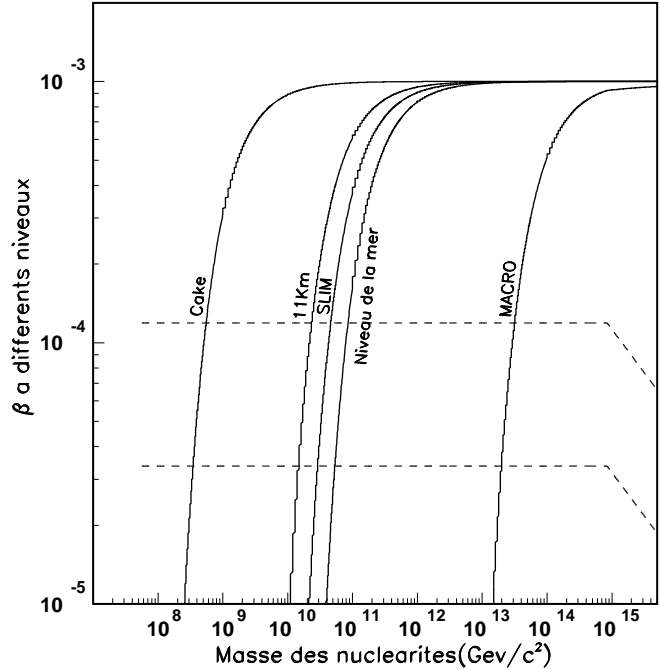,height=10cm}}
\end{center}
\vspace{-0.5cm}
\caption{\em \textbf{Vitesse d'arrivée des nucléarites à différentes profondeurs en fonction de la masse, en considérant que la vitesse d'arrivée à la haute atmosphère $\beta=10^{-3}$ : les lignes solides. Les lignes en pointillés représentent le seuil de détection dans le CR39 et dans le MAKROFOL.}}
\label{acc}
\end{figure}

Pour calculer la région d'accessibilité dans le plan (masse, $\beta$), l'intégration de l'équation (\ref{derujula}) a été résolue analytiquement : \\
\begin{equation}
\int_0^L\rho~dx= abe^{-\frac{H}{b}} \left[e^{\frac{H-h}{b}}-1\right]
\end{equation}
A partir de l'équation (\ref{beta}) on détermine la vitesse minimale que peut avoir les nucléarites de différentes masses pour atteindre les différents niveaux dans l'atmosphère.\\
Sur la figue (\ref{acc}) on montre la vitesse des nucléarites avec laquelle les différentes masses peuvent atteindre les niveaux correspondants à: 
\begin{itemize}
	\item l'expérience de ballon CAKE (40 km) \cite{MAC33},
	\item au possibles expériences portées sur des avions civils (11km),
	\item au laboratoire de Chacaltaya (l'expérience SLIM, 5.29 km) \cite{nuc15}\cite{nuc15r},
	\item au niveau de la mer, 
	\item l'expérience MACRO \cite{nuc16} est aussi inclue. Le seuil de détection dans le CR39 (correspondant à une perte 	d'énergie restreinte REL= $200~MeV~g^{-1}~cm^2$), et dans le MACROFOL (REL=$2500~MeV~g^{-1}~cm^2)$ est représenté.
\end{itemize}
Nous avons considéré les nucléarites ayant une vitesse initial $\beta=10^{-3}$ en haut de l'atmosphère. Un calcul pareil a été fait pour $\beta = 10^{-2},\:10^{-4},\:10^{-5}$ et n'a montré aucun effet de la vitesse initial sur le résultat obtenu. \\
La diminution de la vitesse seuil pour les nucléarites ayant une masse supérieure à $8.4\times10^{14}~GeV$ est due au changement de la section efficace selon l'équation (\ref{sect}).\\
Dans une expérience portée à l'altitude de Chacaltaya la masse minimale des nucléarites détectables s'abaisse d'un facteur 2 par rapport à d'autres expériences installées au niveau de la mer. \par
\begin{figure}[ht]
\begin{center}
\mbox{\epsfig{file=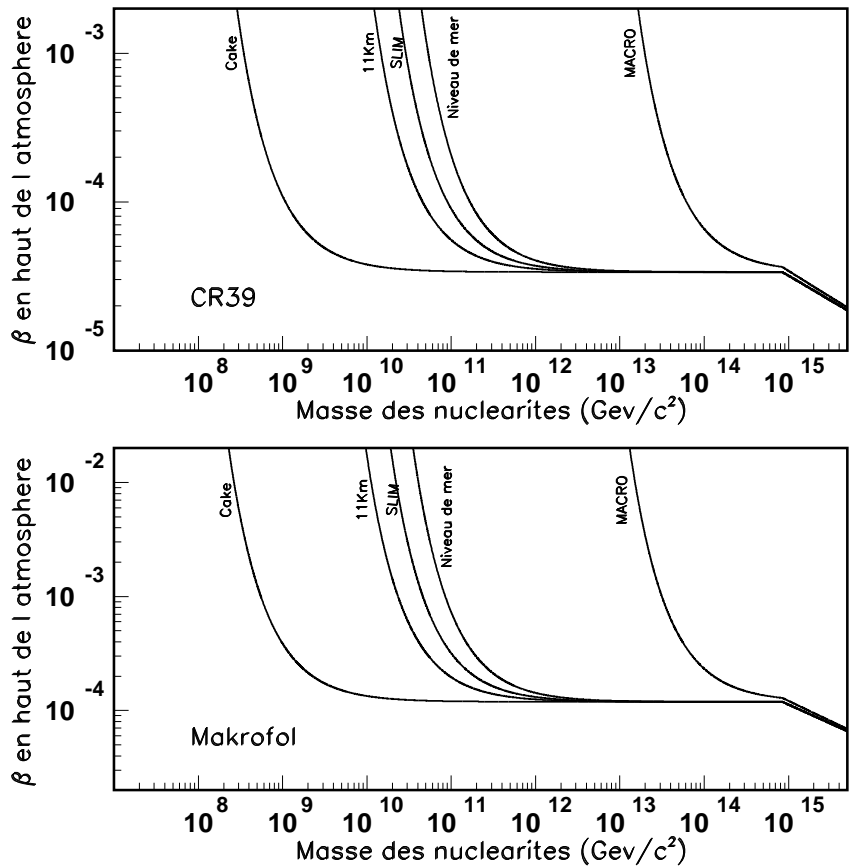,height=13cm}}
\end{center}
\vspace{-0.5cm}
\caption{\em \textbf{Conditions de détection dans le CR39 et le MAKROFOL pour différentes expériences situées à diverses altitudes.}}
\label{beta_min}
\end{figure}

\newpage
Les conditions de détection des nucléarites dans le CR39 sont exprimées par la vitesse d'entrée minimale en haut de l'atmosphère ($50~Km$) en fonction de la masse des nucléarites. Pour les différentes locations d'expériences, on représente sur la figure (\ref{beta_min}) la vitesse d'entrée minimale en fonction de la masse. Dans ce cas, la contrainte est que le nucléarite possède une vitesse minimale au niveau du détecteur afin qu'il puisse produire une trace; notons que pour toutes les expériences le seuil REL de détection est pris égale à $200~MeV~g^{-1}~ cm^2$ pour le CR39 et $250~MeV~g^{-1}~ cm^2$ pour le MAKROFOL. Le changement de l'allure de la courbe est du à la variation de la section efficace en fonction de la masse selon l'équation (\ref{sect}). 
\section{Conclusion}
En se basant sur la formule de De Rûjula (\ref{derujula}) relative à la perte d'énergie des nucléarites et en utilisant la modélisation de Shibata pour le calcul de la densité de l'atmosphère, nous avons réussi à évaluer la perte d'énergie des nucléarites le long de leur parcours dans l'atmosphère. Pour les différentes expériences réalisées à différentes altitudes dans l'atmosphère (CAKE (40km), expériences portées sur des avions civil (11km), SLIM (5.29km), expériences situées au niveau de la mer) nous avons déterminé la perte d'énergie dans le plan des paramètres (masse,$\beta$). En comparant le résultat obtenu dans la figure (\ref{acc}) avec celui du \cite{nuc17}, on remarque une différence de masse d'un facteur de $\approx 3$. Ceci est dû à la façon avec laquelle a été calculée la densité de l'atmosphère.\\
Les nucléarites ayant une masse $>7\:10^{12} GeV/c^2$ ne sentent pas la présence de l'atmosphère. Au dessous de cette masse, et pour une expérience portée à une altitude de 5.29 Km (l'expérience SLIM ), le minimum de masse détectable décroît d'un facteur 2 par rapport aux expériences installées au niveau de la mer. \\
Pour l'expérience souterraine MACRO (3800 m.w.e), les nucléarites de masse supérieure à $1.53\;10^{16} GeV/c^2$ (figure \ref{acc}) ne subissent aucune perte d'énergie, leur vitesse $\beta$ est intacte. Par contre le paramètre $\beta$ décroît exponentiellement pour des masses dans l'intervalle $[1.6\;10^{13},6\;10^{15}]GeV/c^2$. Le CR39 du détecteur MACRO est sensible aux nucléarites de masses $>2\;10^{13}GeV/c^2$ pour lesquelles $3.37\;10^{-4}<\beta<0.9\;10^{-3}$. Dans le cas du MAKROFOL, MACRO est sensible aux nucléarites de masses $>3\;10^{13}GeV/c^2$ pour lesquelles $0.12\;10^{-3}<\beta<0.9\;10^{-3}$).\\
La contribution de l'atmosphère dans le calcul de la perte d'énergie devient considérable pour les nucléarites avec des masses $< 7\:10^{12}GeV/c^2$ et ayant une vitesse $\beta = 10^{-3}$ en haut de l'atmosphère.	

\addcontentsline{toc}{chapter}{Conclusion}
\chapter*{Conclusion}
Le travail que nous avons mené au près de l'expérience MACRO s'est articulé autour de la recherche des corrélations possibles dans le temps d'arrivée des muons cosmiques ainsi que sur la recherche des variations dans le flux de muons. \par
Nous disposons d'un lot de données d'environ $38 \;millions$ de muons d'énergie supérieure à 1.3 TeV au sommet de la montagne du Gran Sasso, collecté à l'aide des tubes à streamer du détecteur MACRO dans sa configuration complète.\par
L'analyse des distributions des temps d'arrivées des muons singuliers, doubles et multiples arrivant de toutes les directions du ciel ainsi que ceux venant des zones sélectionnés formant des cônes, montre qu'elles sont compatibles avec la fonction gamma. Ce qui confirme l'hypothèse d'arrivée aléatoire des RC. Ce résultat a été appuyé par le test de Kolmogorov-Smirnov, dont aucune déviation entre la distribution théorique et celle expérimentale n'a été mise en évidence.\\   
Nos résultats ne sont pas ainsi analogue à ceux obtenus par Bath et al. \cite{dtr7}, ou par Badino et al. \cite{dtr8}, et ceci pour des énergies de l'ordre de 20 TeV ($\mu$ singuliers) et supérieures à 20TeV ($\mu$ doubles et multiples). \par
La recherche de cluster d'événements, dans le flux de muons, par la méthode scan statistics n'a présenté aucune déviation significative de l'hypothèse nulle. Ce qui nous a permis de confirmer l'absence d'éventuel cluster d'événements dans le flux de muons et donc aucune variation du flux n'a été signalé. \\
En analysant les séries de temps des muons par la recherche des périodicités avec la méthode Lomb-Scargle, nous avons observé la modulation saisonnière et des signaux à la position des ondes journalière solaire et sidérale. Aucune autres déviation n'a été signalé. En se basant sur la formule de De Rûjula (\ref{derujula}) relative au calcul de la perte d'énergie des nucléarites et en utilisant la modélisation de Shibata pour le calcul de la densité de l'atmosphère, la contribution de la perte d'énergie dans l'atmosphère a été introduite afin de déterminer la vitesse minimale avec laquelle les nucléarites peuvent atteindre différents niveaux où sont installés les détecteurs. \\
La contribution de l'atmosphère, pour les nucléarites ayant une vitesse $\beta =10^{-3}$ en haut de l'atmosphère, devient considérable pour les masses $<7\:10^{12} GeV/c^2$. Au dessous de cette masse, et pour une expérience portée à une altitude de 5.29 Km (l'expérience SLIM ), la masse minimale détectable décroît d'un facteur 2 par rapport aux expériences installées au niveau de la mer. Pour les expériences souterraines, les masses supérieures à $1.53\;10^{16} GeV/c^2$ ne subissent aucune perte d'énergie.

\evenheading{\hfill Bibliographie}%

\pagebreak
\thispagestyle{empty}%

\vspace{50pt}%
\begin{center}
{\LARGE {Résumé}}
\end{center}
\bigskip
\null
\rule{\textwidth}{2pt}%
\vskip 10 pt%
La radiation cosmique est une composante principale de la galaxie, vu que sa densité d'énergie est comparable, si elle n'est pas supérieure, à celle des autres radiations présentes dans l'univers. Elle joue le rôle des messagers de mécanismes astrophysiques pouvant mettre en jeu des énergies colossales. Elle est la seule matière qui arrive de l'extérieure de notre système solaire et elle constitue l'unique moyen pour l'étude des particules de très haute énergie allant jusqu'à 1019 eV. \par
Parmi les questions de recherche qui présentent un intérêt fondamental, on distingue l'étude de l'anisotropie temporelle du flux de muons cosmiques. Des modulations dans les distributions temporelles du temps d'arrivée de muons ont été observées par quelques expériences. Les résultats de ces expériences n'ont pas été confirmés par d'autres groupes et le problème reste confus. \par
A cet effet, la grande partie de cette thèse a été consacrée à l'étude du caractère aléatoire du temps d'arrivée des rayons cosmiques à travers l'étude des variations du flux des mouns cosmiques collecté avec le détecteur souterrain MACRO.  
Deux techniques ont été utilisées :
\begin{itemize}
	\item L'étude des distributions des temps d'arrivée des muons cosmiques. 
	\item L'étude des variations du flux de muons en utilisant deux approches :
	\begin{enumerate}
		\item méthode Scan-statics.
		\item méthode de Lomb-Scargle.
	\end{enumerate}
\end{itemize}
En s'appuyant sur les deux études cité au-dessus, la recherche de possible corrélations temporelles ou de périodicité dans le flux de muons cosmique de MACRO, nous a permis de confirmer l'hypothèse d'arrivée aléatoire des rayons cosmiques.\par 
La dernière partie de cette thèse a été consacrée à l'étude de la perte d'énergie des nucléarites et leur possibilité pour arriver aux différents niveaux de détection dans l'atmosphère. La contribution de l'atmosphère dans la perte d'énergie pour les nucléarites ayant une vitesse $\beta=10^{-3}$ en haut de l'atmosphère, devient considérable pour les masses $<7\;10^{12}GeV/c^2$. Au dessous de cette masse, et pour une expérience portée à une altitude de 5.29 Km (l'expérience SLIM ), la masse minimale détectable décroît d'un facteur 2 par rapport aux expériences installées au niveau de la mère. Pour les expériences souterraines, les masses supérieures à $1.53\;10^{16} GeV/c^2$ ne subissent aucune perte d'énergie.\\

\end{document}